\documentclass[twocolumn,aps,prd,floatfix,preprintnumbers,a4paper,nofootinbib,superscriptaddress]{revtex4-1}
\usepackage{graphicx}				% Use pdf, png, jpg, or eps§ with pdflatex; use eps in DVI mode
\usepackage{placeins}
\usepackage{amssymb}
\usepackage{amsmath}
\usepackage[dvipsnames]{xcolor}

\usepackage{verbatim}
\usepackage{rotating}
\usepackage{comment}

\usepackage[colorlinks]{hyperref}

\begin{document}

\newcommand\red[1]{{\color[rgb]{0.75,0.0,0.0} #1}}
\newcommand\green[1]{{\color[rgb]{0.0,0.60,0.08} #1}}
\newcommand\blue[1]{{\color[rgb]{0,0.20,0.65} #1}}
\newcommand\cyan[1]{{\color[HTML]{00c3ff} #1}}
\newcommand\bluey[1]{{\color[rgb]{0.11,0.20,0.4} #1}}
\newcommand\gray[1]{{\color[rgb]{0.7,0.70,0.7} #1}}
\newcommand\grey[1]{{\color[rgb]{0.7,0.70,0.7} #1}}
\newcommand\white[1]{{\color[rgb]{1,1,1} #1}}
\newcommand\darkgray[1]{{\color[rgb]{0.3,0.30,0.3} #1}}
\newcommand\orange[1]{{\color[rgb]{.86,0.24,0.08} #1}}
\newcommand\purple[1]{{\color[rgb]{0.45,0.10,0.45} #1}}
\newcommand\note[1]{\colorbox[rgb]{0.85,0.94,1}{\textcolor{black}{\textsc{\textsf{#1}}}}}
\definecolor{brown-ish}{RGB}{201,130,20}
% ***************************************** %
\def\gw#1{gravitational wave#1}
% ***************************************** %
\def\nr#1{numerical relativity
 (NR)#1\gdef\nr{NR}}
 % ***************************************** %
\def\bh#1{black-hole
 (BH)#1\gdef\bh{BH}}
 % ***************************************** %
 \def\bbh#1{binary black hole#1
  (BBH#1)\gdef\bbh{BBH}}
% ***************************************** %
  \def\Bbh#1{Binary black hole#1
   (BBH#1)\gdef\bbh{BBH}}
% ***************************************** %
% quadrupole-aligned (QA)
\def\qa#1{quadrupole-aligned#1
(QA#1)\gdef\qa{QA}}
% ***************************************** %
%\def\oed#1{optimal emission direction#1}
\def\oed#1{coprecessing frame#1}
% ***************************************** %
\def\pn#1{post-Newtonian
 (PN)#1\gdef\pn{PN}}
% ***************************************** %
 \def\qnm#1{Quasinormal Mode
    (QNM)#1\gdef\qnm{QNM}}
% ***************************************** %
   \def\eob#1{effective-one-body
      (EOB)#1\gdef\eob{EOB}}
% ***************************************** %
\def\imr#1{inspiral-merger-ringdown
 (IMR)#1\gdef\imr{IMR}}
 % ***************************************** %
\def\msa#1{multi-scale analysis
 (MSA)#1\gdef\msa{MSA}}
 % ***************************************** %
 \def\Fig#1{Figure~\ref{#1}}
 % ***************************************** %
 \def\fig#1{Fig.~\ref{#1}}
 % ***************************************** %
 \def\cfig#1{Fig.~\ref{#1}}
 % ***************************************** %
 \newcommand{\figs}[2]{Figs.~\ref{#1}-\ref{#2}}
 \newcommand{\Figs}[2]{Figures~\ref{#1}-\ref{#2}}
 % ***************************************** %
 \def\eqn#1{Eq.~(\ref{#1})}
 % ***************************************** %
 \def\ceqn#1{Eq.~\ref{#1}}
 % ***************************************** %
 \newcommand{\Eqns}[2]{Equations~(\ref{#1})-(\ref{#2})}
 % ***************************************** %
 \newcommand{\eqns}[2]{Eqs.~(\ref{#1})-(\ref{#2})}
 % ***************************************** %
 \newcommand{\ceqns}[2]{Eqs.~\ref{#1}-\ref{#2}}
 % ***************************************** %
\def\sect#1{Sec.~\ref{#1}}
% \def\sect#1{Sec.~(\ref{#1})}
% ***************************************** %
\def\csec#1{Sec.~\ref{#1}}
% ***************************************** %
\newcommand{\csecs}[2]{Secs.~\ref{#1}-\ref{#2}}
% ***************************************** %
\def\tbl#1{Table~(\ref{#1})}
% ***************************************** %
\def\ctbl#1{Table~\ref{#1}}
% ***************************************** %
\def\Apx#1{Appendix~\ref{#1}}
% ***************************************** %
\def\apx#1{Appendix~\ref{#1}}
% ***************************************** %
\def\capx#1{Appendix~\ref{#1}}
% ***************************************** %
%
 % ***************************************** %
 \def\lm{{\ell m}}
 \def\tpsi{\tilde{\psi}}
 % ***************************************** %
 % IMRPhenomXP
\def\xp{{\sc{PhenomXP}}}
% SEOBNRv4P
\def\eobnr{{\sc{SEOBNRv4P}}}
% Surrogate 
\def\nrsur{{\sc{NRSur7dq4}}}
% PhenomPv3 
\def\pv3{{\sc{PhenomPv3}}}
% PhenomDCP 
\def\dcp{{\sc{PhenomDCP}}}
% PhenomD
\def\d{{\sc{PhenomD}}}
% Final Model Name (Currently PhenomPNR)
\def\pnr{{\sc{PhenomPNR}}}
% Macro for theta_LS
\newcommand{\tls}{\theta_{\mathrm{LS}}}
\def\fdamp{f_1}
\def\fring{f_0}
\def\J{\vec{J}(t)}
\newcommand{\mlam}{{\boldsymbol{\lambda}}}

\newcommand{\ms}{\text{s}}
\newcommand{\mt}{\text{t}}

\newcommand{\Cardiff}{School of Physics and Astronomy, Cardiff University, Cardiff, CF24 3AA, United Kingdom}
\newcommand{\Zurich}{Physik-Institut, Universit\"at Z\"urich, Winterthurerstrasse 190, 8057 Z\"urich, Switzerland}
\newcommand{\MIT}{MIT-Kavli Institute for Astrophysics and Space Research and LIGO Laboratory, 77 Massachusetts Avenue, 37-664H, Cambridge, MA 02139, USA}
\newcommand{\Rome}{Dipartimento di Fisica, Universit\`{a} di Roma ``Sapienza'', Piazzale A. Moro 5, I-00185, Roma, Italy}
\newcommand{\INFN}{INFN, Sezione di Roma, Piazzale A. Moro 5, I-00185, Roma, Italy}
\newcommand{\Porto}{Centro de Astrof\'{\i}sica e Gravita\c c\~ao  - CENTRA, Departamento de F\'{\i}sica, Instituto Superior T\'ecnico IST, Universidade de Lisboa UL, Avenida Rovisco Pais 1, 1049-001 Lisboa, Portugal}
\newcommand{\Nikhef}{Nikhef -- National Institute for Subatomic Physics, Science Park, 1098 XG Amsterdam, The Netherlands}
\newcommand{\Utrecht}{Institute for Gravitational and Subatomic Physics (GRASP), Utrecht University, Princetonplein 1, 3584 CC Utrecht, The Netherlands}
\newcommand{\Amsterdam}{Institute for High-Energy Physics, University of Amsterdam, Science Park 904, 1098 XH Amsterdam, The Netherlands}

\title{The final twist: \\ A model of gravitational waves from precessing black-hole binaries through merger and ringdown}

\author{Eleanor Hamilton} \affiliation{\Cardiff}\affiliation{\Zurich}
\author{Lionel London} \affiliation{\MIT}
\author{Jonathan E. Thompson} \affiliation{\Cardiff}
\author{Edward Fauchon-Jones} \affiliation{\Cardiff}
\author{Mark Hannam} \affiliation{\Cardiff}
\author{Chinmay Kalaghatgi} \affiliation{\Nikhef}\affiliation{\Utrecht}\affiliation{\Amsterdam}
\author{Sebastian Khan} \affiliation{\Cardiff}
\author{Francesco Pannarale} \affiliation{\Rome}\affiliation{\INFN}
\author{Alex Vano-Vinuales} \affiliation{\Porto}

\begin{abstract}
 We present \pnr{}, a frequency-domain phenomenological model of the gravitational-wave (GW) signal from binary-black-hole 
 mergers that is tuned to \nr{} simulations of precessing binaries. In many current waveform models, 
 e.g., the ``\textsc{Phenom}'' and ``\textsc{EOBNR}'' families that have been used extensively to analyse LIGO-Virgo GW observations, 
 analytic approximations are used to add precession effects to models of non-precessing (aligned-spin) binaries, and it is only 
 the aligned-spin models that are fully tuned to NR results. In \pnr{} we incorporate precesing-binary \nr{} results in two ways: 
 (i) we produce the first \nr{}-tuned model of the signal-based precession dynamics through merger and ringdown, and 
 (ii) we extend a previous aligned-spin model, \d{}, to include the effects of misaligned spins on the signal in the co-precessing
 frame. 
 The \nr{} calibration has been performed on 40 simulations of binaries with mass ratios  between 1:1 and 1:8, where the larger 
 black hole has a dimensionless spin magnitude of 0.4 or 0.8, and we choose five angles of spin misalignment with the orbital angular momentum.
\pnr{} has a typical mismatch accuracy within 0.1\% up to mass-ratio 1:4, and within 1\% up to mass-ratio 1:8. 
\end{abstract}

\maketitle

\section{Introduction}

\Bbh{} mergers are the primary source of \gw{s} observable with current ground-based 
detectors~\cite{LIGOScientific:2014pky, VIRGO:2014yos}; of the 51 detections published by the LIGO-Virgo collaborations, 48 were confirmed as
\bbh{}~\cite{LIGOScientific:2018mvr, Nitz_2020, Zackay:2019btq, Venumadhav:2019lyq, Abbott:2020niy}. 
Measurements of each binary's properties --- the 
\bh{} masses and spins, and the location of the binary --- rely in part on models of the signal predicted by general
relativity. Model development is an active research area, with the aim that the measurement uncertainties due to 
model errors, approximations, and incomplete physics are smaller than statistical errors arising from the strength of the signal above
the detector noise, or parameter degeneracies. Models are informed by analytic approximations for the inspiral of the two \bh{s} and
ringdown of the final \bh{}, and \nr{} solutions of Einstein's equations for the late inspiral, merger and ringdown. 
One key physical effect is the precession of the binary's orbital plane due predominantly to spin-orbit effects, but the two waveform 
families most commonly used for LIGO-Virgo parameter estimates, 
``\textsc{Phenom}''~\cite{Husa:2015iqa,Khan:2015jqa,Hannam:2013oca,London:2017bcn,Khan:2018fmp,Khan:2019kot,Pratten:2020fqn,Garcia-Quiros:2020qpx,Pratten:2020ceb,Thompson:2020nei,Estelles:2020osj,Estelles:2020twz} 
and ``\textsc{EOBNR}''~\cite{Taracchini:2012ig,Pan:2013rra,Taracchini:2013rva,Bohe:2016gbl,Cotesta:2018fcv,Ossokine:2020kjp,Matas:2020wab},
have \emph{not} been tuned to \nr{} simulations of precessing binaries. Instead, precession effects during the strongest part of the
signal have been estimated using simple approximations. These were likely sufficient for observations to date, but, given that they
do not capture several physical features of the merger signal (e.g., Ref.~\cite{Ramos-Buades:2020noq}, plus other effects that we will 
describe in this paper) more accurate models will ultimately be required. 

Here we present the first \textsc{Phenom} model where merger-ringdown precession effects are explicitly tuned to \nr{} simulations.
We show that this model is in general significantly more accurate than previous models, particularly for binaries with large mass ratios,
high spins, and a large spin misalignment.

A \bbh{} system following non-eccentric inspiral is defined by the \bh{} masses, $m_1$ and $m_2$ (we choose $m_1>m_2$),
and the \bh{} spin-angular-momentum vectors $\mathbf{S}_1$ and $\mathbf{S}_2$. As is standard, we choose the alternative 
parameterisation into total mass, $M = m_1 + m_2$, symmetric mass ratio $\eta = m_1 m_2 / M^2$, and the dimensionless spins
$\boldsymbol{\chi}_i = \mathbf{S}_i / m_i ^2$, where $|\boldsymbol{\chi}_i| \in \left[0,1\right]$ respects the Kerr limit. 
It is also convenient to decompose the spins into 
their components parallel and perpendicular to the direction of the Newtonian orbital angular momentum, $\hat{\mathbf{L}}$,
i.e., the magnitudes of the spins parallel to $\mathbf{L}$ are $\chi_i^\parallel = \boldsymbol{\chi}_i \cdot \hat{\mathbf{L}}$, and the 
components that lie in the orbital plane are $\boldsymbol{\chi}_i^\perp = \boldsymbol{\chi}_i - \chi_i^\parallel \hat{\mathbf{L}}$.  

If the spins are parallel to the orbital angular momentum, i.e., $\boldsymbol{\chi}_i^{\perp} = 0$, then the orientation of the binary's orbital plane, 
and the directions of the spin and orbital angular momenta, are all fixed. Waveforms from these aligned-spin, or non-precessing, binaries,
have been modelled with a combination of \pn{} and \eob{} results to describe the insipiral, and
\nr{} results to model the late inspiral, merger and ringdown, to produce \textsc{Phenom} and \textsc{EOBNR}
waveform models~\cite{Husa:2015iqa,Khan:2015jqa,Pratten:2020fqn,Garcia-Quiros:2020qpx,Estelles:2020osj,Estelles:2020twz,Taracchini:2012ig,Bohe:2016gbl}. 
Surrogate models of non-precessing systems have also been constructed purely from \nr{} waveforms, and also from 
\pn{}-\nr{} hybrids~\cite{Blackman:2017dfb,Varma:2018mmi}. 

When $\boldsymbol{\chi}_i^{\perp} \neq 0$, the binary precesses. In most cases the binary undergoes simple 
precession~\cite{Apostolatos:1994mx,Kidder:1995zr}, where the orbital angular 
momentum and spins precess around the binary's total angular momentum, which points in an approximately fixed direction. Precession modulates
the amplitude and phase of the gravitational-wave signal, and leads to a significantly more complicated signal than in 
non-precessing configurations.
However, if we transform to a non-inertial co-precessing frame that tracks the precession, then the signal 
recovers, to a good approximation, the simple form of a non-precessing signal~\cite{Schmidt:2010it}, and, indeed, during the inspiral the 
co-precessing-frame waveform is approximately the signal from the corresponding non-precessing binary defined by setting 
$\boldsymbol{\chi}_i^{\perp} = 0$~\cite{Schmidt:2012rh}. 

This observation has been used to construct current \textsc{Phenom} and \textsc{EOBNR} waveform models, by using a non-precessing 
model as a proxy for the precessing-binary waveform in the co-precessing frame, and then transforming this to the inertial frame via an 
independent model for the precession 
dynamics~\cite{Hannam:2013oca,Khan:2018fmp,Khan:2019kot,Pratten:2020ceb,Pan:2013rra,Taracchini:2013rva,Ossokine:2020kjp}. 
Although some \nr{} information from precessing-binary simulations has been used to model the final state~\cite{Ossokine:2020kjp}, 
the precession effects have \emph{not} been tuned to \nr{} waveforms, and neither
have in-plane-spin contributions to the co-precessing-frame signal. In addition to these models, surrogate models of precessing
binaries have been constructed using \nr{} waveforms that cover roughly 20 orbits 
before merger~\cite{Blackman:2017dfb,Blackman:2017pcm,Varma:2019csw}. This puts an explicit limit on their applicability
to comparatively short signals, i.e., from high-mass binaries with near-equal masses.  

The current work extends the \textsc{Phenom} approach, the development of which has proceeded in order of the most 
measurable physical effects. The most clearly measurable 
binary parameters are the chirp mass, $\mathcal{M} = M \eta^{3/5}$, for low-mass binaries where the detectable signal is dominated
by the inspiral, and the total mass $M$ for high-mass binaries where most of the detectable signal power is in the late inspiral, merger
and ringdown. Hence the first \textsc{Phenom} model considered non-spinning binaries~\cite{Ajith:2007qp,Ajith:2007kx}. 
The next most significant effect is due to a 
mass-weighted combination of the aligned-spin components, and the next set of \textsc{Phenom} models treated aligned-spin 
systems and were tuned to \nr{} simulations that were parametrised by a single effective 
spin~\cite{Ajith:2009bn,Santamaria:2010yb,Husa:2015iqa,Khan:2015jqa}. 
All of these models considered only the dominant contribution to the 
signal, which is from the $(\ell=2, |m|=2)$ multipole moments. Subdominant multipoles become stronger as the mass ratio is increased,
and these were first included through an approximate mapping of the dominant multipole~\cite{London:2017bcn}, and more 
recently with full 
tuning to \nr{} simulations~\cite{Garcia-Quiros:2020qpx}. Individual black-hole spins are unlikely to be measurable for detections with a 
signal-to-noise ratio (SNR) of less than $\sim$100~\cite{Purrer:2015nkh}, but a handful of such detections are likely when the LIGO and 
Virgo detectors reach design sensitivity in the next few years~\cite{Abbott:2020qfu}, and the latest aligned-spin \textsc{Phenom} models 
include \nr{} tuning to unequal-spin \nr{} simulations~\cite{Pratten:2020ceb}. The \textsc{Phenom} approach has been 
predominantly used to produce frequency-domain models, but has recently also been applied in the time 
domain~\cite{Estelles:2020osj,Estelles:2020twz}.

Precession effects are typically difficult to measure~\cite{Fairhurst:2019srr}, and indeed have not yet been definitively 
observed in any single observation~\cite{LIGOScientific:2018mvr,Abbott:2020niy}.
The dominant precession effects follow the phenomenology of single-spin systems, and thus the first precessing \textsc{Phenom} 
models~\cite{Hannam:2013oca}
used a single-spin \pn{} model to estimate the effects of precession. More recent models have included two-spin 
effects~\cite{Khan:2018fmp,Khan:2019kot,Pratten:2020ceb,Estelles:2021gvs}, but, once 
again, individual spin measurements will require SNRs of at least 100, and in most cases likely much higher~\cite{Khan:2019kot}. 
As such, the first
priority for an \nr{}-tuned precession model is the single-spin parameter space. 
Our new \textsc{PhenomPNR} model is tuned to \nr{} simulations that 
cover mass ratios from equal-mass to 1:8 ($\eta \sim 0.1$). The larger black hole has a spin magnitude up to $\chi_1 = 0.8$, and, 
as motivated by the preceding discussion, the smaller black hole has no spin. This is the widest systematic coverage of the 
mass-ratio--spin parameter space to date~\cite{BAM-catalog}.

\subsection{Model approximations, and motivation for a new model}\label{sec:motivation}

Previous \textsc{Phenom} and \textsc{EOBNR} models make use of several approximations. In this section we discuss each of these, 
and illustrate why we remove some of them in our new model, and the effect this has on the waveforms.

One set of approximations applies to the waveforms in the co-precessing frame. 

First, as described above, during the inspiral the co-precessing-frame waveform 
is approximated by an equivalent non-precessing-binary waveform, $h^{\rm{NP}}$. 
 In the most recent \textsc{EOBNR} model, \textsc{SEOBNRv4PHM}~\cite{Ossokine:2020kjp},
the EOB equations of motion are solved for the full precessing system from a chosen starting frequency, and then the approximate 
co-precessing-frame waveform is constructed 
by now solving the non-precessing \pn{} equations of motion, but with time-varying $\chi_i^{\parallel}(t)$ taken from the earlier 
precessing-binary solution. 
In the \textsc{Phenom} models, $h^{\rm{NP}}$ is defined 
by the aligned-spin components of the initial spin configuration, so $\chi_i^{\parallel}$ are constant.
In both families of models, $\chi_i^{\perp}$ contributions to the waveform multipole moment amplitudes are ignored.

Second, the mapping to an equivalent aligned-spin system breaks down at merger. This was already noted in the original presentation of
the aligned-spin mapping~\cite{Schmidt:2012rh}, and is also discussed in Refs.~\cite{Pekowsky:2013ska,Ramos-Buades:2020noq}. 
One reason is that the spin of the final black hole (and therefore the ringdown frequency and damping time) will be different to that in the 
non-precessing case; to first approximation, we must include the contribution from the in-plane spins, $\chi_i^{\perp}$, to the spin of
the final \bh{}. In the \textsc{Phenom} 
models, the merger-ringdown part of the aligned-spin waveform is modified by using this in-plane spin contribution to estimate a modified 
final spin, and hence complex ringdown frequency~\cite{Hannam:2013oca,Khan:2018fmp,Khan:2019kot,Pratten:2020ceb}; 
the recent \textsc{PhenomXP} model~\cite{Pratten:2020ceb} provides a number of optional methods to achieve this. 
In the \textsc{EOBNR} models, the inspiral construction ends at the light ring~\cite{Bohe:2016gbl}, and ringdown modes are 
attached, and in the most recent \textsc{SEOBNRv4PHM} model~\cite{Ossokine:2020kjp} these are based on an \nr{}-tuned final 
spin fit~\cite{Hofmann:2016yih}. 

In \textsc{PhenomPNR}, we retain the mapping to an equivalent aligned-spin system during the early inspiral, but we introduce
the key improvement that in the late inspiral, merger 
and ringdown we explicitly tune the model to \nr{} waveforms in the co-precessing frame. Rather than model the final mass and spin
and use those to estimate the complex ringdown frequency via perturbation theory, we also explicitly model the ringdown frequencies
from \nr{} waveforms in the co-precessing frame. As discussed in \sect{sec:physfeat}, this is necessary because the 
ringdown frequency in the co-precessing frame is shifted with respect to that in the inertial frame.

\begin{figure}[t]
   \begin{tabular}{c}
      \includegraphics[width=0.47\textwidth]{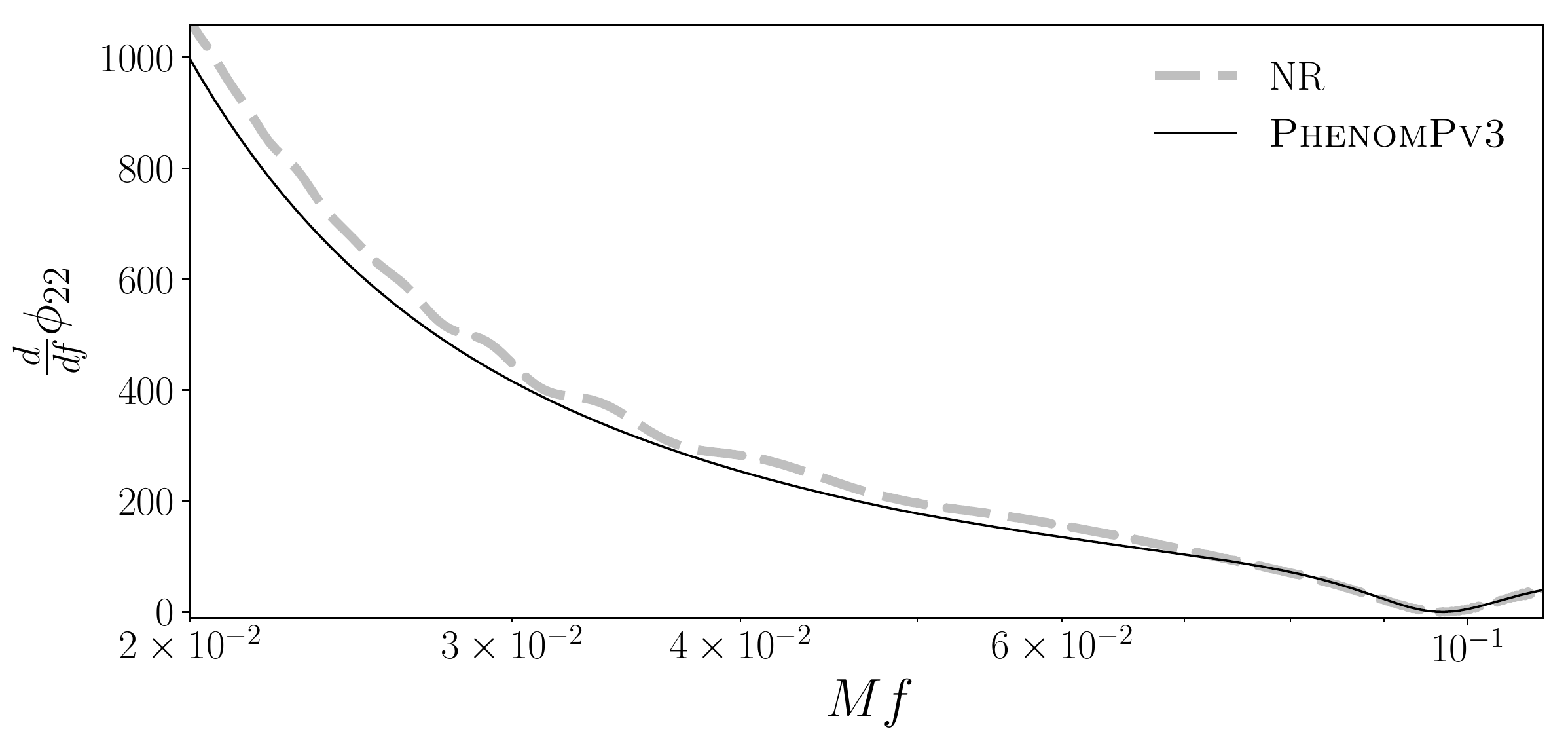} 
      \\   
      \includegraphics[width=0.47\textwidth]{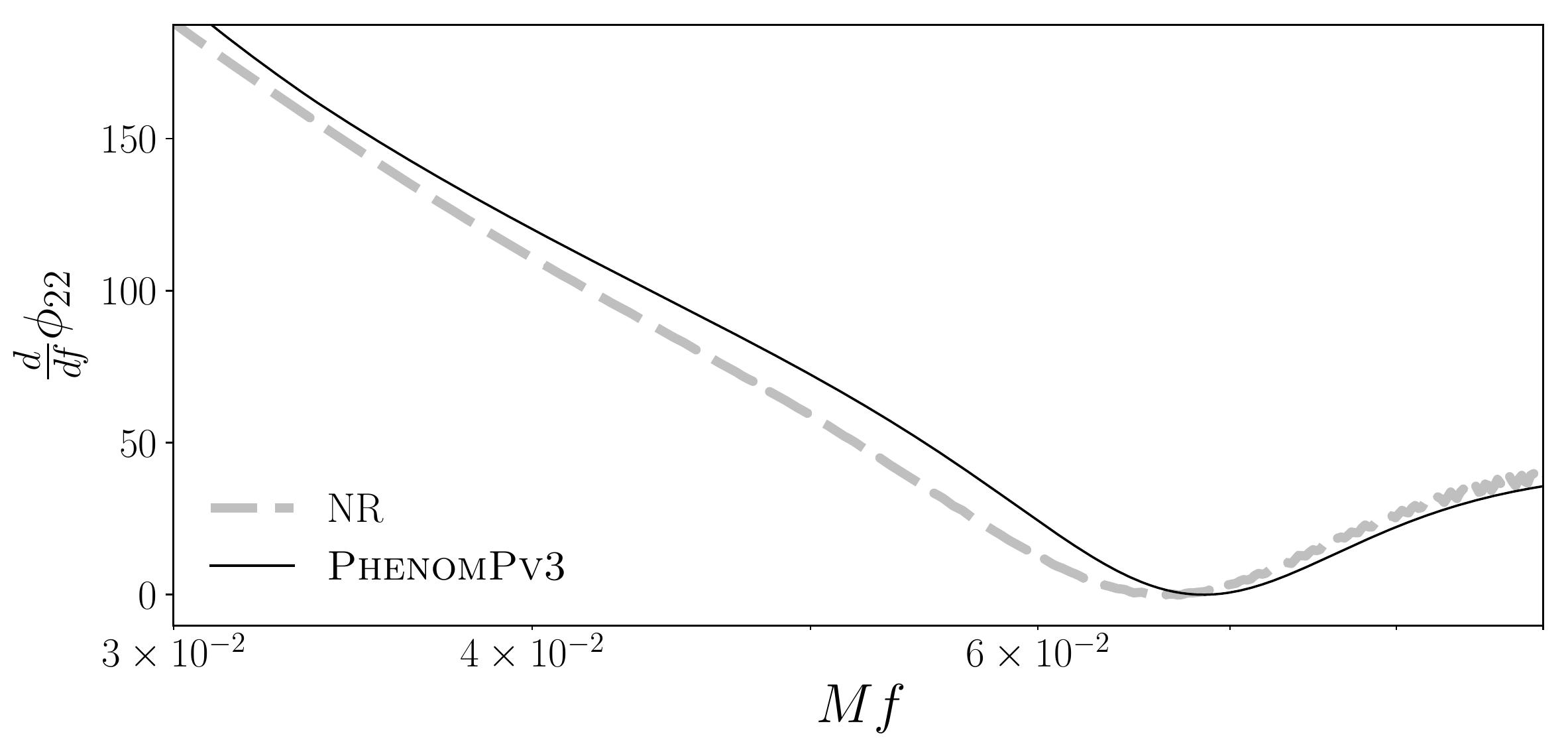} 
      \\
      \includegraphics[width=0.47\textwidth]{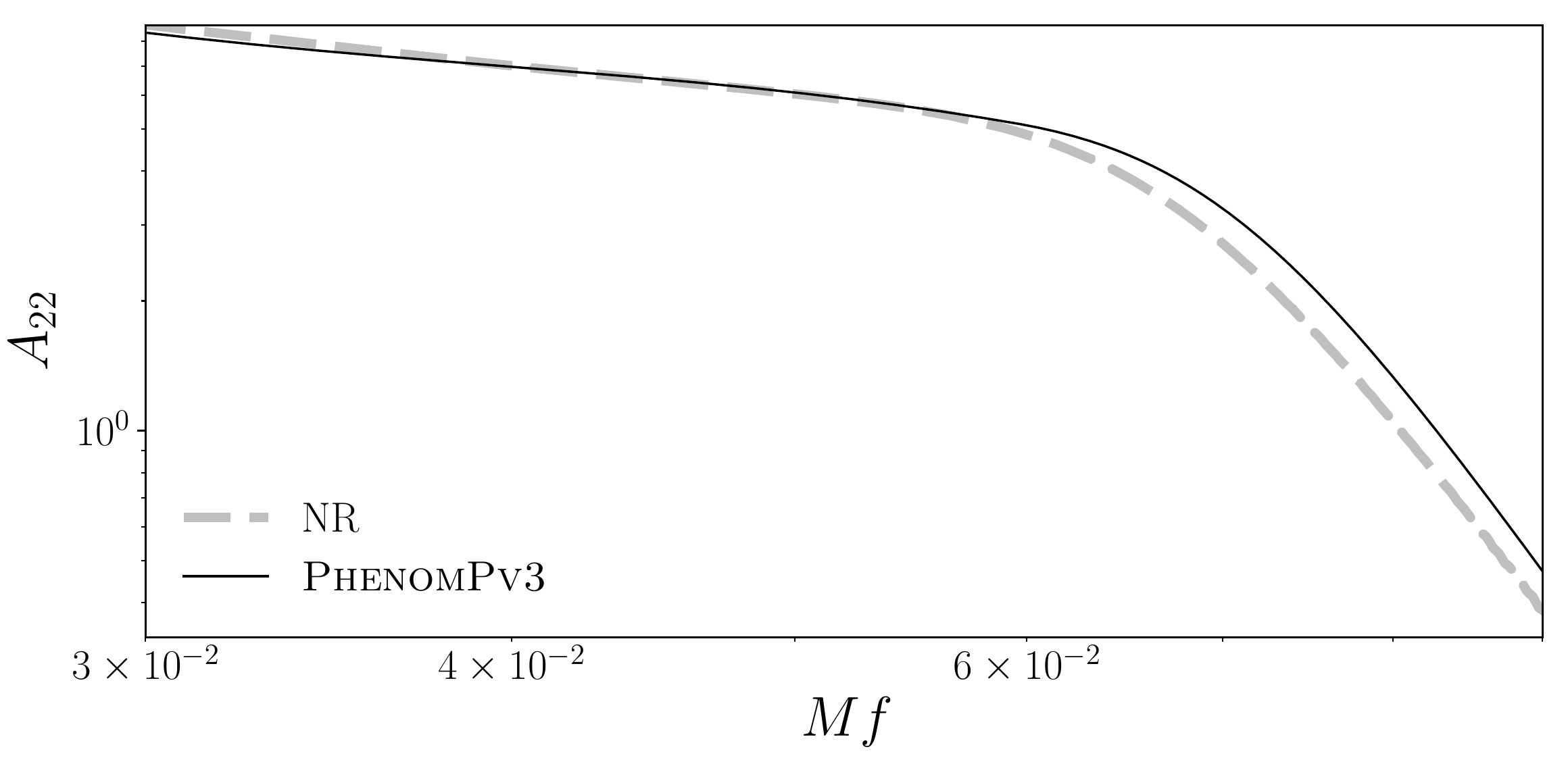} 
   \end{tabular}
    \caption{Frequency domain comparison of \nr{} and model waveforms in the co-precessing frame. 
    (top) phase derivative for the $(q,\chi_1,\tls)=(4,0.8,60^{\circ})$ configuration, which illustrates the variation in the
    inspiral phase.
    (middle and bottom) phase derivative and amplitude for the $(q,\chi_1,\tls)=(4,0.8,150^{\circ})$ configuration, which demonstrate 
    the shift in effective ringdown frequency.
    } 
   \label{fig:coprecessing-example} 
\end{figure}

This issue is illustrated in Fig.~\ref{fig:coprecessing-example}. The top panel shows the frequency-domain co-precessing-frame 
phase derivative $d\phi_{22}/df$
for one of our \nr{} simulations, with mass-ratio $q = m_1/m_2 = 4$, large-black-hole spin $\chi_1 = 0.8$, and spin mis-aligned with the
orbital angular momentum by $\tls = 60^\circ$. The figure also shows the results from the earlier \textsc{PhenomPv3} model.
In the inspiral we see a clear difference between the \nr{} and \textsc{PhenomPv3} results that is largest at low frequencies. 
The middle panel shows a second case, this time with a 
larger misalignment angle of $\tls = 150^\circ$. The location of the minimum can be approximately identified as the ringdown frequency,
and we see that there is a clear shift between the ringdown frequency in the inertial frame (as used in \textsc{PhenomPv3}), and 
the effective ringdown frequency of the \nr{} waveform in the co-precessing frame. This shift is also apparent in the bottom panel,
which shows the amplitude $A_{22}$ in the co-precessing frame. \textsc{PhenomPNR} fixes this problem; see, in particular,
Sec.~\ref{sec:coprecessing model}.

A second set of assumptions apply to the precession. 

In previous models the inertial-frame waveform was constructed via a time- or frequency-dependent rotation of $h^{\rm{NP}}$, 
using the precession angles 
relative to the Newtonian orbital angular momentum, i.e., the normal to the binary's orbital plane. This produces the correct 
inertial-frame multipoles only in the quadrupole approximation. In order to tune the precession angles to \nr{} results, we need a consistent
choice of co-precessing frame that can be applied both to \pn{} and \nr{} data. For \textsc{PhenomPNR} we choose the \qa{} 
frame~\cite{Schmidt:2010it,OShaughnessy:2011pmr,Boyle:2011gg}, which identifies the direction of maximum GW emission. In 
time-domain waveforms, the direction of maximum emission differs depending on 
whether it was defined using GW strain, $h$, the Bondi news function, $\dot{h}$, or the Weyl scalar, $\Psi_4 = \ddot{h}$; and all three
differ from the direction of $\mathbf{L}$~\cite{Schmidt:2010it,Ochsner:2012dj,Boyle:2014ioa,Hamilton:2018fxk}. (The direction of $\mathbf{L}$ 
also depends on whether we use a Newtonian or 
post-Newtonian estimate.) However, we perform our modelling in the frequency domain, where the \qa{} direction is independent of the 
choice of $h$ or $\Psi_4$.
We explain this further in Sec.~\ref{sec:prelims}, where we also describe in detail how we calculate 
the \qa{} frame from the $\ell=2$ multipoles of \nr{} simulations, and in
\sect{sec: beta approx} we discuss the \qa{} frame for \pn{} waveforms. We expect that the latter results would also allow the 
construction of more physically accurate EOBNR waveforms. 

In most previous \textsc{Phenom} models, the precession angles were estimated entirely from \pn{} theory. These angles will not be valid through 
merger, but as a simple approximation, they were used throughout the entire waveform. This approximation was justified by the
 observation that the 
\pn{} angles behave smoothly to arbitrarily high frequencies, and the model gives reasonable agreement to \nr{}
 waveforms~\cite{Hannam:2013oca,Khan:2018fmp,Khan:2019kot,Pratten:2020ceb}. However,
in more extreme parts of parameter space (high mass ratios and large in-plane spins), the inaccuracy of this approximation will become
more serious. In EOBNR models, the inspiral precession dynamics are provided from the solution of the \eob{} equations of motion, and
in the \textsc{SEOBNRv4PHM} model the precession angles are extended through merger and ringdown using an approximation based 
on the quantitative behaviour of \nr{} simulations; the time-domain \textsc{Phenom} model, \textsc{PhenomTPHM}, employs a similar 
approach~\cite{Estelles:2020osj}.

\begin{figure}[htb]
   \centering
   \begin{tabular}{c}
      \includegraphics[width=0.47\textwidth]{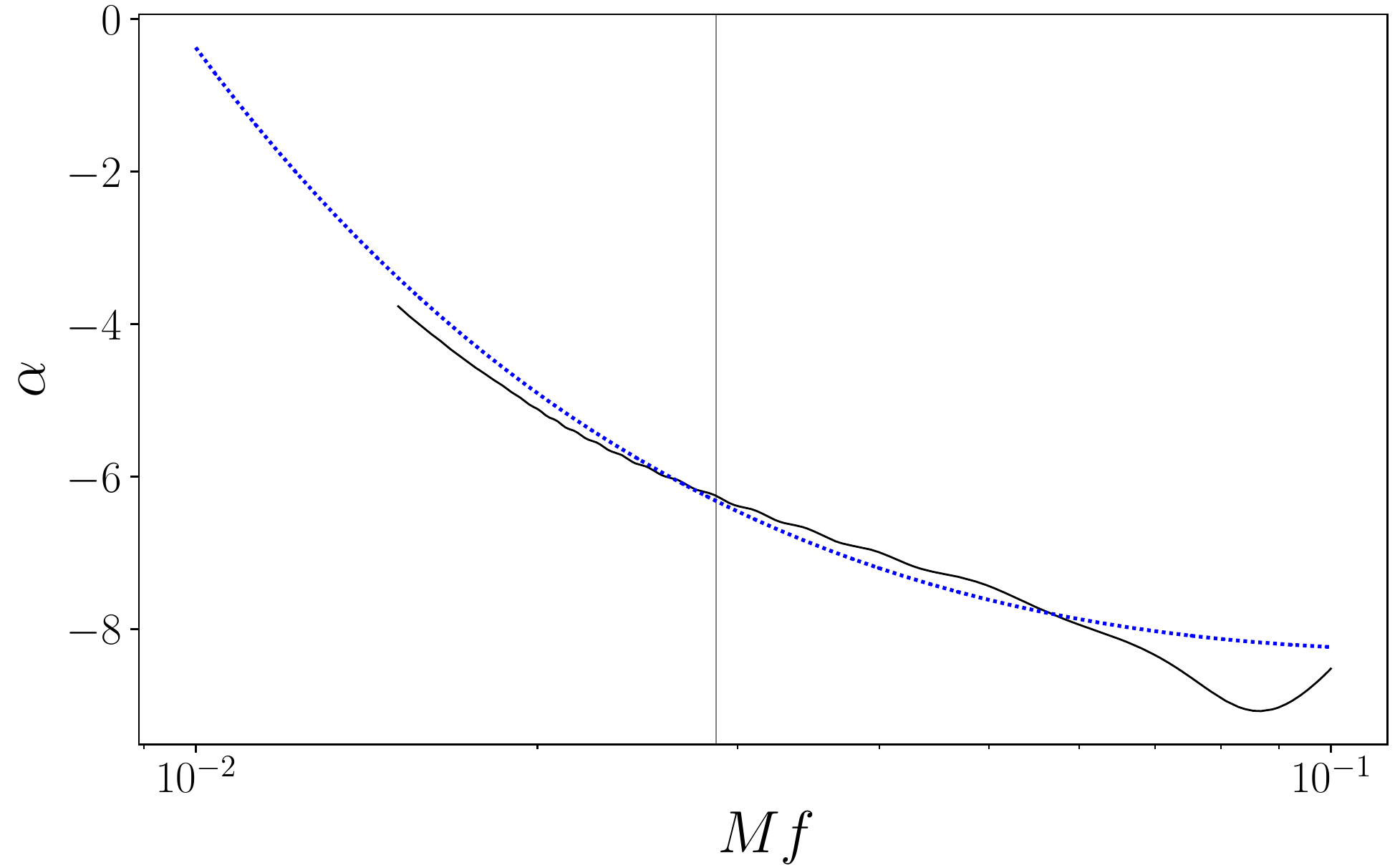}\\
      \includegraphics[width=0.47\textwidth]{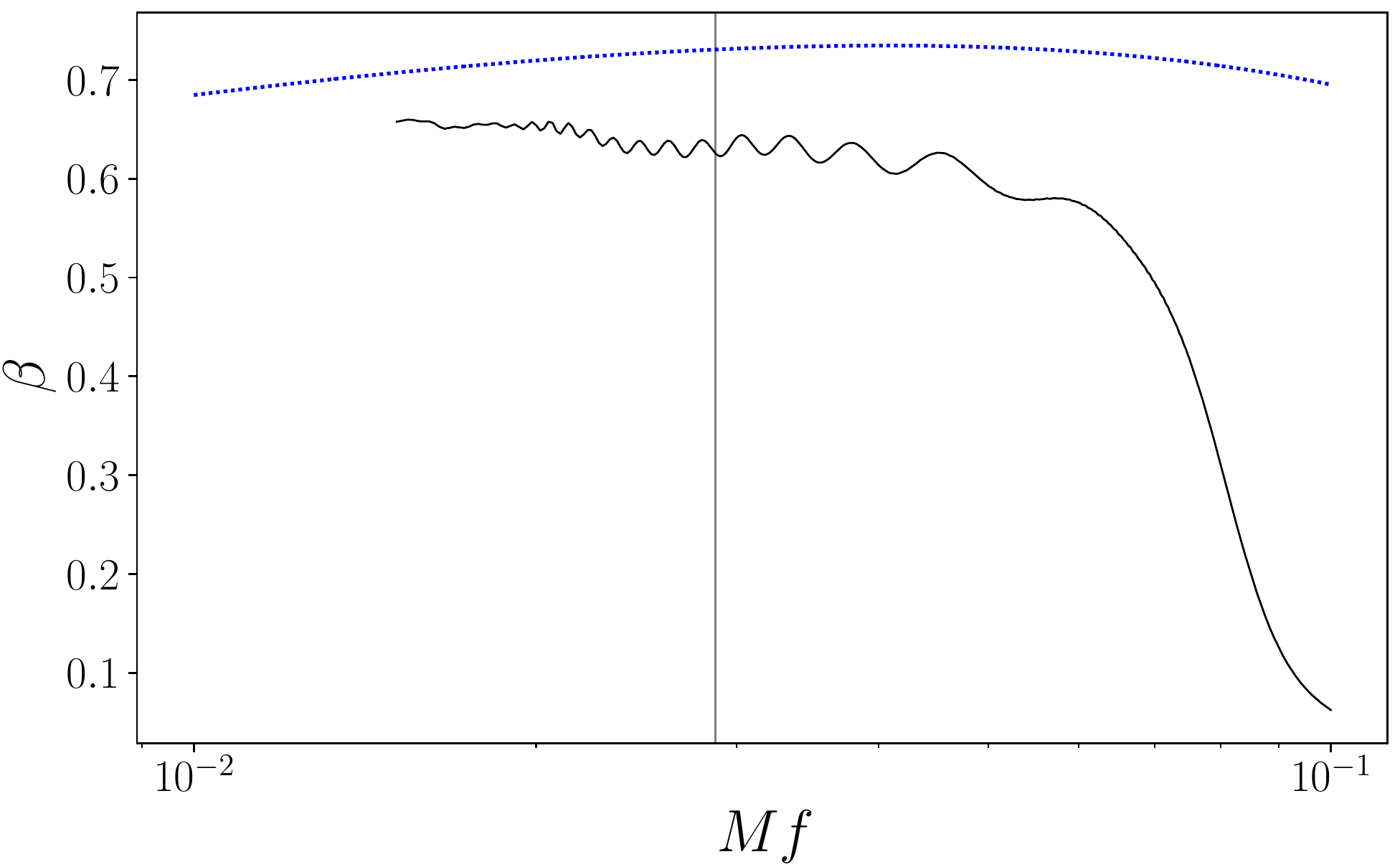}\\
      \includegraphics[width=0.47\textwidth]{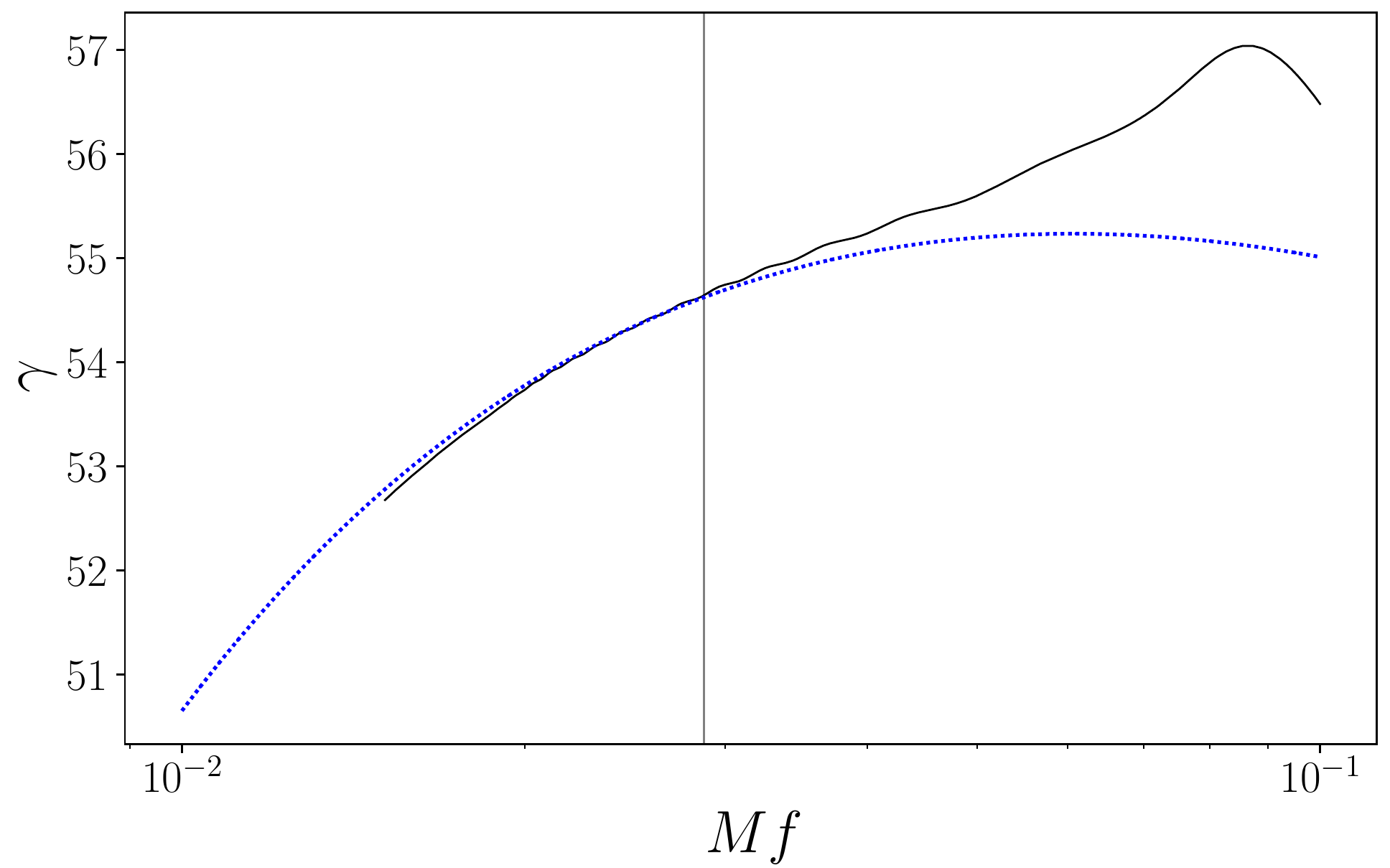}
   \end{tabular}
   \caption{Comparison of the post-Newtonian expressions for each of the precession angles (blue dotted line) with the \nr{} data 
   (black solid line) for the case with $\left(q,\chi,\tls\right)=\left(8,0.8,60^\circ\right)$. The grey vertical lines indicate the ISCO 
   frequency $\left( Mf = 0.0287\right)$ of the final black hole, which has final spin magnitude $\chi_f = 0.799$ and final mass 
   $M_f = 0.981M$.
   }
   \label{fig: Pv3 angles}
\end{figure}

 Fig.~\ref{fig: Pv3 angles} shows the precession angles $(\alpha, \beta, \gamma)$ for a configuration with 
 $\left(q,\chi,\tls\right)=\left(8,0.8,60^\circ \right)$. 
 The figure shows both the \nr{} results, and the \msa{} angles~\cite{2017PhRvL.118e1101C} used 
 in the \textsc{PhenomPv3} and \textsc{PhenomXP} models. We see that at high frequencies that correspond to the merger and
 ringdown, the \msa{} estimates fail to capture the phenomenology of the \nr{} data. The angles $\alpha $ and $\gamma$ both exhibit
 a ``dip'' or ``bump'', reminiscent of the dip in the phase derivative in Fig.~\ref{fig:coprecessing-example}, which is absent in the 
 \msa{} estimates. The \nr{} opening angle $\beta$ drops to close to zero at merger, as we might expect as the two-body inspiral 
 motion terminates and we are left with only a single perturbed black hole. This feature cannot be captured by the \msa{} expressions, 
 which simply extend the inspiral behaviour to higher frequencies. We also find that the \nr{} $\beta$ does not relax to zero, but to some 
 non-zero value, which, if it does decay, typically does so very slowly. (There have been approximate estimates of this asymptotic 
 $\beta$ decay using a toy ringdown model~\cite{OShaughnessy:2012iol,Marsat:2018oam,Estelles:2020osj}, which we discuss and 
 clarify in \sect{sec:physfeat}.) These features must also be modelled.
 
 Finally, we see that at lower frequencies, the \msa{} $\alpha$ and $\gamma$ agree well with the \nr{} results. However, although we 
 expect the \msa{} and \nr{} $\beta$ to also agree at sufficiently low frequencies, they do not agree over the frequency range of our 
 \nr{} data, and would likely require \nr{} simulations that are many times longer. This discrepancy is due to the modelling inconsistency
 discussed earlier: the two estimates are of different quantities. The \msa{} $\beta$ is the orientation of the orbital plane, while the 
 \nr{} $\beta$ is the orientation of the QA direction of the signal, and these are not in general the same. We show how to 
 significantly reduce this discrepancy in~\sect{sec: beta approx}. (The high-frequency oscillations in the \nr{} $\beta$
 are due to a combination of numerical noise and Fourier-transform artifacts. All of our \nr{} $\beta$ results show similar oscillations,
 with varying amplitude and frequency, but in these single-spin cases we will model only a smooth trend through the data, which 
 we expect to represent their relevant physical features.)
 
 The bulk of the results in this paper present a merger-ringdown model for the co-precessing-frame waveforms (\textsc{PhenomDCP}) 
 and a separate model for the precession angles (\textsc{PhenomAngles}). Both modes are tuned to our \nr{} data and capture all of 
 the features described here. We then produce a complete inspiral-merger-ringdown model (\textsc{PhenomPNR}) by connecting 
 our merger-ringdown models to inspiral results. 

There are two remaining assumptions that were made in previous models, which we retain in our new model. 

Non-precessing-binary waveforms satisfy a symmetry between the $m>0$ and $m<0$ multipoles that is broken in precessing
binaries~\cite{Brugmann:2007zj,Ramos-Buades:2020noq,Kalaghatgi:2020gsq}. The ``twisting-up'' construction used by the 
\textsc{Phenom} and \textsc{EOBNR} models neglects these asymmetries. 
Although asymmetries may need to be included in models to allow accurate spin measurements in some GW 
observations~\cite{Kalaghatgi:2020gsq}, in the current \textsc{PhenomPNR} model we retain the approximation that the asymmetries 
in the multipole moments are zero. 

Current \textsc{Phenom} and \textsc{EOBNR} models also assume that the direction of the total angular momentum remains fixed. 
Although the total angular momentum direction changes little through inspiral, there is \emph{some} change due to the loss of 
angular momentum through GW emission. In \textsc{PhenomPNR} we explicitly transform the \nr{} waveforms to a frame where 
$\hat{\mathbf{J}}$ remains fixed along the $z$-axis, and use those waveforms as the basis of the model. 
In this sense the fixed-$\hat{\mathbf{J}}$ approximation is retained in \textsc{PhenomPNR} and
remains valid over the parameter space used to construct the model, which is further discussed in \csec{sec: full model matches}.

This paper is organised as follows. In \sect{sec:NR} we present our \nr{} waveforms. In \sect{sec:prelims} we process 
the raw \nr{} waveforms to produce the frequency-domain co-precessing-frame waveforms and precession angles
that we wish to model. 
Since we limit the \nr{} tuning to 
single-spin binaries, in \sect{sec: spin parameterisation} we specify our procedure to map generic two-spin systems
to approximately equivalent single-spin configurations. With all of these pieces in place, in \sect{sec:coprecessing model}
we present our co-precessing-frame model, \textsc{PhenomDCP}, in \sect{sec:angle model inspiral} our treatment of the 
precession angles during inspiral, and in \sect{sec:angle model MR} our merger-ringdown angle model, \textsc{PhenomAngles}. 
All of these
ingredients are put together into a full inspiral-merger-ringdown model in \sect{sec:fullmodel}. Having modelled precessing-binary 
waveforms, we discuss their physical features in more detail in \sect{sec:physfeat}, and evaluate their accuracy in 
\sect{sec:matches}.

In all of the discussion of \nr{} and \pn{} results, and in all modelling work, we use geometric units, $G = c = 1$. We also choose $M=1$,
although we retain ``$M$'' in plot labels, to make clear that we are dealing with dimensionless quantities. 
Physical masses will only be used in \sect{sec:matches}, where we study the performance of models with respect to a specific 
detector noise curve. 
All of the earlier waveform models used to generate results in this work were called from the software package \texttt{LALsuite}~\cite{lalsuite}. 
The specific model names are \texttt{IMRPhenomD} for \textsc{PhenomD}~\cite{Husa:2015iqa,Khan:2015jqa}, \texttt{IMRPhenomXAS} for \textsc{PhenomXAS}~\cite{Pratten:2020fqn}, \texttt{IMRPhenomPv3} for \textsc{PhenomPv3}~\cite{Khan:2018fmp}, \texttt{IMRPhenomXP} for \textsc{PhenomXP}~\cite{Pratten:2020ceb}, \texttt{SEOBNRv4P} for \textsc{SEOBNRv4P}~\cite{Ossokine:2020kjp}, and \texttt{NRSur7dq4} for \textsc{NRSur7dq4}~\cite{Varma:2019csw}.

\section{Numerical Relativity waveforms}
\label{sec:NR}

In producing the first precessing-binary model tuned to \nr{} waveforms, we wish to capture the dominant precession effects first.
This can be achieved with single-spin systems, i.e., only one of the black holes is spinning, since two-spin effects typically produce only
small modulations of the underlying simple precession~\cite{Buonanno:2004yd,Schmidt:2014iyl}.
We therefore consider single-spin systems that obey simple precession, and the \nr{} catalogue used to tune the model contains 
single-spin configurations where the spin is placed on the larger black hole and neglects two-spin configurations and the impact of the 
azimuthal spin angle. 
This reduces the binary parameter space from seven dimensions (mass ratio, plus the vector components of each black-hole spin), to
three dimensions: the symmetric mass ratio, $\eta$, the magnitude of the spin on the larger black hole, $\chi \equiv \chi_1$, and 
the angle between the spin and the orbital angular momentum of the system, $\tls$. It is important to note that 
these are all defined as part of the initial data of the simulations, since $\tls$ undergoes small oscillations about some mean value during the inspiral. 

We wish our model to extend to the highest mass ratios feasible with current \nr{} simulations.
The earlier tuned 
non-precessing model \textsc{PhenomD}~\cite{Husa:2015iqa, Khan:2015jqa} was based on a catalogue containing systems up to mass ratio 
$q = m_1/m_2 = 18$, or $\eta \sim 0.05$. 
\nr{} simulations at $q=18$ are extremely computationally expensive, and since the mass-ratio of observations is heavily skewed towards
comparable masses~\cite{LIGOScientific:2018mvr, Abbott:2020niy}, for the current model we restrict to $q=8$. We note, however, 
that one recent GW observation, GW190814, was measured with a mass ratio of $q \sim 10$~\cite{Abbott:2020khf}, and therefore extending 
our model to higher mass ratios is an urgent requirement for future work. 

In order to confidently capture the dependence of precession effects on mass ratio, we produced simulations at four different mass 
ratios, approximately equally spaced in symmetric mass ratio $\eta$. Similarly, we chose four equally spaced spin magnitudes $\chi$. 
We already have aligned and anti-aligned waveforms in this range of mass ratios and spin magnitudes, and for non-aligned-spin configurations 
we chose five equally spaced values for the spin angle, $\tls$, excluding $0^\circ$ and $180^\circ$. 

The model is tuned to a subset of this catalogue of 80 waveforms, which was produced using the \texttt{BAM} code~\cite{Brugmann:2008zz}. 
The complete catalogue contains simulations with $q\in[1,2,4,8]$, (or $\eta\in[0.1,0.16,0.22,0.25]$), $\chi\in[0.2,0.4,0.6,0.8]$ and 
$\tls\left(^\circ\right)\in[30,60,90,120,150]$. For tuning we used the 40 waveforms with $\chi = 0.4$ and 0.8. We expect the 
dependence of the precession effects on spin magnitude to be approximately linear, so this is not anticipated to significantly degrade the 
accuracy of the tuned part of the model; and this is borne out in validation of the model against the remaining waveforms in the catalogue,
plus 27 waveforms from the SXS and Maya catalogues~\cite{Boyle:2019kee,SXS:catalog,Jani:2016wkt,GATech:catalog}.

Since our goal is a frequency-domain model, we would like \nr{} waveforms that all cover
a similar frequency range. The majority of the waveforms start at a frequency of $M\Omega = 0.023$.
However, some of the higher mass ratio 
configurations have a higher starting frequency in order to ensure the binary merged in a reasonable time to allow sufficient
accuracy. The highest starting frequencies occur for configurations with a large spin magnitude where the spin is closest to being 
aligned with the orbital angular momentum, due to the hang-up effect~\cite{Campanelli:2006uy}. The highest starting frequency is 
$M\Omega = 0.032$, for the $(q,\chi,\tls) = (8,0.8,30^\circ)$ configuration. We find that these starting frequencies are in 
general sufficient to 
match smoothly to \pn{} results. We will see in Sec.~\ref{sec: matches angles} that there are a few cases for which we would prefer 
\nr{} waveforms with lower 
starting frequencies, but these are actually configurations with large spins and large opening angle, e.g., 
$(q,\chi,\tls) = (8,0.8,150^\circ)$. 
Having identified specific issues with these more challenging regions of parameter space, we will be able to focus on them in 
detail in future iterations of our model.

More details on the production of the \nr{} catalogue, and error analysis of the waveforms, will be given in Ref.~\cite{BAM-catalog}. 
The greatest sources of error in these numerical waveforms are the finite resolution at which we performed the simulations and the finite 
distance from the source at which we extracted the GW data. 
We consider the mismatch (as defined in Sec.~\ref{sec: match definitions}) to be the most useful uncertainty estimate for our purposes.
We make a conservative estimate of the mismatch uncertainty between the waveforms in this \nr{} catalogue and the theoretical `analytical' 
solution of $\mathcal{O}\left(10^{-3}\right)$. For the shorter waveforms in the catalogue, particularly the $q=1$ and $q=2$ cases, the 
mismatch was found to be $\mathcal{O}\left(10^{-4}\right)$. As we will see when validating 
against independent NR data sets (e.g., those from the SXS catalogue, where the finite-extraction-radius error is minimal), the errors in our
model are often an order of magnitude lower than our upper bound, and, where they are comparable or higher, the accuracy limits due to the 
modelling procedure are likely the dominant source of error. 

For each \nr{} simulation, spin weight $-2$ spherical harmonic multipole moment data are stored for the radiative Weyl scalar,
\def\psilm{\psi_{\ell m}}
\begin{align}
  \label{ylm}
  \psi_{\ell m}(t) = \int_{\Omega} \; r \, \Psi_4(t,r,\theta,\phi) {_{-2}}Y^*_{\ell m}(\theta,\phi) \; \text{d}\Omega \;,
\end{align}
where $*$ denotes complex conjugation. The $\psi_{\ell m}$ depend on the choice of decomposition frame, and we provide the
details of our frame choice in Sec.~\ref{sec:prelims}.
Each $\psilm$ time series contains multipole moment data for inspiral, merger and ringdown.

In addition, spurious (``junk'') radiation, due to imperfect initial data~\cite{Cook:1989fb}, is windowed away, 
using a window function that increases
from zero to one over the duration of three gravitational wavelengths. 
It is found that when windowing over more than two wavelengths the choice of (smooth) window function has no significant effect 
on our modelling results.
For simplicity, a standard Hann window is used \cite{hann}.
The window starts at the first peak in the real part of ${\psi_{22}}$ such that the following peak is less than or equal to the largest distance 
between peaks in the time series. This most often results in less than 200$M$ of contaminated inspiral data being tapered away.
The window is applied equally to the real and imaginary parts of $\Psi_4$ for all multipoles.
Similarly, post-ringdown data are windowed such that the Hann window turns off to the right between the point where the 
exponential decay drops below the noise floor, as defined by fitting a constant value to the very end of the timeseries.
The time domain data are also zero-padded to the right such that the frequency domain step size, in geometric units, is less than $5\times 10^{-4}$.

The result of the inspiral and post-ringdown windows is the reduction of frequency-domain power that is broadband and unphysical. 
The result of zero-padding is to enforce that frequency domain features are consistently resolved.

\section{Waveform frames, conventions and approximations}
\label{sec:prelims}

We wish to model the dominant multipoles of the \bbh{} signal. The multipoles depend on the choice of reference frame, and we attempt
to choose a frame that simplifies the modelling. In this section we present the reference frame in which we construct our model, and 
several additional simplifications that we make to the data. 

If we have a set of spin-weighted spherical-harmonic multipoles $q^1_{\ell m}$, and rotate the coordinate system through the Euler 
angles $(\alpha,\beta,\gamma)$, then the multipoles in the new frame, $q^2_{\ell m}$, are given by, \begin{equation}
q^2_{\ell m} =  \sum^{\ell}_{m'=-\ell} e^{i m' \alpha} d^{\ell}_{m'm}\left(-\beta\right) e^{i m \gamma} q^1_{\ell m'},
\label{eq:rotations}
\end{equation} where $d^{\ell}_{m'm}$ are the 
Wigner d-matrices~\cite{WignerEugenePaul1959Gtai, Brugmann:2008zz}. 

We apply these rotations twice to our data. 

First, we retain the approximation that has been used in all \textsc{Phenom} and \textsc{EOBNR} models to date, that the direction of the 
total angular momentum, $\hat{\mathbf{J}}$, is fixed. This convention amounts to a minor modification of the \nr{} data, whose radiative 
$\mathbf{J}(t)$ varies by at most $\sim$$6^\circ$ from its initial direction.
To impose the fixed-\(\hat{\mathbf{J}}\) convention we need to know $\mathbf{J}(t)$ at all times in the original simulation. 
At the beginning of the simulation $\mathbf{J}(0) = \mathbf{J}_{\mathrm{ADM}}$, which can be calculated analytically from
Bowen-York initial data~\cite{Bowen:1980yu}. The angular momentum flux can be calculated from the multipole moments, 
e.g., Ref.~\cite{Ruiz.0707.4654},
and integrating this specifies the time evolution of $\mathbf{J}(t)$. As a consistency check, we compare $\mathbf{J}$ at the end of the 
simulation with the estimate of the final black hole's spin calculated on the apparent horizon~\cite{Campanelli:2006fy}, 
and find a disagreement of at most 5\% in magnitude and 3\% in direction. With $\mathbf{J}(t)$ now in hand, 
we use Eq.~(\ref{eq:rotations}) 
to perform a time-dependent rotation to place the signal in a frame of reference where $\hat{\mathbf{J}}(t) = \hat{z}$ at all times.
The impact of this frame convention is well below the total error budget of the final \pnr{} model, and is discussed in 
more detail in \csec{sec: full model matches}.

Second, we make another time-dependent rotation into a co-precessing frame. We choose the \qa{} frame, 
which was introduced in Ref.~\cite{Schmidt:2010it}, and allows us to define a co-precessing frame using
the gravitational-wave signal, which is the observable quantity we ultimately care about, rather than the orbital dynamics of the two
black holes. The \qa{} method was motivated by the observation that in the quadrupole approximation, if the orbital plane lies in the 
$x$-$y$ plane, then the signal can be represented entirely by the $(\ell=2,|m|=2)$ multipoles. At any other orbital plane orientation, some
signal power will be distributed to the $|m|=1$ and $m=0$ multipoles, therefore reducing the amplitude of the $(\ell=2,|m|=2)$ multipoles. 
It follows that we can always identify the orientation of the orbital plane by locating the direction with respect to which the $(\ell=2,|m|=2)$
multipoles are maximised. In a time-dependent co-precessing frame where this always holds, we can represent the entire signal using 
only the $|m|=2$ multipoles, and, furthermore, precession modulations of the signal amplitude and phase will be significantly reduced. 
In general, i.e., beyond the quadrupole approximation, this direction is only approximately equal 
to the normal to the orbital plane, or to a \pn{} estimate of the direction of the orbital angular 
momentum~\cite{Schmidt:2010it,Boyle:2014ioa,Hamilton:2018fxk}. 
However, although it cannot be
directly related to the dynamics, it does provide us with a convenient 
signal-based definition of a co-precessing frame that suppresses precession modulations.

\par
\begin{figure}[htbp]
   \centering
   \includegraphics[width=0.5\textwidth]{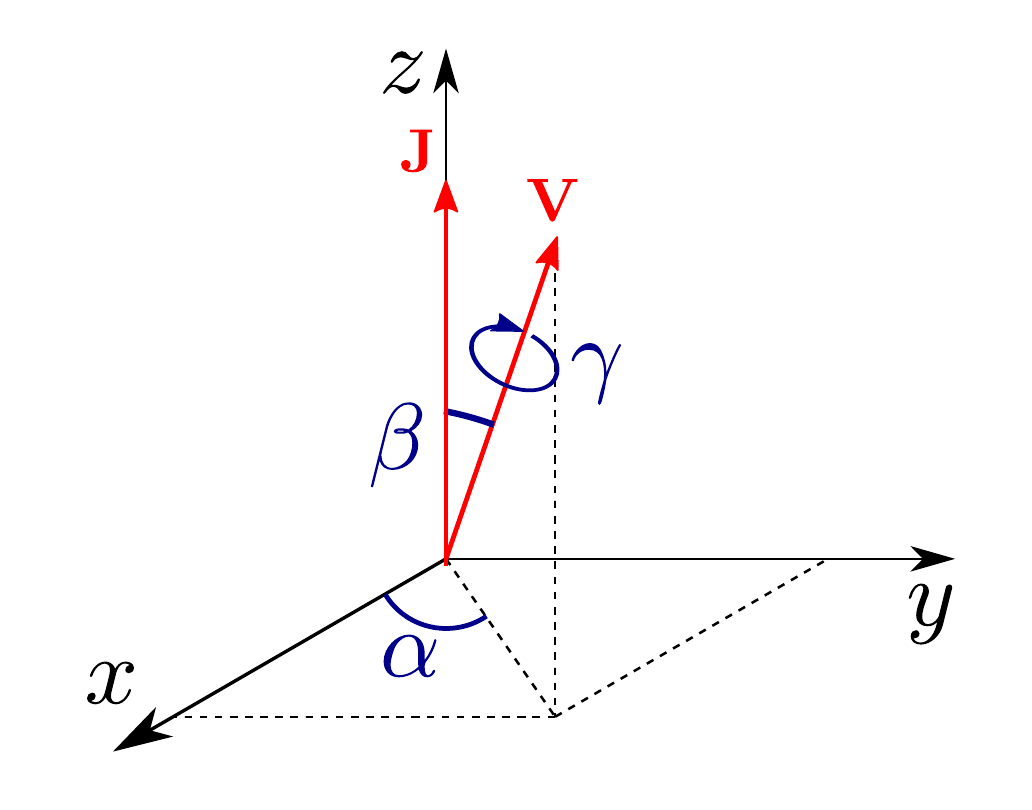}
   \caption{The Euler angles $(\alpha,\beta,\gamma)$
   that make up the precession angles that describe the transformation 
   from the fixed-\(\mathbf{\hat{J}}\) frame into a co-precessing frame. As mentioned in the text, there are different choices for the definition of $\mathbf{V}$; 
   the QA direction, the Newtonian orbital angular momentum and varying orders of the post-Newtonian orbital-angular
   momentum.
   }
   \label{fig: LaboutJ}
\end{figure}

 In the following sections we use the method described in Appendix~\ref{app:calc_angles} to calculate the \oed{}. 
We use the Euler angles $\alpha$, $\beta$ and $\gamma$ to describe the orientation of this direction. 
\Eqns{fda}{fdc} define the angles accordingly, and \fig{fig: LaboutJ} illustrates their geometric meaning.

One potential ambiguity with the \qa{} frame is that it differs depending on whether it is defined using the \gw{} strain, or its
time derivatives, the Bondi news $\dot{h}$ or the Newman-Penrose scalar $\Psi_4$. However, this ambiguity does not exist in
the frequency domain. 

To see this, consider the 
multipoles of the gravitational-wave strain, which can be written as,
\begin{align} 
   h_{\ell m}\left(t\right) = {}& A_{\ell m}(t) e^{-i m \Phi \left(t\right)}.
\end{align}
Our \nr{} data satisfy $\Psi_4 = \ddot{h}$, and so we can write, 
\begin{align}
   \psi_{\ell m}\left(t\right) = {}& A'_{\ell m}(t) e^{-i m \Phi' \left(t\right)},
\end{align}
where the new amplitude and phase are given by,
\begin{align} 
   A'_{\ell m} = {}& \sqrt{ \left( \ddot{A} - m^2 \dot{\Phi}^2 A \right)^2 + m^2 \left( 2 \dot{\Phi} \dot{A} 
   + \ddot{\Phi} A\right)^2 }, \\
   \Phi' = {}& \Phi + \frac{1}{m}\arctan\left( 
   \frac{m \left(2 \dot{\Phi} \dot{A} + \ddot{\Phi} A \right)}{\ddot{A} - m^2 \dot{\Phi}^2 A} \right),
\end{align}
where we have dropped the $(\ell,m)$ subscripts for brevity.
We see that the distribution of power between the multipoles will in general be different for $h$ and for $\Psi_4$ in the time domain, and 
therefore the \qa{} angles $(\alpha,\beta,\gamma)$ will differ. 

By contrast, in the frequency domain we have,
\begin{align} 
   \tilde{\Psi}_4 = {}& \text{F.T.}\left[ \Psi_4 \right] = \text{F.T.}\left[ \ddot{h} \right] = -\omega^2 \tilde{h},
\end{align}
where $\omega=2\pi f$ and $f$ is the gravitational-wave frequency. Since $\omega$ is an overall factor in front of all of the multipoles at a 
given frequency, the direction that maximises both $|\tilde{h}|^2$ and $\omega^4 |\tilde{h}|^2$ will be the same. The \qa{}
precession angles will therefore be the same for $h$ and for $\Psi_4$. Given that the frequency-domain \qa{} angles are independent of
the choice of $\Psi_4$ or strain, we consider this to be the natural regime in which to work. 

Finally, we also retain the standard \textsc{Phenom} and \textsc{EOBNR} approximation that the co-precessing multipole 
moments of our model obey the same symmetry properties as their non-precessing counterparts. This means that we neglect to 
model $\pm m$ asymmetries in the multipole moments. Although the asymmetric 
contributions are weak, there is some evidence that they are necessary for non-biassed measurements of precessing 
systems~\cite{Kalaghatgi:2020gsq}, and they are certainly necessary for measurements of out-of-plane recoil of the 
binary~\cite{Varma:2020nbm}, and we plan to model these contributions in future work. 

Given $\psilm$ that have been transformed first to the fixed-\(\mathbf{\hat{J}}\) and then \qa{} frames in the time domain, we construct the 
symmetric combination, 
\begin{align}
   \label{sym22}
   \psi^\mathrm{sym}_{2,2} \; = \; \frac{1}{2} \, ( \; \psi_{2,2} + \psi_{2,-2}^* \; ).
\end{align}
In \eqn{sym22}, $\psi^\mathrm{sym}_{2,2}$ effects an average of the co-precessing-frame mass-quadrupoles consistent with 
Ref.~\cite{Boyle:2014ioa}.
We then define a symmetrised $(\ell=2,m=-2)$ multipole according to the non-precessing symmetry relationship 
$\psi_{l,-m}=(-1)^{\ell}\,\psi_{\ell m}^*$, thus,
\begin{align}
   \psi^{\mathrm{sym}}_{2,-2} \; = \; (\psi^{\mathrm{sym}}_{2,2})^* \; .
\end{align}
Together, $\psi^{\mathrm{sym}}_{2,-2}$ and $\psi^{\mathrm{sym}}_{2,2}$ encapsulate all waveform information that will be retained 
at this stage. The \qa{}-frame $\ell > 2$ multipoles are discarded, along with the \((\ell=2,|m|<2)\) multipoles; we leave higher 
multipoles to future work. 

The symmetrised multipoles are then rotated back into the fixed-\(\mathbf{\hat{J}}\) frame. We then use these data as our starting point to transform
the multipoles into the frequency domain, and then transform to the \qa{} frame as defined in the frequency domain. 

We separately produce a model (\textsc{PhenomDCP}) of the co-precessing-frame multipole $h^{\rm CP}_{2,2}(f)$,
and another model (\textsc{PhenomAngles}) of the rotation angles $(\alpha(f), \beta(f), \gamma(f))$. Given these two models,
our full intertial-frame model (\textsc{PhenomPNR}) of the $\ell = 2$ multipoles, $h^J_{\ell m}(f; \mlam)$, is given via 
Eq.~(\ref{eq:rotations}),
\begin{equation}
h^J_{\ell m}(f; \mlam) =  \sum^{\ell}_{m'=-\ell} e^{i m' \alpha} d^{\ell}_{m'm}\left(-\beta\right) e^{i m \gamma} h^{\rm{CP}}_{\ell m'} (f; \mlam). \label{eq:hCPtoh}
\end{equation}

\section{Spin parametrisation}\label{sec: spin parameterisation}

Our goal is to model generic non-eccentric black-hole binaries with any physically reasonable values of $M$, $\eta$, 
$\boldsymbol{\chi}_1$ and $\boldsymbol{\chi}_2$. Given NR waveforms that cover only the single-spin parameter space, 
we require a mapping between generic two-spin configurations and approximately equivalent configurations where 
$\boldsymbol{\chi}_2 = 0$. In this section we summarise our spin parameterisation. In Sec.~\ref{sec: full model matches} we 
demonstrate that the resulting model agrees well with a subset of the two-spin precessing-binary NR waveforms that are 
currently available. 

Both our co-precessing-frame model \textsc{PhenomDCP} and angle model \textsc{PhenomAngles} are tuned to the same 40 
single-spin NR waveforms described in Sec.~\ref{sec:NR}. 

In the 
inspiral region \textsc{PhenomD} is based on \pn{} expressions and so parameterised by the masses $m_1$ and $m_2$  and 
dimensionless spins 
$\chi^\parallel_1$ and $\chi^\parallel_2$ of the binary. The leading-order \pn{} spin contribution to the phase is 
$\chi_\text{PN} = \chi_\text{eff} - \frac{38\eta}{113}\left(\chi^\parallel_1 + \chi^\parallel_2\right)$~\cite{Cutler:1994ys,Poisson:1995ef,Ajith:2011ec}, 
in which the main contribution is the symmetric spin combination~\cite{Ajith:2009bn, Santamaria:2010yb} , \begin{equation}
 \chi_\text{eff} = \frac{m_1 \chi^\parallel_1+ m_2 \chi_2^\parallel}{m_1 + m_2}. \label{eqn: chi_eff}
\end{equation} As such, the \nr{} calibrated merger-ringdown region of \textsc{PhenomD} is parameterised by the normalised quantity,
\begin{align} 
   \label{Xeff}
   \hat{\chi} = {}& \left(1-\frac{76\eta}{113}\right)^{-1}\chi_\text{PN}.
\end{align}
The final black hole is parameterised by the final mass $M_f$ and spin $a_f$, which are estimated using independent fits to the \nr{}
data~\cite{Husa:2015iqa}. 

Although \textsc{PhenomD} is tuned to equal-spin or single-spin \nr{} waveforms, and is often described as a single-spin model, 
the use of both spins in the underlying  inspiral \pn{} phase expressions, and the two different single-spin parameterizations $\hat{\chi}$ and $a_f$ in the 
merger-ringdown calibration, mean that the model also incorporates some two-spin effects, and indeed has been shown in some cases to 
describe two-spin configurations to high accuracy~\cite{Kumar:2016dhh}. 
 
\textsc{PhenomDCP} is constructed such that \textsc{PhenomD} is explicitly recovered in the absence of precession. 
To this end, \textsc{PhenomD}'s phenomenological parameters, which we will generically refer to as $\lambda_k$, are modified according to, 
\begin{align}
   \label{dcp0}
   \lambda'_k \; = \; \lambda_k + \chi_\perp \, \nu_k \, , 
\end{align}
where $\nu_k$ is the new phenomenological parameter to be modelled across the intrinsic parameter space and $\chi_\perp$
quantifies the in-plane spin component and as such gives a measure of the degree of precession in the system. 
In \eqn{dcp0} it is manifestly evident that, when $\chi_\perp=0$, \textsc{PhenomDCP} reduces to \textsc{PhenomD}.
The parameter $\chi_\perp$ is defined as part of our treatment of the precession angles, which we will now describe. 

As with previous precessing-binary \textsc{Phenom} models, we will also use \pn{} results to describe the precession angles 
through inspiral. Ref.~\cite{Chatziioannou:2017tdw,Khan:2018fmp} provide complete two-spin expressions, and as such 
are parameterised by the masses $m_1$ and $m_2$ and the dimensionless spins $\boldsymbol{\chi}_1$ and 
$\boldsymbol{\chi}_2$ of the binary.

Conversely, for the merger-ringdown we will construct phenomenological expressions for the angles, parameterised according to 
the parameters of the single-spin \nr{} simulations, $(\eta, \chi, \tls)$.
Although the \nr{}-calibrated merger-ringdown angle model is a model of single-spin systems, we can estimate the angles for generic
two-spin systems by making an approximate mapping from two-spin systems to our single-spin angle model. Our mapping is defined as follows. 

We first map the spin components to the two effective spin parameters used in previous \textsc{Phenom} models. For the aligned-spin components
we use the combination $\chi_{\rm eff}$, as defined in Eq.~(\ref{eqn: chi_eff}). Although $\chi_{\rm{PN}}$ is the appropriate aligned-spin 
parameter from \pn{} theory, in precessing systems $\chi_{\rm{eff}}$ is a constant of the \pn{} equations of motion without radiation 
reaction~\cite{Racine:2008qv}, and can be seen to vary less during inspiral than $\chi_{\rm{PN}}$. 

Following Ref.~\cite{Schmidt:2014iyl}, we also define the effective precession spin, $\chi_{\rm p}$, based on the leading-order \pn{} precession
dynamics,
\begin{equation} 
   \chi_{\rm p} =  \frac{S_{\rm p}}{m_1^2}, \label{eqn: chi_p}
\end{equation}
where $S_{\rm p} = \frac{1}{A_1} \text{max} \left( A_1 S_1^\perp, A_2 S_2^\perp \right)$, $A_1 = 2 + 3 m_2 /(2 m_1)$, and 
$A_2 = 2 + 3 m_1/(2 m_2)$.
$\chi_\text{eff}$ parameterises the spin parallel to the 
orbital angular momentum while $\chi_{\rm p}$ parameterises the spin perpendicular to the orbital angular momentum, i.e., in the plane of the binary.

This definition was motivated by the observation that the vectors $\mathbf{S}^\perp_1$ and $\mathbf{S}^\perp_2$ rotate in the plane at 
different rates, and over
the course of the inspiral the magnitude of their vector sum will oscillate between the sum and difference of their two magnitudes. 
As shown in Ref.~\cite{Schmidt:2014iyl}, 
the average value of the in-plane spin contribution to the precession dynamics can be approximated well by $\chi_{\rm p}$ for
mass ratios $q \gtrsim 1.5$. 
However, at mass ratios very close to one the spins precess in the plane at approximately the same rate, and so add or 
cancel in the same way at all times, and $\chi_{\rm p}$ does not provide an ideal single-spin mapping. (This is illustrated in more detail 
in Ref.~\cite{Gerosa:2020aiw}.) Extreme examples are the ``superkick'' 
configurations~\cite{Brugmann:2007zj},
where the black holes are of equal mass, and $\chi^\parallel_1 = \chi^\parallel_2 = 0$ and $\chi^\perp_1 = -\chi^\perp_2$. 
From the symmetry of the configuration, the two spins rotate at the same rate at all times, and therefore the total in-plane spin is zero, 
and the system does not precess. For a superkick configuration $\chi_{\rm p}$ clearly does not provide the appropriate ``single-spin'' mapping, 
which in this case should be to a system with zero in-plane spin.

To deal with such cases, we also introduce $\chi_\text{s}$, which is constructed from the vector sum of the in-plane spin vectors at a 
single reference time/frequency of the waveform. In our construction these are the in-plane components of the spin vectors input to the 
waveform generation. We define $\chi_\text{s}$ as,
\begin{align}
   \chi_\text{s} = \frac{ \left| \mathbf{S}_1^\perp + \mathbf{S}_2^\perp \right| }{m_1^2}.
 \end{align} 
 
Given a two-spin system defined by $\mathbf{S_1}$ and $\mathbf{S_2}$, we model the precession angles through the merger and 
ringdown by mapping to a corresponding single spin, which is placed on the larger black hole. This single spin has magnitude 
$\chi_\parallel$ in the direction parallel to the orbital angular momentum and $\chi_\perp$ in the orbital plane, where,
\begin{align} 
   \chi_\parallel = {}& \frac{M \chi_\text{eff}}{m_1},  \label{eq:singleparallel} \\
   \chi_\perp = {}& \begin{cases}
   \cos^2\left(\theta_q\right) \chi_\text{s} + \sin^2\left(\theta_q\right) \chi_{\rm p}, 
   & 1 \le q \le 1.5 \\
   \chi_{\rm p}, & q > 1.5, 
   \end{cases} \label{eq:singleperp}
\end{align}
where $\theta_q= \left(q-1\right)\pi$. This combination of $\chi_\text{s}$ and $\chi_{\rm p}$ given for $1 \le q \le 1.5$ is designed to provide a smooth 
transition between the regimes where $\chi_\text{s}$ and $\chi_{\rm p}$ are most appropriate. We note that for systems with $q < 1.5$, the precession
effects are weak, and so the error incurred from this approximation is small, and we expect that different choices for $\chi_\text{s}$, or for the 
transition to $\chi_{\rm p}$, would have an impact on GW measurements smaller than the other approximations used in our model. 
(Alternative choices of single-spin mapping are suggested in Refs.~\cite{Gerosa:2020aiw,Thomas:2020uqj}; since we use a 
single-spin mapping only to connect our single-spin merger-ringdown model to a generic-spin inspiral model, we expect that there are 
many reasonable choices of mapping that would work equivalently well.) 
This expression for $\chi_\perp$ is also used to parameterise the in-plane spin effects in the co-precessing model, as described 
in Eq.~(\ref{dcp0}).
 
The total spin magnitude $\chi$ and the angle between the orbital and spin angular momenta are given by
\begin{align} 
   \chi = {}& \sqrt{\chi_\parallel^2 + \chi_\perp^2}, \label{eqn: chi_mag} \\
   \cos\tls = {}& \frac{\chi_\parallel}{\chi} \label{eqn: costheta}.
\end{align}
These reduce to the correct values for the cases to which we tuned the model and also correctly re-weight two-spin cases and cases 
where the spin is predominantly on the smaller black hole.

In the $\mathbf{J}$-aligned frame, in which we have constructed our model, the spin placed on the larger black hole has the components
\begin{align}\label{eqn: single spin components}
   \mathbf{S'} = {}& 
   \begin{pmatrix}
   \cos\alpha \left( \chi_\perp \cos\beta + \chi_\parallel \sin\beta \right) \\
   \sin\alpha \left( \chi_\perp \cos\beta  + \chi_\parallel \sin\beta \right) \\
   -  \chi_\perp \sin\beta + \chi_\parallel \cos\beta
   \end{pmatrix}
\end{align}
where $\alpha$ and $\beta$ are the values of the precession angles introduced in Sec.~\ref{sec:prelims}, 
here evaluated at the reference frequency.

% ~--~--~--~--~--~--~--~--~--~--~--~--~ %
\section{Co-precessing-frame model}
% ~--~--~--~--~--~--~--~--~--~--~--~--~ %
\label{sec:coprecessing model}

A key assumption of most precessing signal models has been that the coprecessing multipole moments are largely devoid of 
precession related effects~\cite{Schmidt:2010it,Hannam:2013oca,Pratten:2020ceb,Ossokine:2020kjp}.
This assumption is motived by the \pn{} description of inspiral, where in-plane spin components do not impact the coprecessing 
waveforms' phase, and so can be disregarded~\cite{Arun:2008kb,Kidder:1995zr}.
In this sense, most precessing signal models have used un-modified non-pressing inspiral waveforms in the coprecessing frame.
Because the \pn{} motivation is only well suited for inspiral, for the waveforms' immediate pre-merger and merger, additional assumptions 
must be made~\cite{Schmidt:2012rh,Pekowsky:2013ska,Ramos-Buades:2020noq}.
For example, all previous precessing-binary \textsc{Phenom} models use an estimate of the 
precessing system's final mass and spin to compute the remnant \bh{}'s \qnm{} frequencies.
In turn, these \qnm{} frequencies allow the frequency-domain waveforms' features at merger to be shifted such that they occur 
near physically appropriate values.
In ~\sect{sec:motivation} we illustrated deviations from the simplifying assumptions made in both the inspiral and merger-ringdown,
and in this section we refine those assumptions by constructing a tuned coprecessing waveform model.

We introduce \dcp{}, a model for the $\ell=|m|=2$ coprecessing \gw{} multipole moment tuned to \nr{}.
\dcp{} is tuned to the 40 late inspiral, merger and ringdown \nr{} simulations discussed in Sec.~\ref{sec:NR}.
By construction, \dcp{} reduces to \d{} for non-precessing \bbh{} systems. 
We could have instead adapted the more recent \textsc{PhenomXAS} model~\cite{Pratten:2020fqn}, which is tuned also to two-spin 
systems, but since two-spin effects are unlikely to be measurable in most observations~\cite{Purrer:2015nkh,Khan:2019kot}, and we have
tuned to NR results only from single-spin precessing systems, we will leave two-spin extensions of the co-precessing-frame model 
to future work.

We consider \dcp{} to be a first step towards a high accuracy coprecessing waveform model.
Here we briefly review the structure of \d{}, and how this structure is extended by \dcp{}.
Physical features of the \nr{} waveforms and \dcp{} are provided and discussed in detail in \sect{sec:physfeat}.
Plots showing fits of model parameters across the space of initial binary masses and spins are provided in \apx{app:fitsurfaces}.

\begin{figure}[t]
   \begin{tabular}{c}
      \includegraphics[width=0.47\textwidth]{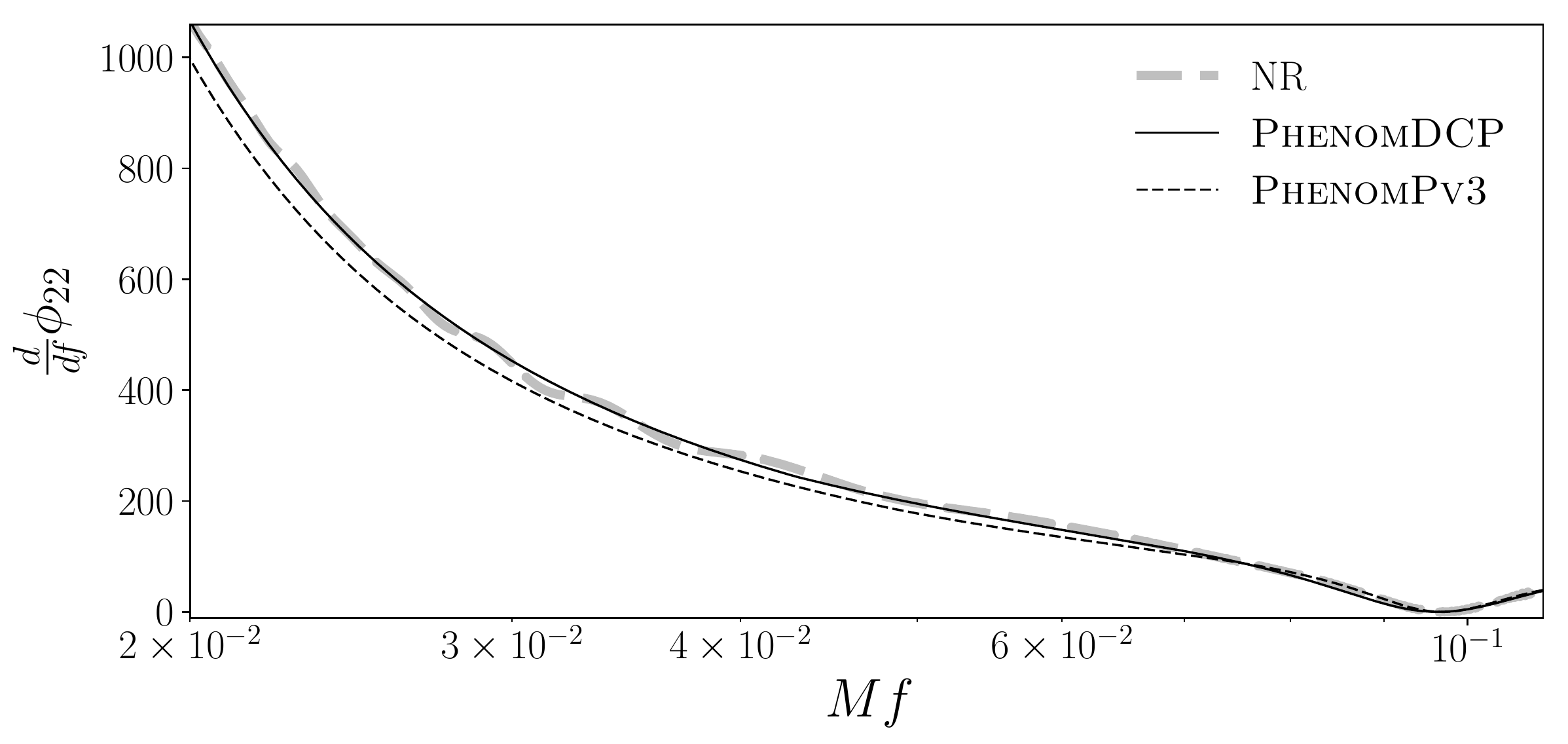} 
      \\   
      \includegraphics[width=0.47\textwidth]{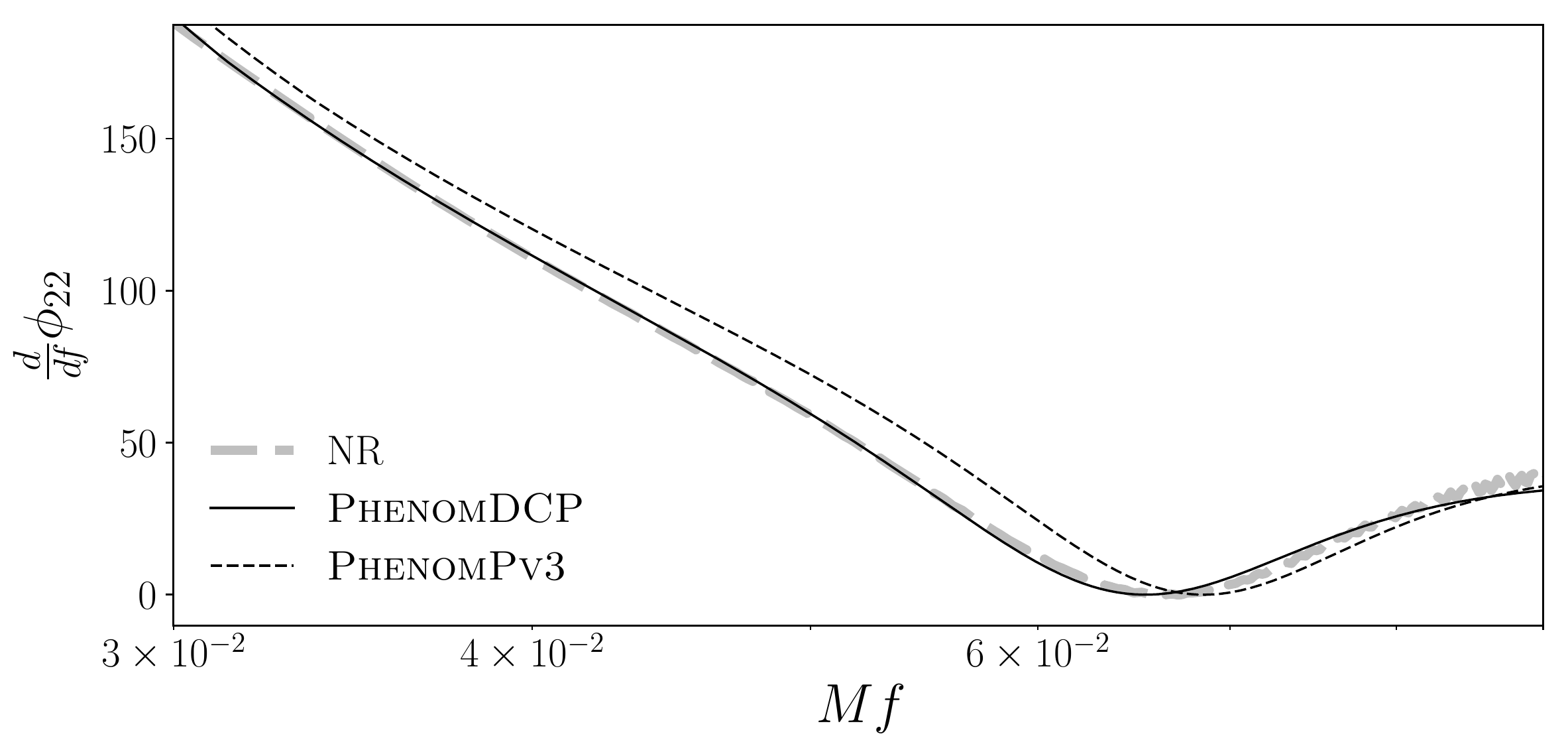} 
      \\
      \includegraphics[width=0.47\textwidth]{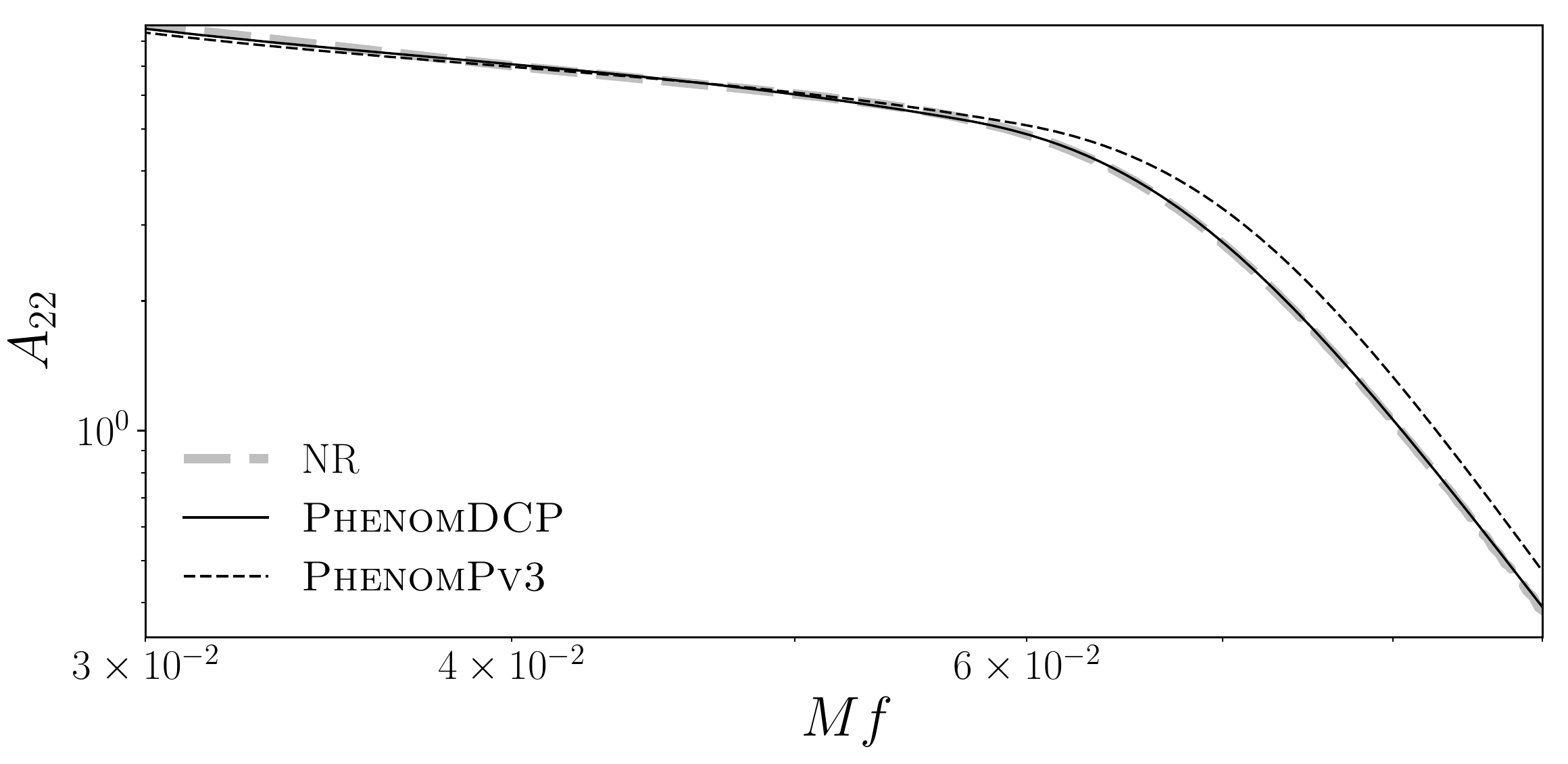} 
   \end{tabular}
    \caption{Frequency domain comparison of \nr{} and model waveforms in the co-precessing frame. 
    (top) phase derivative for the $(q,\chi,\tls)=(4,0.8,60^{\circ})$ configuration, which llustrates the variation in the
    inspiral phase.
    (middle and bottom) phase derivative and amplitude for the $(q,\chi,\tls)=(4,0.8,150^{\circ})$ configuration, which demonstrates 
    the shift in effective ringdown frequency.
    } 
   \label{fig:coprecessing-example-2} 
\end{figure}

\subsection{Briefly on the structure of \d{}}

\d{}~\cite{Khan:2015jqa,Husa:2015iqa} is a phenomenological model for the $\ell=|m|=2$ frequency-domain multipole 
moments of \gw{s} from non-precessing \bbh{s}.
The morphology of each multipole moment is organized into three regimes: (1) inspiral, where \pn{} theory applies, (2) intermediate, 
where the time domain evolution of the black holes is near merger, and (3) merger-ringdown, where the time domain evolution corresponds 
to the final coalescence and formation of a stationary remnant \bh{}.
\d{} models each of these regimes with different ans\"atze.
The coefficients of each \d{} ansatz are functions of the initial binary's masses and aligned spins.
In \dcp{} these coefficients are modified to depend on information about the in-plane spins.

\textsc{PhenomD} was calibrated to 19 \nr{} waveforms between $q=1$ and $q=18$. 
For unequal-mass systems, \d{} is calibrated to $\chi_{\mathrm{eff}} \in [-0.85, 0.85]$, and for equal-mass
systems is it calibrated to $\chi_{\mathrm{eff}} \in [-0.98,0.98]$.
In each \nr{} simulation the black-hole spins were either equal, $\chi_1 = \chi_2$, or the smaller black hole was non-spinning. 
The calibration waveforms were hybrids of \textsc{SEOBNRv2} (without \nr{} tuning) and \nr{} waveforms. Over the model's 
calibration region, its typical deviations (mismatches) from \nr{} are less than 1\%~\cite{Khan:2015jqa}.

\subsection{Construction of \dcp{}}

In the \textsc{PhenomP} models~\cite{Hannam:2013oca,Khan:2018fmp,Khan:2019kot} \d{} is used as an approximate 
co-precessing-frame model, with the ringdown frequency modified according to an estimate of the final black hole's spin. 
In \dcp{} we instead use \nr{} waveforms to tune in-plane-spin deviations to a subset of the model coefficients. 
Here we briefly overview the modifications of \d{} that result in \dcp{}.
\def\hcp{h^{\rm CP}_{22}}

As in previous models, \dcp{} assumes that in the coprecessing frame only the $(\ell,m)=(2,\pm 2)$ multipole moments are 
needed, and that the $m=2$ and $m=-2$ strain moments are related by conjugation~(\csec{sec:prelims}).
Under these assumptions we only need model the amplitude and phase of $\hcp{}$,
\begin{align} 
   \label{dcp1}
   \hcp{} \left(f;\mlam \right) = {}& A\left(f;\mlam \right) e^{-i \phi\left(f;\mlam \right)}. 
\end{align}
In \eqn{dcp1}, $A\left(f;\mlam \right)$ is the frequency domain amplitude of $\hcp{}$, $\phi\left(f;\mlam \right)$ is its phase, 
$f=\omega / 2 \pi$ references a frequency bin in geometric units, and $\mlam$ encapsulates the system's initial 
parameters~(\csec{sec: spin parameterisation}),
\begin{align}
   \label{mlam}
   \mlam \in \left( \eta, \chi, \tls{} \right) \; ,
\end{align}
where, as described in Sec.~\ref{sec: spin parameterisation}, the total spin $\chi$ consists of the aligned-spin component
$\chi_{\rm eff}$ and the in-plane component $\chi_\perp$, and for 
 our single-spin calibration waveforms, $\chi_\perp = \chi_{\rm p} = \chi^{\perp}_1$.

Given the system's initial parameters $\mlam$, \dcp{} is defined by a series of polynomials between $\mlam$ and phenomenological model parameters.
\dcp{}'s model parameters are based directly on those of \d{}~(\ceqn{dcp0}). 
Specifically, \dcp{} uses the \d{} amplitude and phase ansatz with model parameters offset by a term proportional to $\chi_\perp$.
Thus, when $\chi_\perp=0$, \dcp{} reduces to \d{}.

Precession effects are known to be most relevant in the late inspiral and merger-ringdown~\cite{Khan:2019kot,Hannam:2013oca}. 
Thus \textsc{PhenomDCP} is made to be equivalent to \d{} in the early inspiral.
Modified versions of \d{} are used for the waveforms' late-inspiral phase, merger-ringdown phase, and merger-ringdown amplitude:
\begin{equation}
   \label{dcp1b}
   \phi_{\text{Int}} = \frac{1}{\eta} \left(\beta_{0} + \beta_{1} f
                 + \beta'_{2} \, \ln(f)
                 - \frac{\beta_{3}}{3}f^{-3}\right) \; ,
\end{equation}

\begin{equation}
   \label{dcp2}
   \begin{split}
      \phi_{\text{MR}} & = \frac{1}{\eta} \left\{ \alpha_{0}
                  + \alpha_{1} f
                  - \alpha_{2} f^{-1}
                  + \frac{4}{3} \alpha_{3} f^{3/4} \right. \\
                  & \left. + \, \, \alpha'_{4}
                     \tan^{-1}\left(\frac{f-\alpha_{5} f^{(\phi)}_0}
                     {f^{(\phi)}_1}\right)\right\} \;,
   \end{split}
\end{equation}

\begin{equation}
   \label{dcp3}
   \frac{A_{\text{MR}}}{A_0} = \gamma_{1}
                   \frac{\gamma_{3} \fdamp^{(A)}}
                   {( f-\fring^{(A)} )^2+(\gamma_{3}\fdamp^{(A)})^2}               
   e^{-\frac{ \gamma'_{2} (f-\fring^{(A)}) }{\gamma_{3}\fdamp^{(A)}}}
   \; .
\end{equation}
In \eqns{dcp1b}{dcp3} Greek symbols denote model parameters defined in Ref.~\cite{Khan:2015jqa}, and of those, 
primed symbols, such as $\alpha'_4$, denote parameters modified for \dcp{}. Please note that these Greek symbols 
should not be confused with the Euler angles that define the coprecessing frame. 
In \eqn{dcp2}, $f_0^{(\phi)}$ is an ``effective ringdown frequency'' that is particular to the phase. 
Similarly, $f_1^{(\phi)}$ corresponds to the ringdown decay rate. 
In the setting of \d{}, $f_0^{(\phi)}$ and $f_1^{(\phi)}$ are simply refereed to as $f_{\rm RD}$ and $f_{\rm damp}$. 
In \eqn{dcp3}, $f_0^{(A)}$ is an effective ringdown frequency particular to the amplitude, and $f_1^{(A)}$ is equivalent to the 
ringdown decay rate used in \d{},
\begin{align}
   \label{dcp3b}
   f_1^{(A)} = f_{\rm damp} \; .
\end{align}

Our notation for the effective ringdown frequencies signals that we will not assume a direct relationship between the 
ringdown frequencies predicted by \bh{} perturbation theory, and those relevant for coprecessing waveforms. 
This point is discussed further in \sect{sec:physfeat}.

In constructing \dcp{} it was found that only a subset of \d{}'s parameters needed to be modified. 
These parameters are those needed to address the disconnect between \d{} and the coprecessing frame \nr{} data discussed in \sect{sec:NR}. 
The modified parameters correspond to the late inspiral behavior of the frequency domain phase,
\begin{align}
   \label{dcp4}
   \beta'_{2} \; = \; \beta_{2} + \chi_\perp \zeta_2 \; ,
\end{align}
the merger-ringdown phase,  
\begin{align}
   \label{dcp5}
   \alpha'_{4} \; &= \; \alpha_{4} + \chi_\perp \nu_4 
   \\
   \label{dcp6}
   f^{(\phi)}_0 \; &= \; \fring{} + \chi_\perp \nu_5 
   \\
   \label{dcp7}
   f^{(\phi)}_1 \; &= \; \fdamp{} + \chi_\perp \nu_6 \; ,
\end{align}
and the merger-ringdown amplitude,
\begin{align}
   \label{dcp8}
   \gamma'_2 \; &= \; \gamma_2 + \chi_\perp \, \mu_2 
   \\ 
   \label{dcp9}
   \fring^{(A)} \; &= \; \fring + \chi_\perp \, \mu_4 \; .
\end{align}
In \eqns{dcp1b}{dcp3b}, all parameters not defined in \eqns{dcp4}{dcp9} are defined in Ref.~\cite{Khan:2019kot}.
Similarly, in \eqns{dcp4}{dcp9}, $\{\alpha_4,\fring{},\fdamp{},\gamma_2\}$ are defined in Ref.~\cite{Khan:2019kot}.

The calibration of \dcp{} has been performed by fitting \eqns{dcp1b}{dcp3b} to each \nr{} waveform in our calibration set.
This yields a collection of calibration points for each model parameter. 
For each of \dcp{}'s model parameters, these points were modeled as polynomials in $\mlam$ using \verb|gmvpfit|, 
which uses multidimensional least-squares regression driven by a greedy algorithm~\cite{London:2018nxs,positive:2020}.
\par \Figs{fig:dcpmu}{fig:dcpnu} show the behavior of the \dcp{} model parameters as functions of symmetric mass-ratio and $\tls{}$ 
over the calibration space.
The parameter surfaces shown in \figs{fig:dcpmu}{fig:dcpnu} correspond to percent root-mean-square errors of {$3.42\%$} in amplitude 
and {$2.53\%$} in phase.

\Fig{fig:coprecessing-example-2} compares evaluations of \dcp{} to \nr{} and \pv3{} for the cases discussed in \sect{sec:motivation}.
The top panel of \fig{fig:coprecessing-example-2} highlights the effect of modifying the phase.
The middle and bottom panels highlight the effect of modifying the effective ringdown frequency and damping times. We see that
\dcp{} successfully corrects for the discrepancies in the modified-\d{} co-precessing-frame model used in \pv3{}; see 
Sec.~\ref{sec:matches} for quantitative accuracy results.

\section{Precession angle model: inspiral} 
\label{sec:angle model inspiral}

Our model of the precession angles consists of two parts. The first describes the precession during inspiral, and is based on the 
\msa{} angles presented in Ref.~\cite{2017PhRvL.118e1101C}, and used in previous \textsc{Phenom} 
models~\cite{Khan:2018fmp,Khan:2019kot,Pratten:2020ceb}. The second part is a 
phenomenological model of the precession angles during merger and ringdown, tuned to the NR waveforms presented in \sect{sec:NR}. 
We discuss the inspiral angles in this section, the merger-ringdown angles in \sect{sec:angle model MR}, 
and the combined \imr{} angle model in \sect{sec:fullmodel}.

\subsection{MSA angles}\label{sec: PhenomPv3 angles}

The precession angles in the inspiral regime are calculated using \pn{} theory.
In Ref.~\cite{2017PhRvL.118e1101C,Chatziioannou:2017tdw} the authors
derived a closed-form analytic approximation to the inspiral precession
dynamics.
To achieve this GW driven radiation-reaction was introduced into an
analytic solution to the conservative precession
dynamics~\cite{2015PhRvL.114h1103K} by exploiting the
hierarchy of timescales in the binary inspiral problem
using a mathematical technique called multiple scale analysis
~\cite{2013PhRvD..88l4015K,2013PhRvD..88f3011C}.
The hierarchy of timescales are $t_{\rm{orb}} \ll t_{\rm{prec}} \ll t_{\rm{rr}}$,
where $t_{\rm{orb}}$, $t_{\rm{prec}}$ and $t_{\rm{rr}}$ are the
orbital, precession and radiation-reaction timescales respectively.
This model is a function of all 6 spin components (two $3$-vectors for each \bh{}) 
and incorporates spin-orbit and spin-spin
effects to leading order in the conservative dynamics and up to 3.5PN
order in the dissipative dynamics, ignoring spin-spin terms.
The \msa{} angles are shown for an example configuration in Fig.~\ref{fig: Pv3 angles}. We can see that the agreement is poor for
all three angles at high frequencies, which correspond to the merger and ringdown. 
At lower frequencies, the PN and NR values for
$\alpha$ and $\gamma$ agree well, but for $\beta$ do not. As noted earlier, this is because the \pn{} $\beta$ describes the 
inclination of the orbital plane with respect to $\hat{\mathbf{J}}$, which differs from the inclination of the QA direction. 

In the next section we apply higher-order \pn{} information to improve the \pn{} estimate of $\beta$.

\subsection{Higher-order \pn{} corrections to $\beta$}
\label{sec: beta approx}

As discussed in \sect{sec:prelims}, in the quadrupole approximation the maximum GW signal power is emitted perpendicular to the orbital plane, 
and therefore the angles that describe the precession dynamics of the orbital plane are the same as those associated with the \qa{} frame
 of the GW signal~\cite{Schmidt:2010it,OShaughnessy:2011pmr,Boyle:2011gg}; this motivated the original \qa{} procedure
 presented in Ref.~\cite{Schmidt:2010it}. For the full signal, this identification is only
 approximate~\cite{Schmidt:2010it,Ochsner:2012dj,Boyle:2014ioa,Hamilton:2018fxk}, and we expect the approximation to be
 less accurate at higher frequencies. 
 Our modelling approach is based on applying a frequency-dependent rotation to a model of the waveform in the co-precessing \qa{} frame,
 and as such the rotation angles should be those associated with the \emph{signal}. However, all current
 models~\cite{Hannam:2013oca,Pan:2013rra,Taracchini:2013rva,Khan:2018fmp} use the angles associated with the \emph{dynamics}.

As we saw in Fig.~\ref{fig: Pv3 angles}, the \msa{} dynamics $\alpha$ and $\gamma$ provide a good approximation to the 
corresponding \nr{} signal angles at low frequencies, but the \msa{} $\beta$ does not. 
Fortunately, we have access to \pn{} signal amplitudes beyond the quadrupole approximation, and can use these to calculate a more accurate
estimate of the signal $\beta$.
One way to do this would be to calculate a full \pn{} waveform, e.g., from the model in Ref.~\cite{Chatziioannou:2017tdw}, and apply
the quadrupole-alignment
procedure to calculate $\beta$. However, this will be much more computationally expensive than the current \msa{} approximant,
and it is possible to obtain a sufficiently accurate result with a simpler approach.

In this calculation we will refer to the opening angle of the orbital plane with respect to $\mathbf{J}$ as $\iota$, and continue to denote 
the opening angle of the \qa{} frame by $\beta$.

To illustrate our approach, consider the rotation from a co-precessing signal that contains only the $(\ell=2,|m|=2)$ multipoles,
$h_{2,\pm2}^{\rm NP}$, to produce a precessing-binary signal in the inertial frame. We begin in the quadrupole approximation, where
the inertial frame is identified with the precession of the orbital plane, and so we use the opening angle $\iota$. 
We will focus on only the resulting $(2,2)$ and $(2,1)$ multipoles in the inertial frame,
and only the angles $\iota, \alpha$ (since the additional phase rotation $\gamma$ will not affect our argument). The 
precessing-binary signal in the inertial frame, $h^{\rm P}$, is now, \begin{eqnarray}
h^{\rm P}_{2,2} & = &  e^{- 2 i \alpha} \left( \cos^4\left(\frac{\iota}{2}\right) h_{2,2}^{\rm NP} + \sin^4\left(\frac{\iota}{2}\right) h_{2,-2}^{\rm NP}\right), \\
h^{\rm P}_{2,1} & = & - 2 e^{- i \alpha} \left(  \cos^3\left(\frac{\iota}{2}\right) \sin\left(\frac{\iota}{2}\right) h_{2,2}^{\rm NP} \right. \nonumber \\
& & \left. \ \ \ \ \ \ \ \ \ \ - \cos\left(\frac{\iota}{2}\right) \sin^3\left(\frac{\iota}{2}\right) h_{2,-2}^{\rm NP} \right) .
\end{eqnarray}

The non-precessing multipoles can be written as,
\begin{equation}
h_{2,\pm2}^{\rm NP} = A e^{\mp 2 i \Phi},
\end{equation} where $A$ and $\Phi$ are the time/frequency-dependent amplitude and orbital phase. When $\iota$ is small, $h_{2,2}^{\rm NP}$
makes the strongest contribution to the precessing-waveform multipoles, and we see that $\iota$ determines the relative amplitude of
$h^{\rm P}_{2,2}$ and $h^{\rm P}_{2,1}$. We can isolate the $e^{-2 i \Phi}$ term as follows,
\begin{eqnarray}
\bar{h}^{\rm P}_{2,2} & = & \frac{1}{2\pi} \int_0^{2\pi} h^{\rm P}_{2,2} e^{2 i \Phi} d\Phi \\
& = & A e^{-2 i \alpha} \cos^4\left(\frac{\iota}{2}\right), \\
\bar{h}^{\rm P}_{2,1} & = & - 2 A e^{-i \alpha} \cos^3\left(\frac{\iota}{2}\right) \sin\left(\frac{\iota}{2}\right).
\end{eqnarray} From these we can readily calculate that the inclination $\iota$ is \begin{equation}
\iota = 2 \tan^{-1} \left( \frac{|\bar{h}^{\rm P}_{2,1}|}{2 |\bar{h}^{\rm P}_{2,2}|} \right). \label{eqn:easybeta}
\end{equation} At leading (quadrupole) order, $\iota$ is the precession angle $\beta$.

If we now use higher-order \pn{} amplitude expressions~\cite{Arun:2008kb}, then the angle $\beta$ that identifies the frame in which 
the $(\ell=2,|m|=2)$ multipoles are maximised will not necessarily be the same as the inclination angle $\iota$, 
but the expression above \emph{will} still give us an estimate of the orbit-averaged $\beta$. Note that the \msa{} angles
in Ref.~\cite{Chatziioannou:2017tdw} are also orbit-averaged (i.e., nutation effects are absent),
so this is a consistent treatment.

The multipole expressions in
Ref.~\cite{Arun:2008kb} are given in terms of the orbital phase $\Phi$, the precession angles $\alpha$ and $\iota$, and the spin components.
For the spin components, we make an approximate reduction
to our single-spin systems as follows. The inclination of the spin from the $z$-axis is the spin's inclination from the orbital angular momentum
vector, $\tls$, minus the inclination of the orbital angular momentum from the $z$-axis, $\iota$. The azimuthal angle of the spin vector
is $(\alpha + \pi)$, because, since $\mathbf{L} = \mathbf{J} - \mathbf{S}$, the $x$-$y$-plane components of $\mathbf{L}$ and $\mathbf{S}$ 
will be in opposite directions, and so their azimuthal angles will differ by $\pi$. 
The final result, for a given configuration,
depends only on the dynamics inclination $\iota$ as a function of frequency; we use the \msa{} expression for $\iota(f)$.

In Ref.~\cite{Arun:2008kb} the amplitudes are expanded in powers of $v = (\pi f)^{1/3}$.
We define $\delta = m_1 - m_2$, where $m_1> m_2$, and so $\delta > 0$;
$\eta = m_1 m_1 / (m_1+m_2)^2$, $\chi_{\rm s} = (\chi_1 + \chi_2)/2$, $\chi_{\rm a} = (\chi_1 - \chi_2)/2$,
and so,
\begin{eqnarray}
\chi_{{\rm s/a},x} & = & \chi \sin(\tls - \iota) \cos(\alpha + \pi)/2, \nonumber \\
\chi_{{\rm s/a},y} & = & \chi \sin(\tls - \iota) \sin(\alpha + \pi)/2, \nonumber \\
\chi_{{\rm s/a},z} & =  & \chi \cos(\tls - \iota)/2. \label{eq:chixyz}
\end{eqnarray} If we substitute these into the \pn{} multipole expressions for $h_{2,2}^\text{P}$ and $h_{2,1}^\text{P}$, and then
apply Eq.~(\ref{eqn:easybeta}), we obtain the relatively simple expression,
\def\markSec{\sec}
\begin{equation}
   \beta =  2 \tan^{-1} \left( \frac{\markSec{}(\iota/2) \left( c_0 + c_2 v^2 + c_3 v^3 \right) }{d_0 + d_2 v^2 + d_3 v^3} \right), \label{eqn:betafix}
\end{equation} where \begin{eqnarray}
c_0 & = & 84 \sin \iota, \\
c_2 & = & ( 110 \eta  -214 ) \sin \iota, \nonumber \\
c_3 & = &  - 7  (6 + 6 \delta + 5 \eta) (2 \cos \iota -1) \chi \sin \tls, \nonumber \\
& & + 56 \left( 3 \pi - (1 + \delta - \eta) \chi \cos \tls \right)  \sin \iota,  \nonumber \\
d_0 & = & 84 \cos \left(\frac{\iota}{2} \right)  , \nonumber \\
d_2 & = & (110 \eta -214) \cos \left(\frac{\iota }{2}\right)  , \nonumber \\
d_3 & = & 14  (6+ 6 \delta +5 \eta) \chi \sin \tls  \sin \left(\frac{\iota }{2}\right) \nonumber \\
   & & + 56 \cos \left(\frac{\iota}{2}\right) \left( 3\pi - (1 + \delta - \eta  ) \chi  \cos \tls  \right).
\end{eqnarray}

\begin{figure}[htbp]
   \centering
   \includegraphics[width=0.47\textwidth]{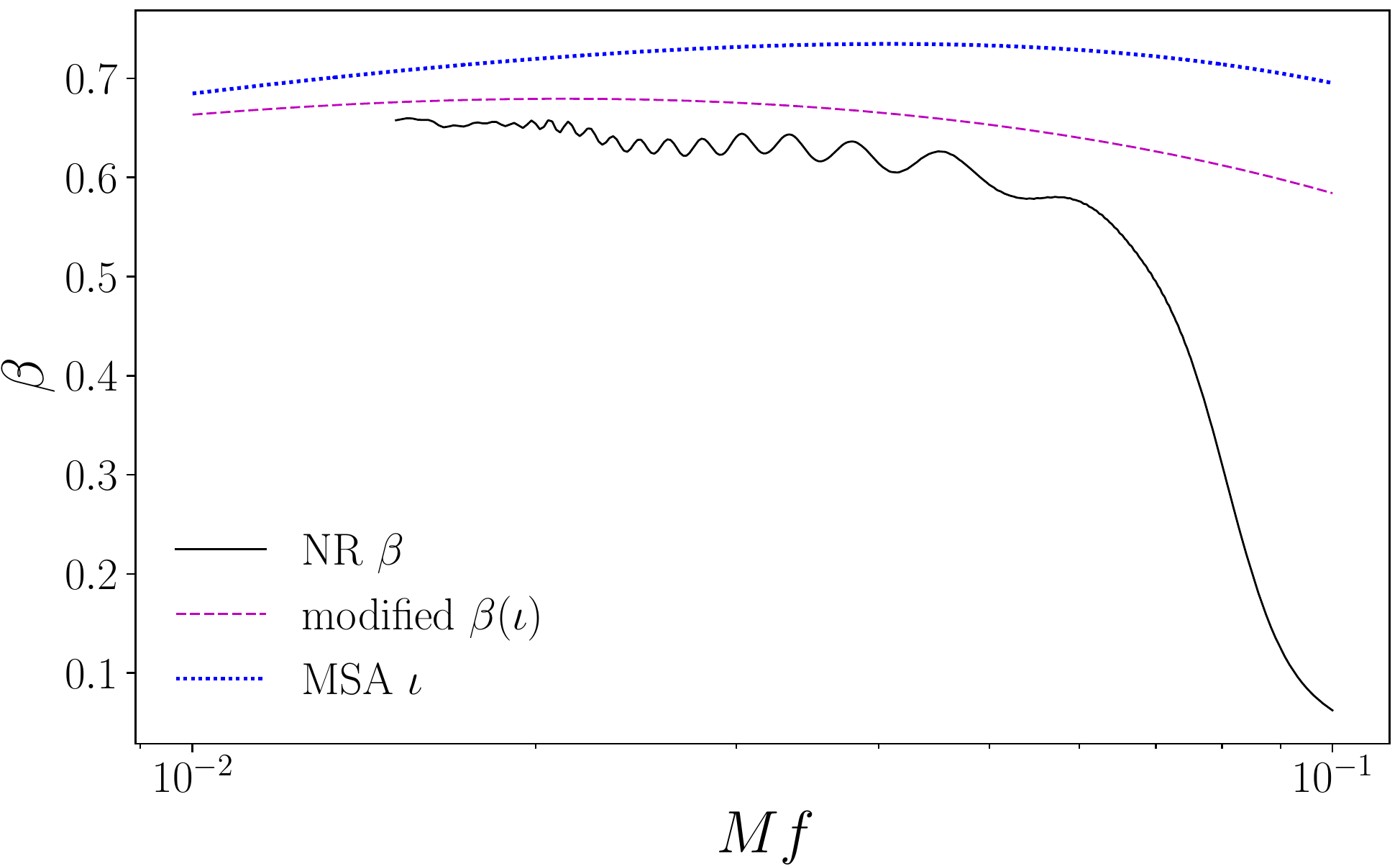}
   \caption{Opening angles for the \((q,\chi,\tls)=(8,0.8,60^\circ)\) configuration. Solid black: the \nr{} opening angle of the \qa{} frame, 
   $\beta$. Dotted blue: the \pn{} opening angle of the orbital plane, $\iota$. Dashed magenta: Approximate \qa{} angle $\beta$ as a function of 
   $\iota$; see text for details.}
   \label{fig:betterbeta}
\end{figure}

Fig.~\ref{fig:betterbeta} also shows the modified $\beta(\iota)$ for the \((q,\chi,\tls)=(8,0.8,60^\circ)\) configuration.
We see the \pn{} inspiral $\beta(\iota)$ now shows much better agreement with the \nr{} result at low frequencies. 
We find similar results across the parameter space that we have considered, and therefore
to calculate $\beta$ in our model, we use Eq.~(\ref{eqn:betafix}) in conjunction with the \msa{} $\iota$ as calculated
in Refs.~\cite{Chatziioannou:2017tdw,Khan:2018fmp}, to construct $\beta$ through the inspiral. The features of the \nr{} 
$(\alpha,\beta,\gamma)$ at higher frequencies, which are not captured at all by the \pn{} expressions, will be explicitly modelled in 
\sect{sec:angle model MR}.

\subsection{Two-spin $\beta$}\label{sec: two spin beta}

\begin{figure*}[htbp]
   \centering
   \includegraphics[width=\textwidth]{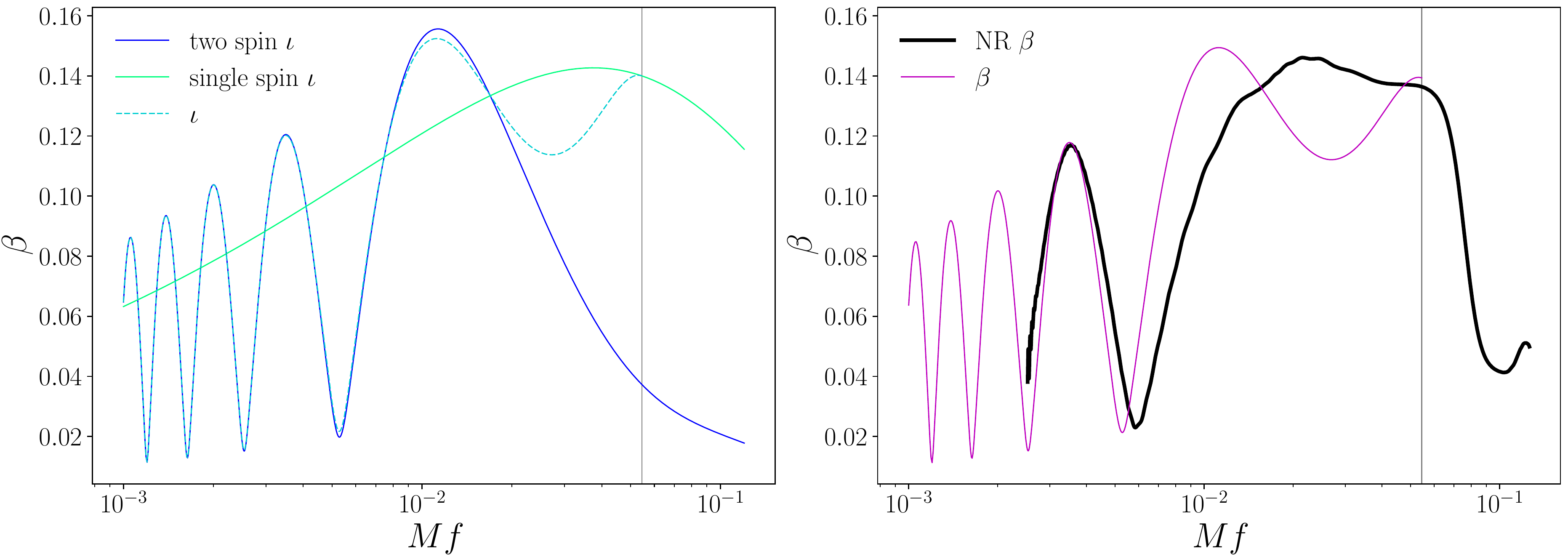}
   \caption{Various options for the \pn{} expression for the opening angle. The left-hand panel shows the \pn{} value of $\iota$ for a two-spin system (blue) and for the equivalent single-spin system (green) calculated using the expressions used in \textsc{PhenomPv3}. In light blue is shown the effect of tapering the two-spin oscillations to the single-spin value at the connection frequency $f_\text{c}$, shown as a grey vertical line. In the right-hand panel the value for $\beta$ used in the model (pink) is compared with the \nr{} value of $\beta$ found for this case. We only show $\iota$ and $\beta$ up to $f_\text{c}$, since the merger-ringdown model is used at higher frequencies. The configuration shown is SXS1397 
 in Table~\ref{tab: SXS and GT cases}.
   }
   \label{fig: two spin beta}
\end{figure*}

The \msa{} $\iota$ for a two spin system shows oscillations that become unphysically large through late 
inspiral and towards merger and which are not seen in the precession angles calculated for two-spin \nr{} systems, as can be seen in 
Fig.~\ref{fig: two spin beta}. These oscillations 
also complicate connecting the inspiral expression to the single-spin-tuned merger-ringdown ansatz. We therefore taper these oscillations 
to recover the value and gradient of $\beta$ for an equivalent single-spin system at the point at which we wish to connect the inspiral and merger-ringdown parts of the model.

For a system described by two spins $\mathbf{S_1}$ and $\mathbf{S_2}$ we use the mapping to the appropriate single spin system 
defined in \sect{sec: spin parameterisation}: $\mathbf{S'_1}$ is given by Eq.~(\ref{eqn: single spin components}) and 
$\mathbf{S'_2} = \left(0,0,0\right)$. We evaluate the 
PhenomPv3 expression for $\iota$ for both of these configurations and identify the oscillations introduced by the two-spin effects 
as,
\begin{align} 
   \iota_\text{osc} = {}& \iota\left(\mathbf{S_1},\mathbf{S_2}\right) - \iota\left(\mathbf{S'_1},\mathbf{S'_2}\right).
\end{align}
We then apply a taper to these oscillations that ensures $\iota$ will tend to the single spin value and gradient at a given frequency $f_\text{c}$
and add the oscillations back to the single-spin function. The final two-spin expression for $\iota$ is then given by
\begin{equation} 
   \iota =%
   \begin{cases} 
   \iota\left(\mathbf{S'_1},\mathbf{S'_2}\right) + \cos^2\left( \frac{2\pi f}{4 f_\text{c}} \right) \times \iota_\text{osc} & f \le f_\text{c} \\
   \iota\left(\mathbf{S'_1},\mathbf{S'_2}\right) & f > f_\text{c},
   \end{cases}
\end{equation}
where $f_\text{c}$ is the frequency at which the inspiral expression for $\beta$ is connected to the merger-ringdown expression defined below 
in Eq.~(\ref{eqn: fc def}). 

Given an estimate for the dynamics $\iota$, we now wish to rescale it to produce an estimate for the signal $\beta$, as described in
\sect{sec: beta approx}. To do this we also need an estimate of the frequency-dependent in-plane spin component, and therefore
$\chi$ and $\tls$, as required in Eqs.~(\ref{eq:chixyz}). 
We assume that the component of the spins parallel to the orbital angular momentum, \(S_\parallel\), remains fixed. 
We further approximate that the frequency dependence of the magnitude of \(\mathbf{J}\) is dominated by changes to the magnitude of \(\mathbf{L}\), 
\begin{equation}
J(f) = J_0 + L(f) - L_0,
\end{equation}
where the magnitude \(L\) is given by the 3PN expression for the orbital angular momentum used by \pv3{} to calculate \(\iota\) and the 0-subscript denotes quantities specified at the reference frequency.
As such, we may write the frequency-dependent in-plane spin component \(S_\perp\) as
\begin{equation}
S_\perp(f) = J(f)\sin\iota
\end{equation}
Substituting this expression for $S_{\rm p}$ in Eq.~(\ref{eqn: chi_p}) we get a value for 
$\chi_{\rm p}$. The quantities $\chi$ and $\cos\tls$ are then calculated as described in Eqs.~(\ref{eqn: chi_eff})-- (\ref{eqn: costheta}) and 
these values are used to rescale $\iota$ to produce $\beta$, according to Eq.~(\ref{eqn:betafix}).

The effect of this treatment can be seen in Fig.~\ref{fig: two spin beta}, which shows $\beta$ for SXS1397 (the intrinsic properties of which are given in Tab.~\ref{tab: SXS and GT cases}). The \pn{} expression for the angle captures the oscillations seen at low frequency very well. However, these oscillations do not continue to high frequency and are greatly over-estimated by the full two-spin \pn{} expression. Tapering the oscillations to the single spin value at the connection frequency resolves this issue well. For $f>f_\text{c}$ the \pn{} expression is replaced by the merger-ringdown expression described in the following section, so the behaviour of the \pn{} angles here are not an issue. In the rare event where the merger-ringdown contributions are not attached (see \csec{sec: angles beyond calibration}), only the effective single-spin beta is used beyond \(f > f_\text{c}\).

\section{Precession angle model: merger-ringdown}
\label{sec:angle model MR}

The \pn{} expressions for the precession angles cannot be reliably extended through merger and ringdown and when compared with 
the \nr{} angles do not capture the features present at high frequency, as was clear in Fig.~\ref{fig: Pv3 angles}. 
We therefore present a phenomenological description of the precession angles $\alpha$ and $\beta$ in the merger-ringdown regime; the
remaining angle $\gamma$ can then be calculated via Eq.~(\ref{fdc}).
We describe the functional form of the angles and produce a global fit for each of the co-efficients of the ansatz. This provides a frequency domain description of the precession angles across the parameter space.

\begin{figure*}[htbp]
   \centering
   \includegraphics[width=\textwidth]{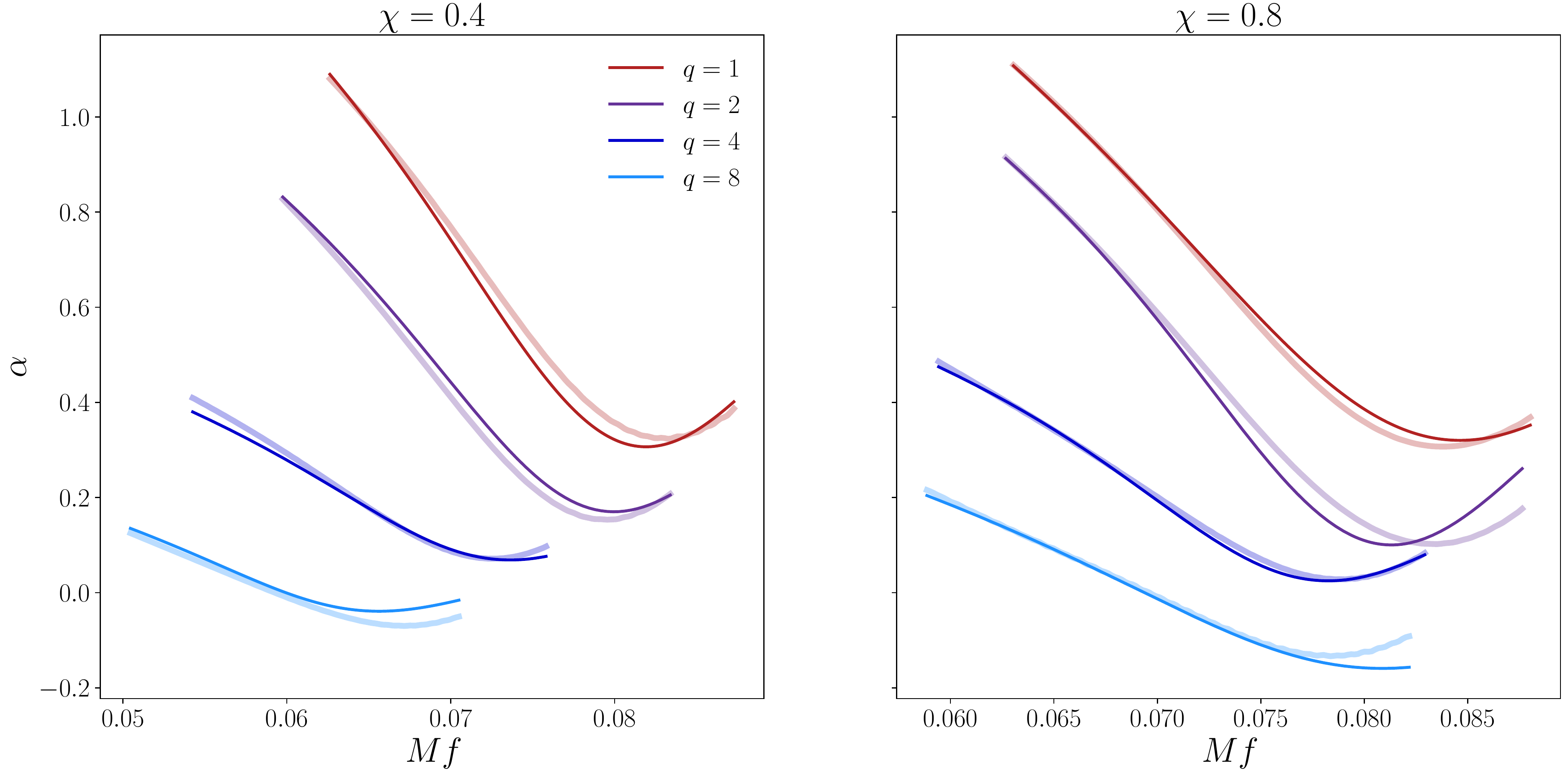}
   \caption{Comparison of the phenomenological ansatz presented in Eq.~(\ref{eqn: alpha}) (solid lines) with the \nr{} data (translucent lines) over the frequency range to which the co-efficients in the ansatz were tuned for a selection cases in the \nr{} catalogue with $\tls=90^\circ$ at varying mass ratios. We have made use of the freedom to choose a constant offset in $\alpha$ in order to offset the curves shown here to make them easier to distinguish.
   }
   \label{fig: alpha fits}
\end{figure*}

\begin{figure*}[htbp]
   \centering
   \includegraphics[width=\textwidth]{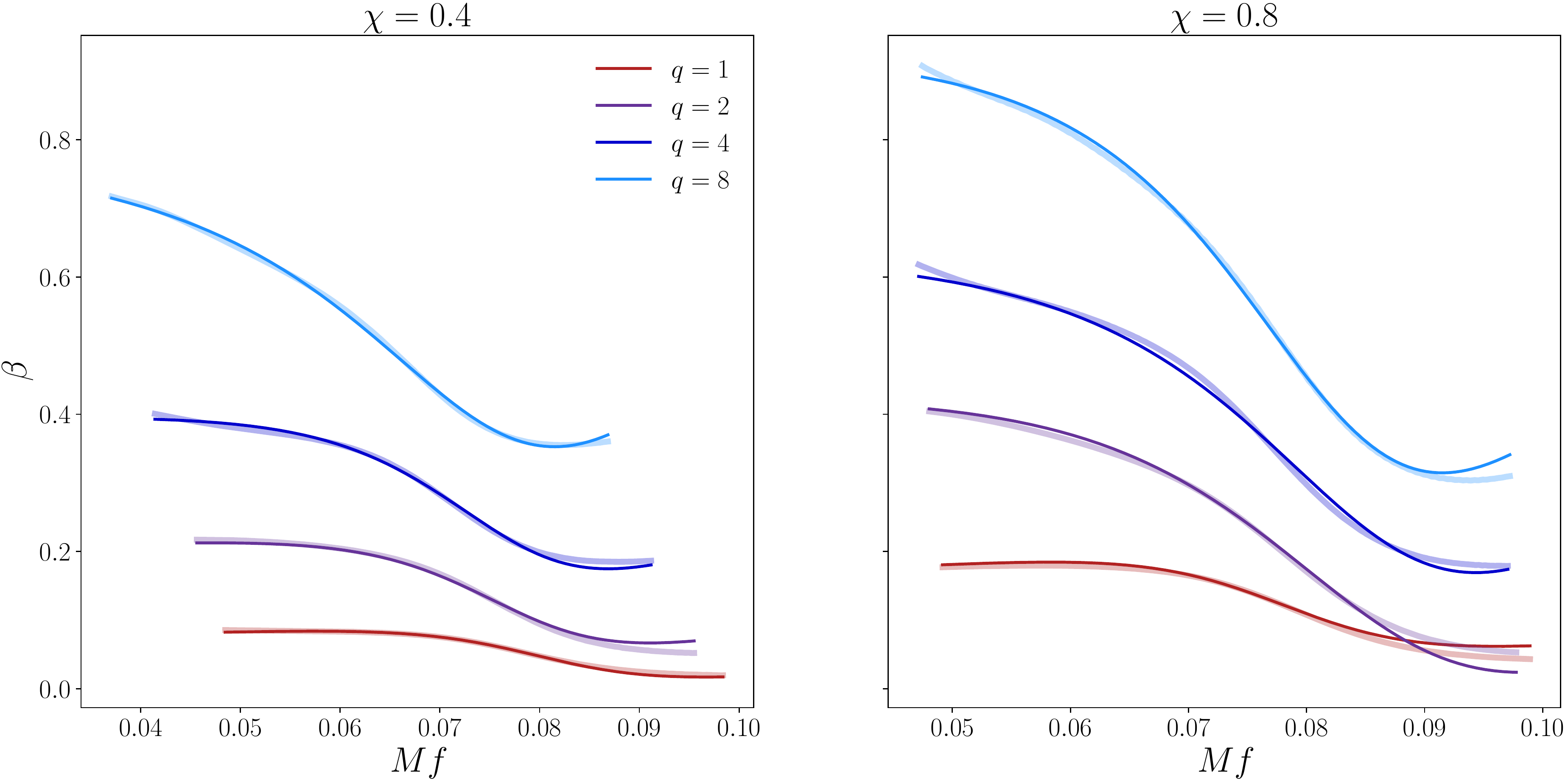}
   \caption{Comparison of the phenomenological ansatz presented in Eq.~(\ref{eqn: beta}) (solid lines) with the \nr{} data (translucent lines) over the frequency range to which the co-efficients in the ansatz were tuned for a selection of cases in the \nr{} catalogue with $\tls=90^\circ$ at varying mass ratios.}
   \label{fig: beta fits}
\end{figure*}
 
\subsection{Functional forms of $\alpha$ and $\beta$}

The morphology of the merger-ringdown part of $\alpha$ is qualitatively very similar to that of the phase derivative, seen in
Ref.~\cite{Husa:2015iqa, Khan:2015jqa}. $\alpha$ shows a $1/f$ fall-off with a Lorentzian dip centred around what is approximately the ringdown 
frequency of the \bbh{} system. This prompts the ansatz,
\begin{align}\label{eqn: alpha}
   \alpha \left( f \right) - \langle \alpha \left( f \right) \rangle = {}& \frac{A_1}{f} + \frac{A_2 \sqrt{A_3}}{A_3 + \left( f - A_4 \right)^2},
\end{align}
where $A_1$, $A_2$, $A_3$ and $A_4$ are free co-efficients.

The fitting region is based around the Lorentzian dip; it is defined to be the range $ f_\text{dip} - 0.0225 \le f \le f_\text{dip} + 0.0075$,
where $f_\text{dip}$ is the frequency at which $\alpha$ reaches its minimum, and recall that we have chosen $M=1$.  
The global fit for $\alpha$ within this fitting region has a root mean square error of $4.80\times10^{-5}$, averaged across the 40 waveforms.
Some example comparisons of the result of these fits with the \nr{} value for $\alpha$ are shown in Fig.~\ref{fig: alpha fits}.

During merger and ringdown, $\beta$ drops rapidly as the dominant emission direction relaxes to its final direction, as discussed in more
detail in \sect{sec:physfeat}.  The ansatz used to describe $\beta$ is therefore chosen to grow at low frequencies (as seen in the \pn{} expressions), 
turnover at the correct frequency, capture the drop and finally tend asymptotically towards the constant value to which the dominant emission direction 
relaxes. The ansatz we chose to describe this behaviour is,
\begin{align}\label{eqn: beta}
   \beta \left( f \right) - \langle \beta \left( f \right) \rangle = {}& \frac{B_1 + B_2 f + B_3 f^2}{1 + B_4 \left( f + B_5 \right)^2},
\end{align}
where $B_1$, $B_2$, $B_3$, $B_4$ and $B_5$ are free co-efficients. 

The fitting region for $\beta$ is centred around the inflection point in the turnover $f_\text{inf}$; $f  \in  f_\text{inf} \pm 0.03$. The global fit for 
$\beta$ within this fitting region has a root mean square error of $7.47\times10^{-6}$, averaged across the 40 waveforms. 
Some example comparisons of the result of these fits with the \nr{} value for $\beta$ are shown in Fig.~\ref{fig: beta fits}.

It should be noted that a key feature of the above ansatz is that it does \emph{not} fall to zero after merger. This feature can be seen in both the time 
and frequency domain values of $\beta$, as shown in Figs.~\ref{fig: beta fits} and \ref{fig:rdcomp2}. We discuss this in more detail
in \sect{sec:physfeat}.

\subsection{The phenomenological co-efficients}\label{sec: map to phys}

The two ans{\"a}tze given above, which describe the merger-ringdown behaviour of $\alpha$, Eq.~(\ref{eqn: alpha}), and $\beta$, 
Eq.~(\ref{eqn: beta}), have 10 free co-efficients between them. Each of these co-efficients was fit across the three-dimensional 
parameter space described by the symmetric mass ratio, $\eta$, the dimensionless spin magnitude, $\chi$, and the cosine of the angle 
between the orbital angular momentum and the spin angular momentum, \(\cos\tls\).

The optimum value of each of the co-efficients for each waveform in the calibration set was found by fitting the relevant ansatz to the \nr{} 
data using the non-linear least-squares fitting function \verb|curve_fit| from the python package \verb|Scipy|~\cite{2020SciPy-NMeth}. 
This function uses the Levenberg-Marquardt algorithm to perform the least-squares fitting. We then performed a three-dimensional fit of 
each of the co-efficients using the fitting algorithm \verb|mvpolyfit|~\cite{London:2018nxs,positive:2020}.  
This gives each of the co-efficients as a 
polynomial expansion in $\eta$, $\chi$, $\cos\tls$. We specify the terms that appear in the expansion and the algorithm finds the co-efficients of 
these terms that optimise the fit as well as a measure of how good the fit is. Since we have 40 calibration waveforms, the maximum possible 
number of terms 
that can appear in these expressions is 39 in order to avoid over fitting. The fits are restricted so that the highest order term in each 
dimension is one less than the total number of data points in that dimension. Since the value of each of the co-efficients in the ansatz is to 
some extent dependent on 
the value of each of the other co-efficients, we found a global fit for each co-efficient in turn, re-fitting the ansatz to the data while 
keeping fixed the co-efficients that had already been fit. 
We first fitted the co-efficients that varied most smoothly across the parameter space and those for which the general behaviour across the 
parameter space was already understood. For $\alpha$ this meant we first fitted the location of the dip, $A_4$, followed by the other co-efficients 
in the order $A_1$, $A_2$ and $A_3$. For $\beta$ we fitted the value of $\langle \beta \left( f \right) \rangle$ separately as this had a clear parameter 
space trend. We then fitted the co-efficients in the order $B_1$, $B_2$, $B_3$, $B_5$ and $B_4$ since the co-efficients in the numerator were generally 
better behaved than those in the denominator.

The general expression for each co-efficient is
\begin{align}\label{eqn: global fit expression}
   \Lambda^{i} = {}& \sum_{p=0}^3 \sum_{q=0}^1 \sum_{r=0}^4 
   \lambda^i_{pqr} \eta^p \chi^q \cos^r\tls,
\end{align}
where $\Lambda \in \left[A, B\right]$ are the co-efficients in the ansatz describing $\alpha$ and $\beta$ respectively and $i \in [1,2,3,4]$ and 
$[0,1,2,3,4,5]$ respectively. The $\lambda^{i}_{pqr}$ give the co-efficients of the polynomial expansion of the multi-dimensional fits of $\Lambda_{i}$. 
This expression has a maximum of 40 terms. Not all of these terms are used in the expressions for each of the co-efficients; the co-efficient with the 
fewest number of terms has only 25 while that with the greatest number of terms contains 39.

The co-efficients for $\alpha$ and $\beta$ vary smoothly across the parameter space, as can be seen in Figs.~\ref{fig: alpha coefficients} 
and \ref{fig: beta coefficients} in Appendix~\ref{app:fitsurfaces} respectively. The residual plots above the fit surfaces show that the global fits 
agree closely with the values 
of the co-efficients found from fitting the ansatz to each individual simulation.

\section{Full inspiral-merger-ringdown angle model}
\label{sec:fullmodel}

The expressions for the precession angles for the two distinct inspiral and merger-ringdown regions are connected so that the 
connection is smooth and the full \imr{} expression for the angles agrees with the \nr{} data over the entirety of the region for which 
it is available. The method used to connect the two regions was different for each angle.

\subsection{Connection method for $\alpha$}

For $\alpha$, the regions are connected using an interpolating function of the form
\begin{align}\label{eqn: alpha interp}
   \alpha_{\text{interp}}\left(f\right) = {}& a_0 f^2 + a_1 f + a_2 + \frac{a_3}{f},
\end{align}
defined over the frequency range $\left[f_1, f_2\right]$. 
This range was chosen to be as small as possible. 
The lower frequency limit was chosen to be the highest frequency for which the inspiral expressions agreed with the \nr{} data while the 
upper frequency limit was chosen to be the lower limit for which the fitted merger-ringdown expressions still agreed well with the \nr{} data. 
Since the \msa{} \pn{} expressions for the angles agree well with the \nr{} data over most of the waveform, there is a wide range of frequency 
values over which the interpolation could be performed. 
We choose the frequency range to be defined in terms of the location of the Lorentzian dip, $A_4$: $f_1 = 2 A_4/7$ and $f_2 = A_4/3$. 

The co-efficients of Eq.~(\ref{eqn: alpha interp}) are chosen so that
\begin{enumerate}
   \item{$\alpha_{\text{interp}}\left(f_1\right) = \alpha_\text{PN}\left(f_1\right)$ and $\alpha_{\text{interp}}\left(f_2\right) = \alpha_\text{MR}\left(f_2\right)$, 
   since there is freedom in an overall constant offset in $\alpha$,}
   \item{$\alpha'_{\text{interp}}\left(f_1\right) = \alpha'_\text{PN}\left(f_1\right)$ and $\alpha'_{\text{interp}}\left(f_2\right) = \alpha'_\text{MR}\left(f_2\right)$ 
   in order to ensure the two parts are connected continuously.} 
\end{enumerate}
$\alpha_\text{PN}$ is the \msa{} \pn{} expression used for $\alpha$ in the inspiral regime. $\alpha_\text{MR}$ is the merger-ringdown ansatz given in 
Eq.~(\ref{eqn: beta}). The co-efficients are given by
\begin{widetext}
\begin{align}
   a_0 = {}& \frac{1}{D} \left[ 2\left( f_1 \alpha_1 - f_2 \alpha_2 \right) - \left( f_1 - f_2 \right) \left( \left( f_1 \alpha'_1 + f_2 \alpha'_2 \right) + \left( \alpha_1 - \alpha_2 \right) \right) \right], \nonumber \\
   a_1 = {}& \frac{1}{D} \left[ 3 \left( f_1 + f_2 \right) \left( f_1 \alpha_2 - f_2 \alpha_1 \right) + \left( f_1 - f_2 \right) \left( \left( f_1 + 2 f_2 \right) \left( f_1 \alpha'_1 + \alpha_1 \right) + \left( 2 f_1 + f_2 \right) \left( f_2 \alpha'_2 + \alpha_2 \right) \right) \right], \nonumber\\
   a_2 = {}& \frac{1}{D} \left[ 6 f_1 f_2 \left( f_1 \alpha_1 - f_2 \alpha_2 \right) + \left( f_1 - f_2 \right) \left( f_2 \left( 2 f_1 + f_2 \right) \left( f_1 \alpha'_1 + \alpha_1 \right) + f_1 \left( f_1 + 2 f_2 \right) \left( f_2 \alpha'_2 + \alpha_2 \right) \right) \right], \nonumber\\
   a_3 = {}& \frac{1}{D} \left[ f_1 f_2^2 \left( f_2 - 3 f_1 \right) \alpha_1 - f_1^2 f_2 \left( f_1 - 3 f_2 \right) \alpha_2  + f_1 f_2 \left( f_1 - f_2 \right) \left( f_2 \left( f_1 \alpha'_1 + \alpha_1 \right) + f_1 \left( f_2 \alpha'_2 + \alpha_2 \right) \right) \right],
\end{align}
\end{widetext}
where $\alpha_i$ and $\alpha'_i$, \(i=1,2\), are the value of $\alpha$ and its derivative at the limits of the frequency range and $D = (f_2 - f_1)^3$. 

\subsection{Connection method for $\beta$}

For $\beta$, the agreement between the \pn{} expression and the \nr{} data is insufficient to employ the interpolation method described above. 
Even including the higher order amplitude corrections described in \sect{sec: beta approx}, the starting frequency of the \nr{} simulations is not low 
enough in order to cover the region in which the \pn{} expression closely matches the data for all cases. Instead, we employ a rescaling function that 
leaves the \pn{} expression invariant at low frequencies but ensures it smoothly connects with the merger-ringdown value of $\beta$ at the connection 
frequency $f_\text{c}$. This rescaling function is given by
\begin{align}
   k\left(f\right) = {}& 1 + b_1 f + b_2 f^2,
\end{align}
which tends to one at low frequencies thus leaving the \pn{} expression unchanged. 
In order to ensure the value of $\beta$ and its derivative match at the connection frequency, the co-efficients $b_1$ and $b_2$ are given by
\begin{align}
   b_1 = {}& -\frac{1}{\beta_1^2 f_\text{c}}\left[-2 \beta_1 \left(\beta_2 - \beta_1\right) + \left(\beta_1\beta'_2 - \beta_2\beta'_1\right)f_\text{c} \right], \\
   b_2 = {}& -\frac{1}{\left(\beta_1 f_\text{c}\right)^2}\left[ \beta_1\left(\beta_2 - \beta_1\right) - \left(\beta_1 \beta'_2 - \beta_2 \beta'_1\right) f_\text{c} \right],
\end{align}
where the $\beta_i$ and $\beta'_i$ are the value of $\beta$ and its derivative evaluated at the connection frequency. The subscript 1 indicates that 
this is the value of $\beta$ given by the original \pn{} expressions while 2 indicates the values from the merger-ringdown expression.

The definition of the connection frequency depends on the morphology of the merger-ringdown ansatz for $\beta$ for a particular case.  As can be 
seen in Fig.~\ref{fig: beta fits}, in some parts of the parameter space $\beta$ rises gently until just before merger then turns over and drops rapidly. However, 
in other parts of the parameter space this turnover is much more gradual and begins at much lower frequencies. Our ansatz for $\beta$ captures both of 
these morphologies well. In cases where the turnover occurs within the fitting region, we define the connection frequency $f_\text{c}$ as the frequency at 
which the merger-ringdown part has a particular gradient $\text{d}\beta_{\text{c}}$. The value of this gradient varies across the parameter space. We define 
it to be 
\begin{align}
\label{eqn: grad beta c}
   \text{d}\beta_{\text{c}} = {}& 2.5 \times 10^{-4} \times \text{d}\beta_{\text{inf}}^2,
\end{align}
where $\text{d}\beta_{\text{inf}}$ is the gradient at the inflection point. The connection frequency is then found by expanding the gradient of the curve 
about the maximum as a Taylor series. We find the connection frequency is given by
\begin{align}\label{eqn: fc def}
   f_{\text{c}} = {}& f_\text{max} + \frac{1}{\beta'''}\left[-\beta''+
   \sqrt{\beta''^2+2\beta'''\text{d}\beta_{\text{c}}}\right],
\end{align}
where $f_\text{max}$ is the frequency at which the maximum occurs
and $\beta''$ and $\beta'''$ are the second and third derivatives of $\beta$ evaluated at $f_\text{max}$, respectively. 

In cases where the turnover is not present within the fitting region we instead define the connection frequency to be the lower frequency limit of the fitting region, 
thus ensuring $\beta$ is still falling at this frequency. In this case, 
\begin{equation}
f_\text{c} =
\begin{cases}
f_\text{inf} - 0.03, & f_\text{inf} \ge 0.06\\
3 f_\text{inf} / 5, & f_\text{inf} < 0.06,
\end{cases}
\end{equation} 
where $f_\text{inf}$ is the inflection point.

\begin{figure*}[htbp]
   \centering
   \includegraphics[width=0.49\textwidth]{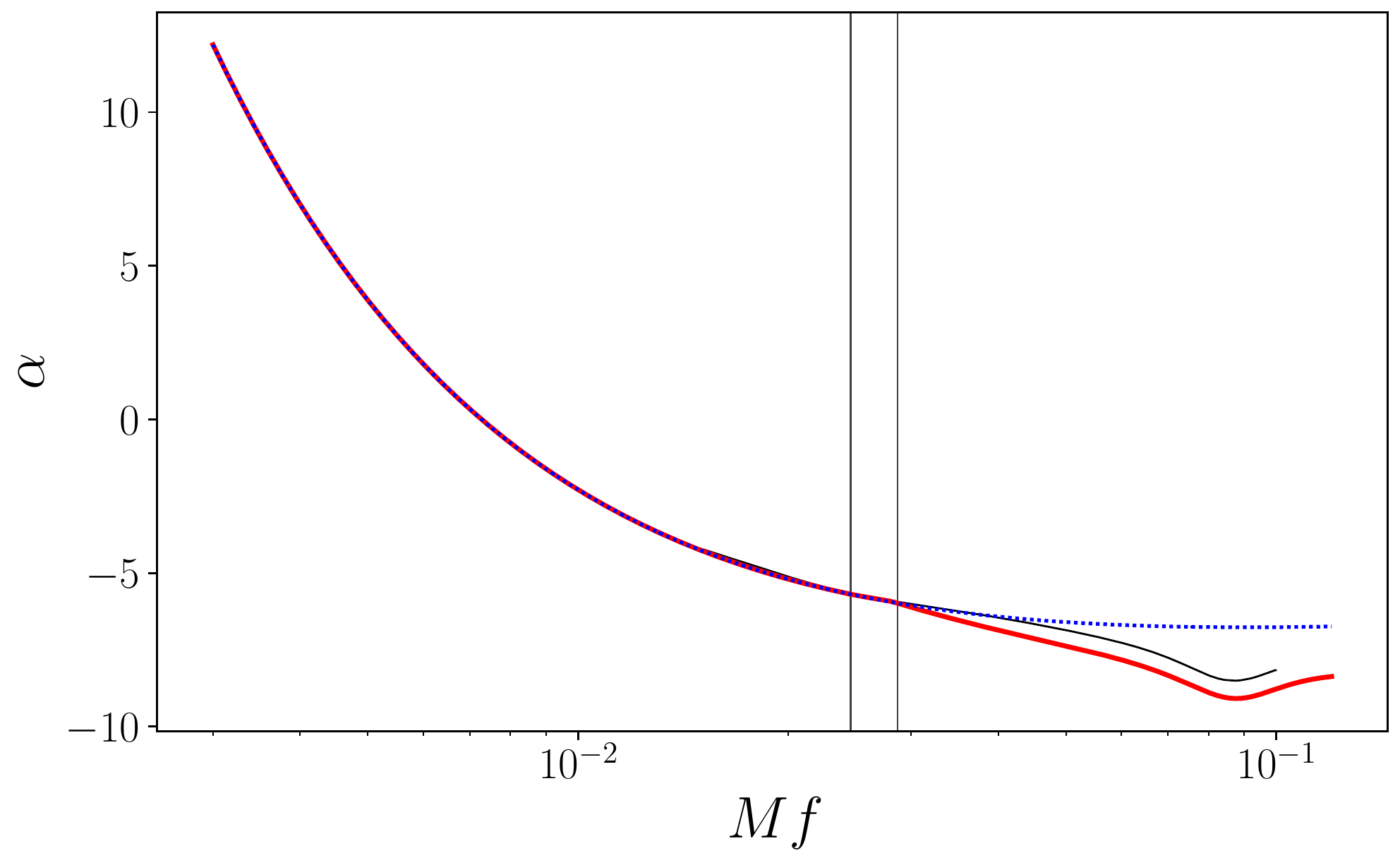} 
   \includegraphics[width=0.49\textwidth]{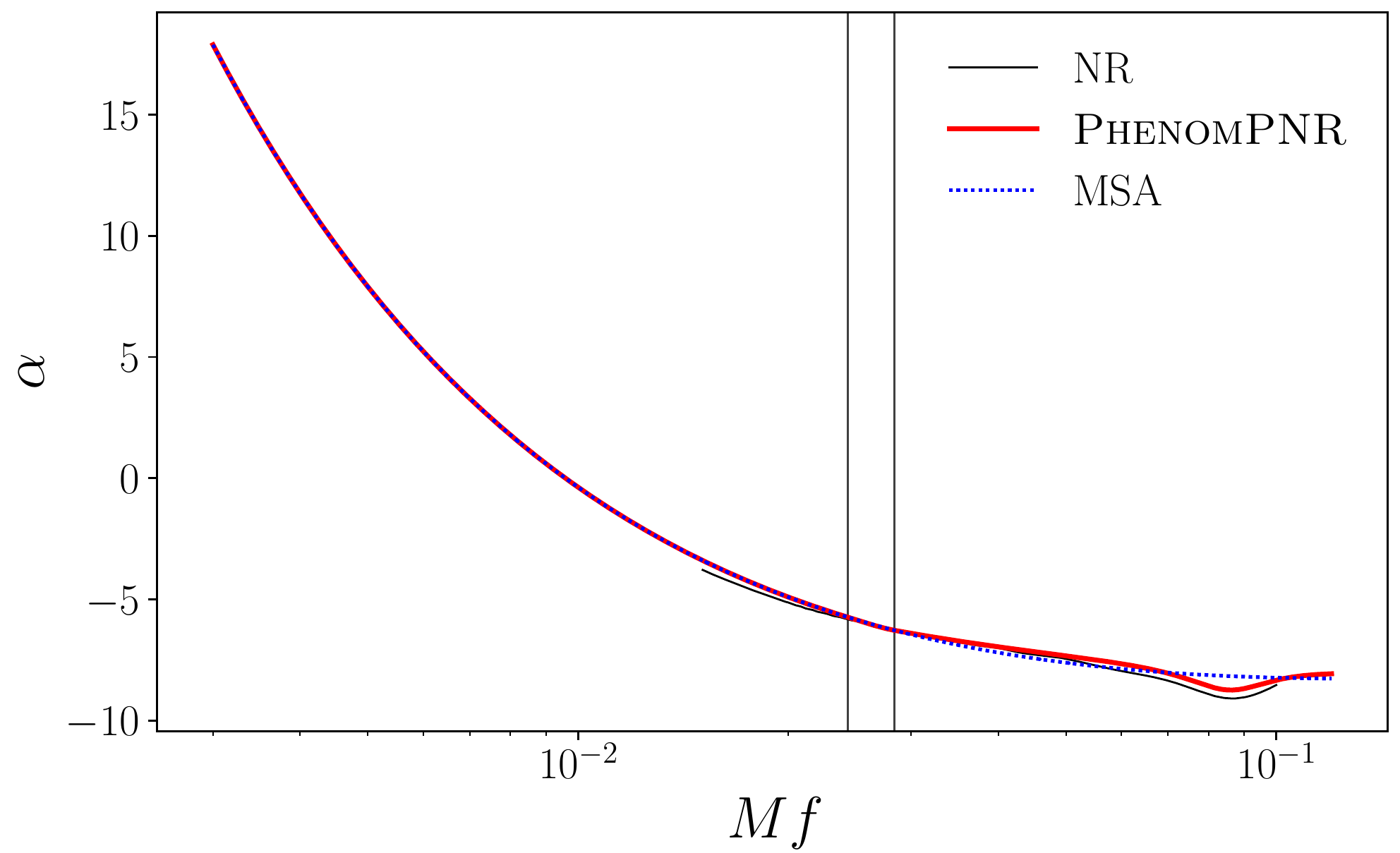} \\
   \includegraphics[width=0.49\textwidth]{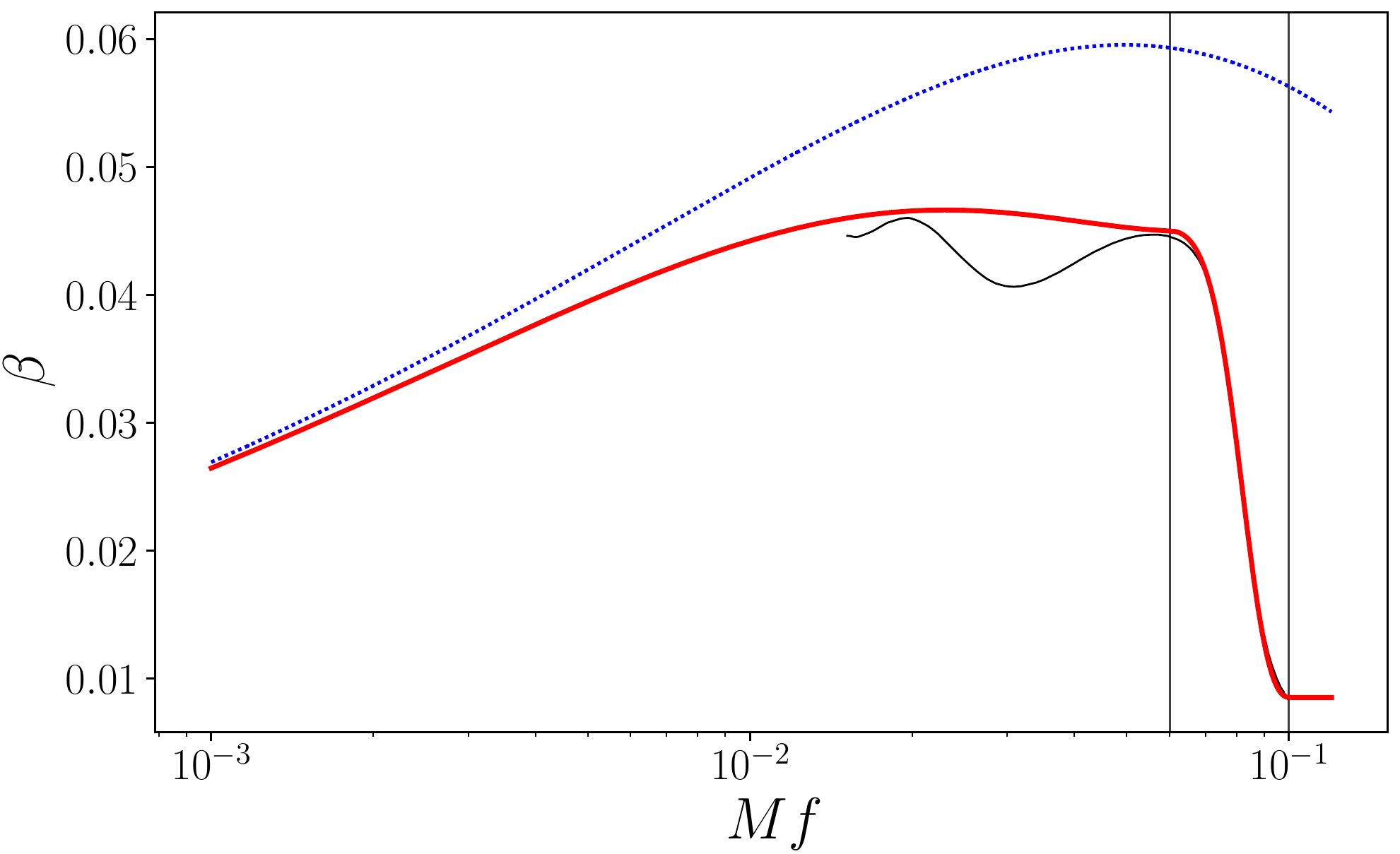} 
   \includegraphics[width=0.49\textwidth]{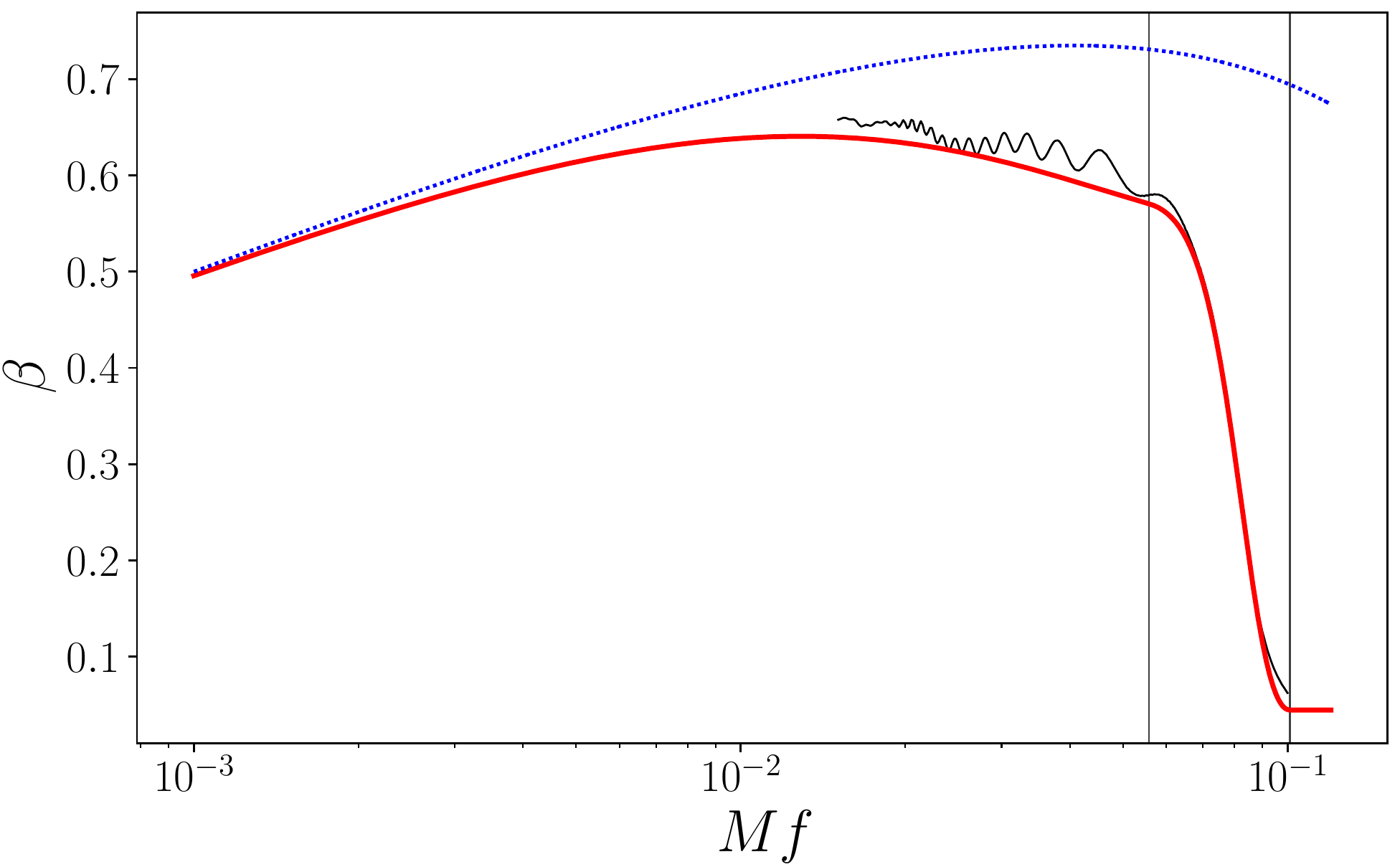}\\
   \includegraphics[width=0.49\textwidth]{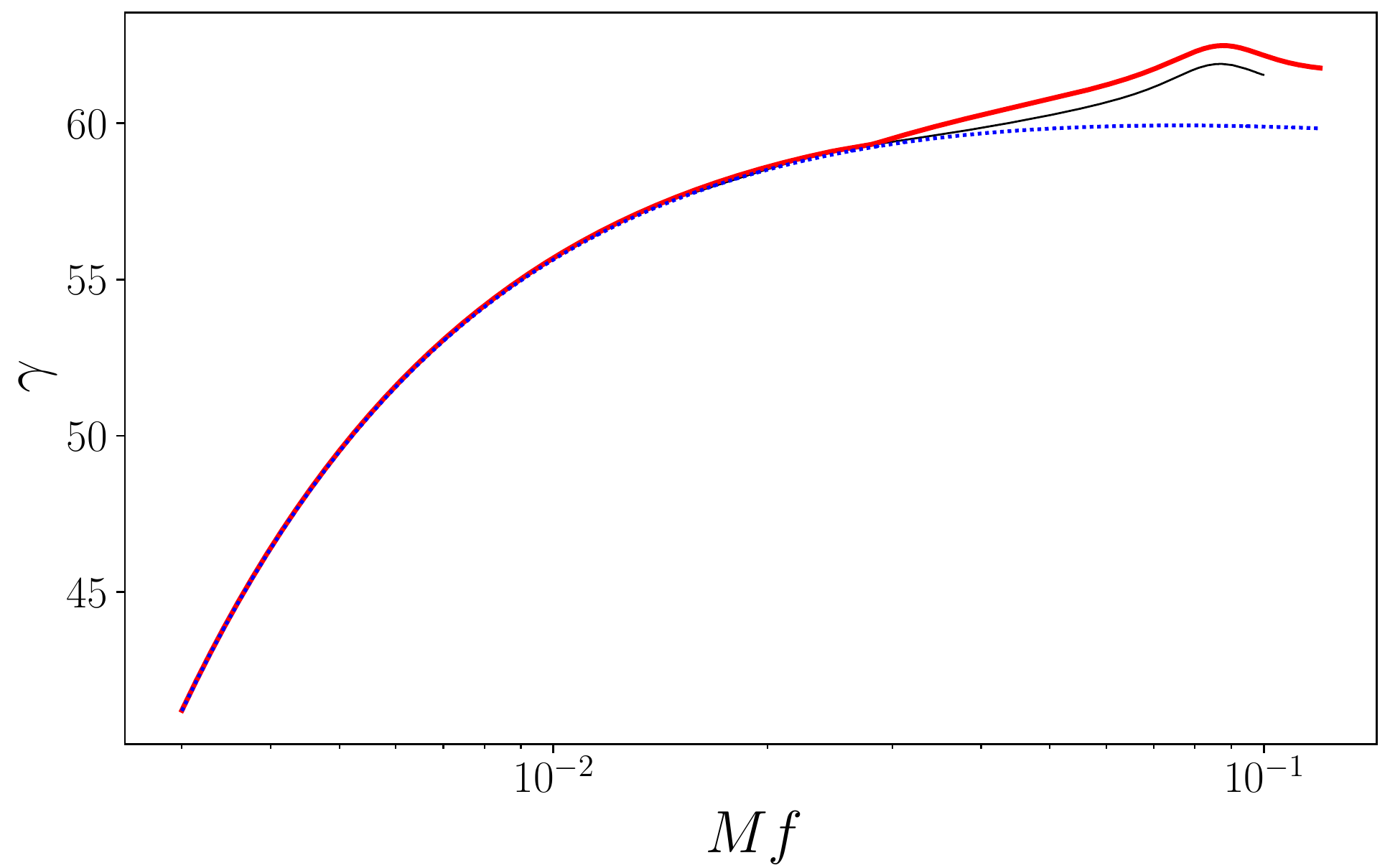}
   \includegraphics[width=0.49\textwidth]{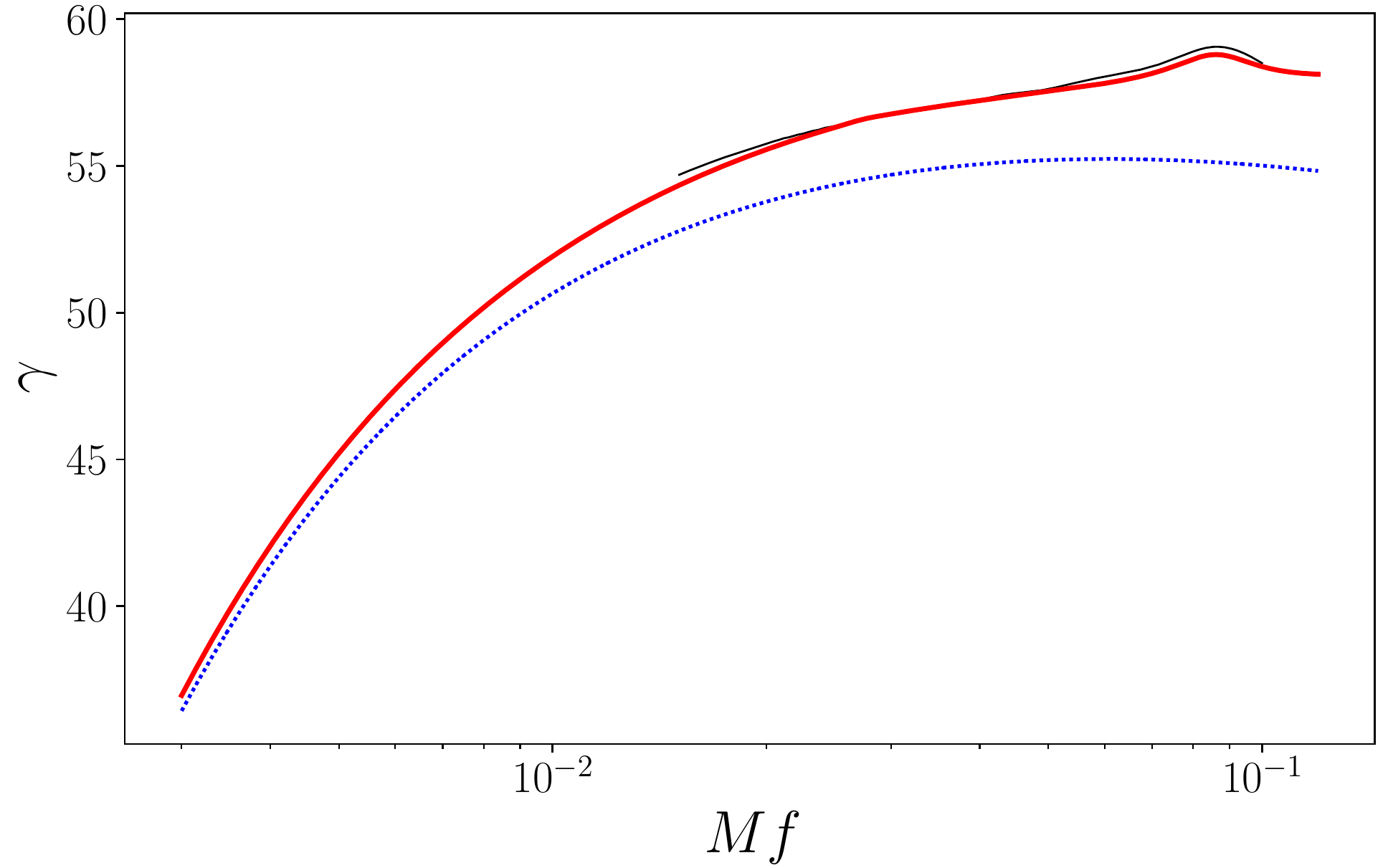}
   \caption{Comparison of the complete model for each of the precession angles (thick red line) with the \nr{} data (thin black line). 
   The \msa{} angles (blue dotted line) are shown for reference. The left hand column shows the case with $\left(q,\chi,\theta\right)=\left(1,0.4,30^\circ\right)$. 
   The right hand column shows the case with $\left(q,\chi,\theta\right)=\left(8,0.8,60^\circ\right)$. The vertical black lines show the connection frequencies for 
   $\alpha$ and $\beta$.}
   \label{fig: model all angles}
\end{figure*}

\subsection{Full \imr{} expressions}

The expressions describing the precession angles in each of the different regions are connected using piece-wise \(C^1\)-continuous functions.

The full \imr{} expression for $\alpha$ is

\begin{equation}
\alpha_{\text{IMR}}\left(f\right) =
\begin{cases}
      \alpha_{\text{PN}} & 0 \leq f < f_1 \\
      \alpha_{\text{interp}} & f_1\leq f < f_2 \\
      \alpha_{\text{MR}} & f_2 \leq f
\end{cases}
\end{equation}
where $\alpha_{\text{PN}}$, $\alpha_{\text{interp}}$ and $\alpha_{\text{MR}}$ are the \pn{} expression used to describe $\alpha$ during inspiral, the 
interpolating function used to describe the late inspiral angles in the region $f_1$ to $f_2$ and the phenomenological ansatz used which has been 
tuned to \nr{} to describe the merger-ringdown angles respectively.

Across the majority of the parameter space, the merger-ringdown ansatz for $\beta$ has a minimum immediately following the inflection point (as 
shown in the central panel of Fig.~\ref{fig: beta morphology}). In these cases, the full \imr{} expression for $\beta$ is
\begin{equation}
\beta_{\text{IMR}}\left(f\right) =
\begin{cases}
      k\beta_{\text{PN}} & 0 \leq f < f_\text{c} \\
      \beta_{\text{MR}} & f_\text{c}\leq f < f_\text{f} \\
      \beta_{\text{RD}} & f_\text{f} \leq f
\end{cases},
\end{equation}
where $\beta_{\text{PN}}$ is the \pn{} expression for $\beta$ including the higher-order amplitude corrections discussed in \sect{sec: beta approx}, 
$k$ is the rescaling function applied to these expressions as outlined above, $\beta_{\text{MR}}$ is the phenomenological ansatz which has been 
tuned to \nr{} in the merger-ringdown regime, and $\beta_{\text{RD}}$ is the constant value of $\beta$ to which the system settles down after merger, 
as discussed in \sect{sec: ringdown beta}. We model this quantity by the minimum value of $\beta$ in the merger-ringdown expression. $f_\text{f}$ is 
correspondingly given by the frequency at which the minimum occurs. 

In cases where $\beta$ tends towards an asymptote immediately following the inflection point (which occur in some regions of parameter space 
beyond the fitting region),
the full \imr{} expression for $\beta$ is
\begin{equation}
\beta_{\text{IMR}}\left(f\right) =
\begin{cases}
      k\beta_{\text{PN}} & 0 \leq f < f_\text{c} \\
      \beta_{\text{MR}} & f_\text{c}\leq f
\end{cases}.
\end{equation}

We would physically expect $\beta$ to be bounded by 0 and $\pi$ across the parameter space. In order to enforce this requirement, we pass the 
resulting \(\beta_\text{IMR}\) through a windowing function $w\left(\beta\right)$ given by
\begin{equation}
   w\left(\beta\right) = 
   \text{sgn}\!\left(\beta-\frac{\pi}{2}\right)\!\!\left(\frac{\pi}{2}\right)^{1-p}\!
   \arctan^p\!\!\left[\left(\frac{\beta-\frac{\pi}{2}}{\left(\frac{\pi}{2}\right)^{1-p}}\right)^\frac{1}{p}\right] +
   \frac{\pi}{2},
\end{equation}
where $p=0.002$. This function is linear with $w\left(\beta\right)=\beta$ over the range $\beta \in \left[0.01, \pi-0.01\right]$ to within 0.045\%. This 
ensures that the fits for $\beta$ are unaffected within the calibration but that $\beta$ is bounded by 0 and $\pi$ across the whole of parameter space.

The precession angle $\gamma$ is then calculated over the entirety of the frequency range for which the waveform is produced by enforcing the 
minimal rotation condition given in \eqn{fdc}. The decision to do this rather than produce a separate model for $\gamma$ was made as it was found 
that $\gamma$ must be very accurate in order to consistently transform between an inertial frame and the co-precessing frame. The very small 
discrepancy between the expression for $\gamma$ presented in~\cite{Chatziioannou:2017tdw} and the numerically calculated value is sufficient to 
seriously degrade the model. This discrepancy is exacerbated here since we are no longer using the dynamical expression for $\beta$ presented 
in~\cite{Chatziioannou:2017tdw}. (We note that independently integrating Eq.~(\ref{fdc}) was also found to be more accurate in the 
\textsc{SEOBNRPv4HM} and \textsc{PhenomTPHM} models~\cite{Ossokine:2020kjp,Estelles:2020osj}.)

The full model of these angles is shown for two examples in very different parts of the parameter space in Fig.~\ref{fig: model all angles}.

\subsection{Behaviour beyond calibration region}
\label{sec: angles beyond calibration}

\begin{figure*}
   \centering
   \includegraphics[width=\textwidth]{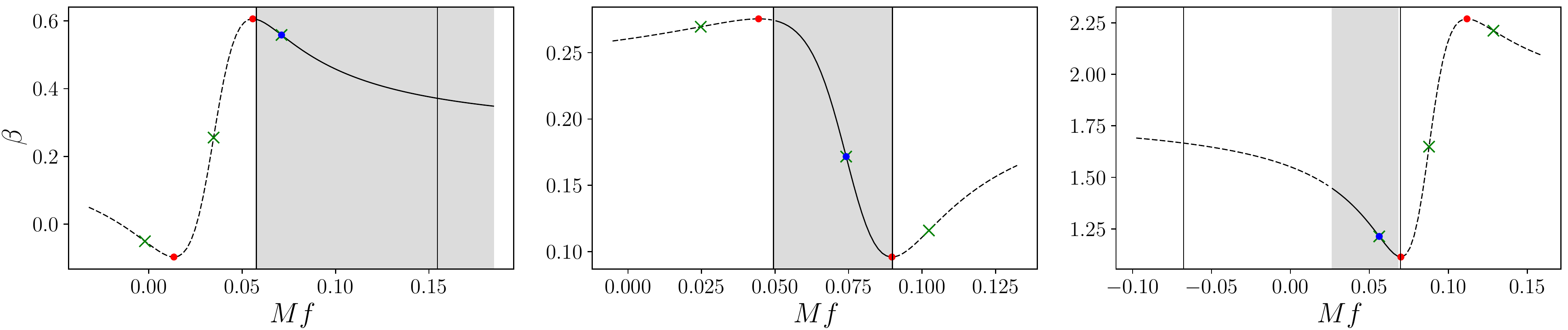}
   \caption{Possible morphologies of the ansatz given by Eq.~(\ref{eqn: beta}) depending on the values taken by the co-efficients in different regions 
   of the parameter space. From left to right the panels show systems with $\left(q,\chi,\tls\right) = \left(8,0.2,155^\circ\right)$, $\left(2.5,0.4,90^\circ\right)$ 
   and $\left(5,0.8,160^\circ\right)$. The red dots mark the extrema, the green crosses show the inflection points and the blue dot indicates the inflection 
   point chosen as described in \sect{sec: angles beyond calibration}. The points of maximum curvature around this inflection point are shown by the black lines, 
   which give a measure of the width of the turnover. The solid black line in the shaded region indicates the frequency region that will be used as the 
   merger-ringdown portion of the full angle model. All cases within our calibration region will have the morphology shown in the middle panel; the outer panels 
   show that a reasonable choice is made outside the calibration region.
   }
   \label{fig: beta morphology}
\end{figure*}

As with any tuned model, beyond the calibration region there is no guarantee of the accuracy of the model for the angles. However, we want to ensure 
that they do not display pathological or obviously physically incorrect behaviour. 

For $\alpha$ there are a number of possibilities inherent in the ansatz to see either pathological or physically incorrect behaviour. We have implemented 
restrictions on the values taken by the co-efficients to ensure this does not occur and a visual inspection of the waveforms shows that we do not see any 
pathological features. We would see pathological behaviour for $A_3 < 0$ and physically incorrect behaviour for $A_1 < 0$ ($\alpha$ would decrease as 
a function of frequency) or $A_2 > 0$ (the dip in $\alpha$ would have the wrong sign). As it is only a small region of parameter space in which this might 
happen, we enforce the conditions that $A_1, A_3 > 0$ and $A_2 < 0$ by taking the absolute value of the co-efficients with the appropriate sign. For $A_2$ 
we replace any positive values with zero. $A_1$ and $A_2$ take the wrong sign for systems with $q<10$ only at very small spins ($\chi < 0.1$) or large 
anti-aligned spins ($\chi\sqrt{-\cos\theta} \sim 0.7$). For $A_2$ there is an additional region for $q>7$ around $\chi = 0.4$ for anti-aligned spins 
($\cos\theta > 0.75$). $A_3$ does not go negative within the calibration region, though this does start to occur for $q>10$.

We see pathological behaviour for $B_4 \lesssim 0$. Physically incorrect behaviour starts to emerge when $B_4$ drops below $\mathcal{O}\left(10^{2}\right)$. 
In order to avoid such behaviour we require $B_4 \ge 175$ and replace the fitted value of $B_4$ by 175 where it falls below this value. Since $B_4 \sim 10^{3}$ 
across the majority of the parameter space this concern only arises for very extreme configurations ($\chi \approx 1$) where the accuracy of the model cannot 
be guaranteed anyway.

The morphology of the merger-ringdown ansatz of $\beta$ also changes in some parts of the parameter space outside the calibration region, as shown 
in Fig.~\ref{fig: beta morphology}. We can ensure we always employ the correct part of the expression (for which $\beta$ displays a drop at merger) in 
our model by selecting the correct inflection point. The inflection points of an expression occur at the roots of the second derivative of the expression. 
The second derivative of Eq.~(\ref{eqn: beta}) takes the form
\begin{align}
   \beta''\left(f\right) = {}& \frac{a f^3 + b f^2 + c f + d}{\left(1+B_4\left(B_5+f\right)^2\right)^3},
\end{align}
where $a$, $b$, $c$ and $d$ are functions of the fitting co-efficients $B_1$, $B_2$, $B_3$, $B_4$ and $B_5$. In order to find the roots of this cubic 
we re-write it in the form of a depressed cubic
\begin{align}
   x'^3 + p x' + q = {}& 0,
\end{align}
where
\begin{align}
   x' = {}& x + \frac{b}{3a}, \\
   p = {}& \frac{3ac - b^2}{3a^2}, \\
   q = {}& \frac{2b^3-9abc+27a^2d}{27a^3}.
\end{align}
In the case where this expression has three real roots, these are given by
\begin{align}
   x' = {}& 2\sqrt{-\frac{p}{3}}\cos{\left[\frac{1}{3}\arccos{\left(\frac{3q}{2p}\sqrt{-\frac{3}{p}}\right)}-\frac{2n\pi}{3}\right]},
\end{align}
where $n=0,1,2$.

We want to be able to define a single, smoothly varying inflection point that tracks the location of the turnover in $\beta$ during merger across the 
parameter space. As the co-efficients of the cubic vary, the morphology of Eq.~(\ref{eqn: beta}) changes, as shown in Fig.~\ref{fig: beta morphology}. 
For $a<0$ we have the morphology shown in the central panel of the figure. We therefore select the central root, which is the only one with a negative gradient. 
For $a>0$, we have the morphology shown in the outer panels. For this morphology we need to distinguish between the two outer roots, which both have a 
negative gradient.  This is determined by the ``shift'' of the roots, $b/3a$. In cases where 
\begin{align} 
   \frac{b}{3a} > \frac{B_5}{2} + \frac{\lambda^{B_2}_{004}}{4\lambda^{B_3}_{004}},
\end{align}
where the $\lambda^i_{pqr}$ are the co-efficients given in Eq.~(\ref{eqn: global fit expression}),
we choose the first root (as seen in the left-hand panel), otherwise we choose the final root (as seen in the right-hand panel). 
This condition was found to select the correct root across the entire calibration region for the model as well as most of the extended regions encompassing 
the validation waveforms. 

In the case where we have complex roots, two of the roots will be in the complex plane while one will be on the real axis. In this case we select the only real root.

We also consider the case where $a=0$ and the second derivative is a quadratic. In this case we have only one root with a negative gradient, which is the 
desired root. Finally, we consider the case where both $a=0$ and $b=0$. Here we have only one root which gives us the desired inflection point.

Enforcing these conditions gives us a smoothly varying value of the inflection point across the parameter space and ensures our expression for $\beta$ 
always has the correct morphology, dropping off at merger.

\section{Physical features of the waveforms} 
\label{sec:physfeat}

In motivating, constructing and presenting the \textsc{PhenomPNR} model, we have observed several features of precessing-binary
waveforms that deserve more detailed discussion.

\subsection{Ringdown frequency} 

\begin{figure}[htb]
   \begin{tabular}{cc}
      \includegraphics[width=0.47\textwidth]{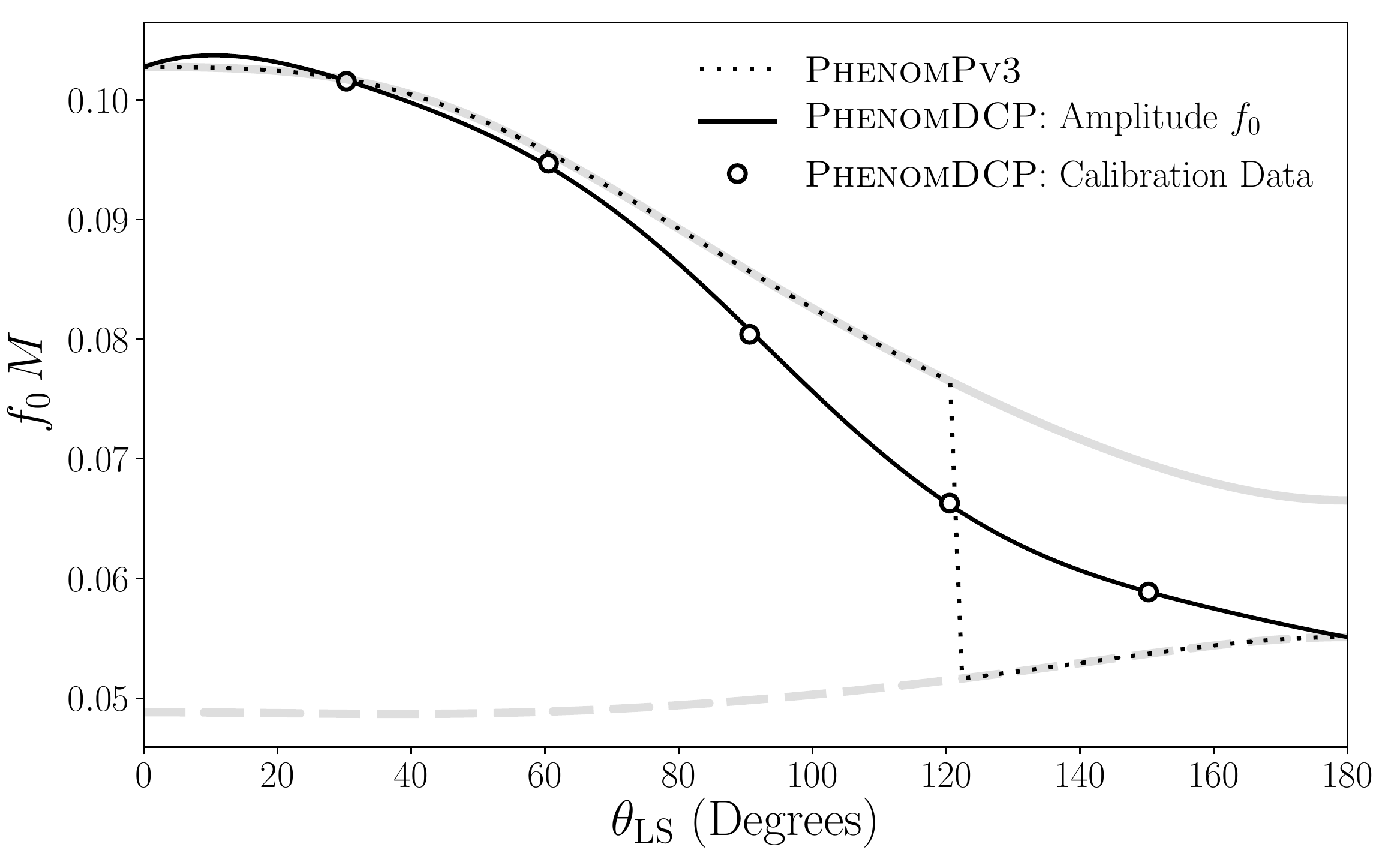}
   \end{tabular}
   \caption{Effective frequency-domain ringdown frequencies for $(q,\chi)=(8,0.8)$, as modelled by \pv3{} and \dcp{}.
   Additional lines show \qnm{} frequencies predicted from standard perturbation theory methods using the remnant \bh{}'s mass and 
   spin~\cite{leaver85}. The solid thick grey line traces prograde \qnm{} frequencies, and the dashed thick grey line traces the 
   retrograde \qnm{} frequencies. All curves are bound between pro- and retrograde \qnm{} frequencies. \pv3{} displays a discontinuity 
   near $\tls{}=120^\circ$, while \nr{} data and \pnr{} do not.} 
   \label{fig:rdcomp1} 
\end{figure}

As discussed in \sect{sec:coprecessing model}, in previous \textsc{Phenom} models, the co-precessing-frame model consists 
of an aligned-spin model, with ringdown
frequency and damping time adjusted according to the values predicted for the full precessing configuration. This prediction was made
by using approximate \nr{} fits for the final mass and spin, which then imply, via perturbation theory, the ringdown frequencies. This 
prediction of the ringdown frequency was then used in the co-precessing-frame model. 

One interesting feature of this approach is that in some parts of the parameter space it leads to a discontinuity in the ringdown-frequency
estimate. This arises as follows. There are two choices of ringdown frequency for a given \bh{} spin, depending on
whether the \bh{} perturbations were generated by orbits that were prograde or retrograde with respect to the final \bh{} spin; this 
can be represented as choosing either a positive or negative final spin. As an example, consider configurations with mass ratio $q=8$ 
and a spin on the larger \bh{} of $\chi=0.8$. If the spin is aligned with the orbital angular momentum, we predict that after merger the final
\bh{} will have a spin of 0.86, and a ringdown frequency of $\sim$0.1. If the large \bh{} spin is anti-aligned to the orbital angular 
momentum, i.e., $\theta_{\rm LS} = 180^\circ$, then the final \bh{} spin is $-0.275$, and the ringdown frequency is $\sim$0.06.

We can now ask, what happens for other values of $\theta_{\rm LS}$? In previous \textsc{Phenom} models, the final spin was estimated
as follows. We first estimate the final spin for an equivalent aligned-spin binary, $\chi_{\rm AS}$, and then calculate the vector sum of this
aligned spin with the in-plane spin contribution $\chi_{\rm p}$, which, in our single-spin example above, would take the value
$\chi \sin(\theta_{\rm LS})$. The final spin is then estimated as,
\begin{equation}
\chi_f = \sqrt{ \chi_{\rm AS}^2 + (m_1/M_f)^2 \chi_{\rm p}^2 }.
\end{equation} 
When we use this final-spin estimate to calculate the ringdown frequency, we must choose a sign. In previous \textsc{Phenom} models,
the same sign was chosen as $\chi_{\rm AS}$, but in some cases (as in the example above), this means that $\chi_f$ swaps sign at
some value of $\theta_{\rm LS}$, and the resulting estimate of the ringdown frequency is discontinuous. This is illustrated by the dashed line in 
Fig.~\ref{fig:rdcomp1} for our $q=8$, $\chi = 0.8$ series of configurations. As an estimate of the ringdown frequency in
the ($\ell=2,|m|=2$) multipoles in the $J$-aligned frame, this approach appears to be quite accurate, including the sharp transition
from prograde to retrograde branches.

One issue with this approach is that the transformation from the co-precessing to inertial frame will introduce a shift in
the GW frequency, and therefore a change in the ringdown frequency. If we apply the correct inertial-frame ringdown frequency
to our co-precessing-frame model, it will be changed when the angle model is applied, and the final model will have the \emph{wrong}
ringdown frequency. This is what happens in previous \textsc{Phenom} models. We could take this shift into account when we
prescribe the ringdown frequency in the co-precessing frame, but instead we simply produce a phenomenological fit to the 
ringdown frequency in the construction of the co-precessing-frame model \textsc{PhenomDCP}. This is also shown in 
Fig.~\ref{fig:rdcomp1}, in comparison with the effective co-precessing-frame ringdown frequency that we find from the \nr{} data.

\subsection{The collapse of $\beta$ through merger}\label{sec: ringdown beta}

During the inspiral, the angle $\beta$ is related to the opening angle between the total and orbital angular momenta, i.e., the
opening angle of the precession cone. At merger the orbital motion ceases, and we are left with a ringing black hole, and would
expect that the corresponding optimal emission direction would relax to the $\hat{\mathbf{J}}$ direction of the final black hole. 
However, we may also consider an alternative picture. A stationary \bh{} does not radiate. We may perturb a non-spinning \bh{} 
such that we completely determine the dominant emission direction as the perturbation rings down. Adding spin to the \bh{}, 
either small or large, does not change this freedom. Thus the optimal emission direction after merger, and in particular, the final values of 
$\alpha$ and $\beta$, may encode information about how the remnant \bh{} was perturbed through merger, and the relationship to 
$\hat{\mathbf{J}}$ is not so clear.

In Ref.~\cite{OShaughnessy:2012iol} an attempt was made to describe the late-time precession behaviour using results from 
perturbation theory. We know the general form of the ringdown signal, \begin{equation}
h_{\ell m}(t) \approx A_{\ell m} e^{i \omega_{\ell m} t} e^{- t/\tau_{\ell m}},
\end{equation} where the $A_{\ell m}$ are unknown constants, and $\{\omega_{\ell m}, \tau_{\ell m}\}$ are determined by the mass and
spin of the final \bh{} through perturbation theory. Given this general form, we can predict the general behaviour of the precession angles
$\alpha$ and $\beta$ in the ringdown regime, similar to the approximate approach followed in \sect{sec: beta approx}. 
Ref.~\cite{OShaughnessy:2012iol} note that, if we consider only the dominant $\ell = 2$ 
modes, the QA 
direction precesses around $\hat{\mathbf{J}}$ with a frequency $\omega_{22} - \omega_{21}$, and $\beta$ either falls
exponentially to zero at a rate given by $\tau_{22} - \tau_{21}$, or grows exponentially to $\pi$, depending on the relative magnitude 
of the two damping times. A similar calculation was later discussed in Refs.~\cite{Marsat:2018oam,Ossokine:2020kjp,Estelles:2020osj}. 

Several points are worth noting. (1) For much of the parameter space, although the decay of $\beta(t)$ is exponential, it is nonetheless
extremely slow, and on a much longer timescale than the decay of the signal amplitude. (2) We can consider non-zero 
$(\ell =2, |m|=2)$ and $(\ell=2,|m|=1)$ multipoles where the QA direction does not 
precess at all, for example all aligned-spin binaries. (3) Just as \bh{} perturbation theory cannot tell us how much each \qnm{} is 
excited~\cite{Kamaretsos:2012bs,Kamaretsos:2011um}, this analysis cannot tell us the magnitude of $\beta$ at whatever point we
wish to designate as the beginning of the ringdown regime. 

\begin{figure}[t]
   \begin{tabular}{c}
      \includegraphics[width=0.47\textwidth]{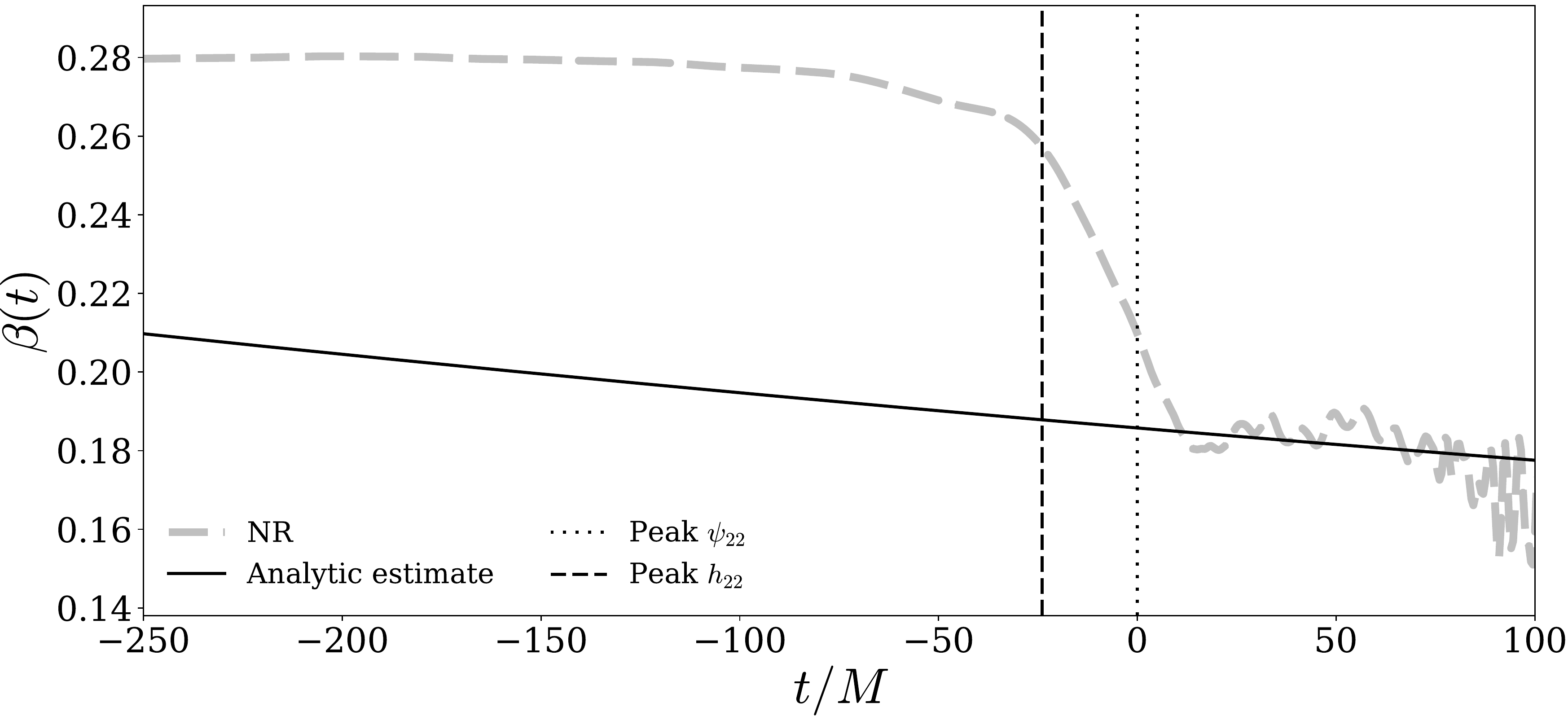}
      \\
      \includegraphics[width=0.47\textwidth]{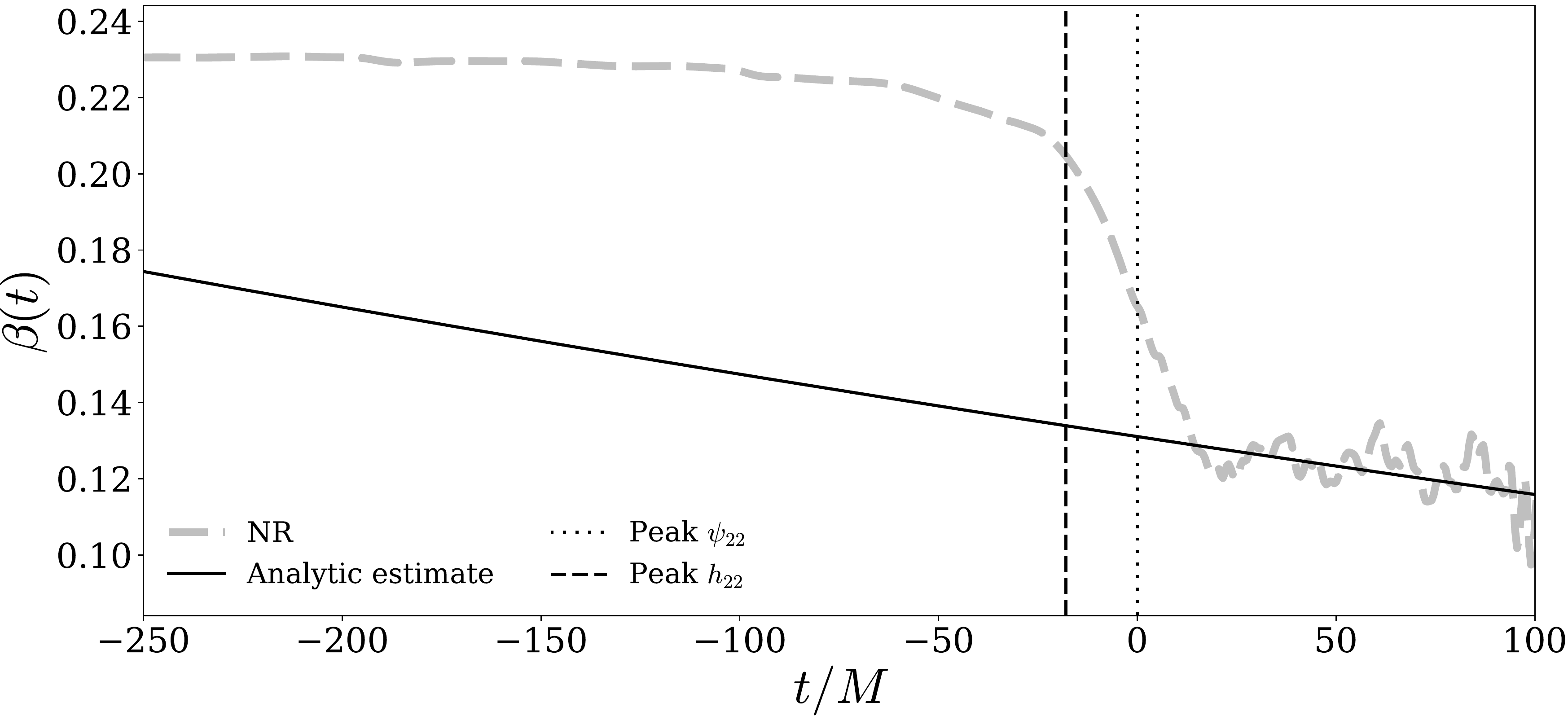}
   \end{tabular}
   \caption{Comparison of analytic ringdown estimate and numerical relativity for (top) $ (q,\chi,\tls)=(4,0.4,60^{\circ}) $ and (bottom) $ (q,\chi,\tls)=(8,0.4,30^{\circ}) $.} 
   \label{fig:rdcomp2} 
\end{figure}

We now turn to our \nr{} data to address these points. Fig.~\ref{fig:rdcomp2} shows the late-time behaviour of \nr{} $\beta$ for the 
$(q,\chi,\theta_{\rm LS}) = (4, 0.4, 60^\circ)$ and $(8, 0.4, 30^\circ)$ configurations, as well as approximate fits to the $\beta$ 
decay rate predicted by the ringdown toy model discussed above. In these fits the decay rate is prescribed by the toy model and only 
the overall amplitude is fit to the numerical data. The data are not clean enough to conclusively show that 
late-time $\beta$ follows the decay rate predicted by the toy model, but the data are certainly consistent with that model. What is
worth highlighting is that the decay rate is indeed very slow; we expect $\beta$ to be greater than, say, 10\% of its peak value, 
for several hundred $M$ after merger, at which point the total signal amplitude will have decayed by several orders of magnitude. 
In this context, our simple approximation in \textsc{PhenomPNR}, that late-time $\beta$ is constant, appears to be justified. 

The other important observation is that this late-time ringdown behaviour begins \emph{after} $\beta$ has dropped significantly
through merger. This strongly suggests that ringdown begins significantly after the peak in both strain and $\psi_4$, which is possibly 
at tension with recent efforts to apply \bh{} perturbation theory at those points~\cite{Giesler:2019uxc}.

Although a PN treatment can approximately describe $\beta$ during the inspiral, and a simple ringdown analysis can
describe the decay rate of $\beta$ during ringdown, neither can capture the rapid drop in $\beta$ through merger, or predict
the value of $\beta$ at the point where the ringdown behaviour takes over. This feature, which is included in \textsc{PhenomPNR}, 
was not explicitly modelled in previous \textsc{Phenom} and \textsc{EOBNR} models; \textsc{PhenomP/Pv2/Pv3/XP} used the \msa{} angles at all
frequencies, and both \textsc{SEOBNRv4PHM} and \textsc{PhenomTP} use a constant late-time value of $\beta$ determined by its 
value near merger.

\subsection{Hierarchy in the turnover frequency of the $\ell = 2$ multipoles}\label{sec: hierarchy}

\begin{figure}[t]
   \centering
   \includegraphics[width=0.47\textwidth]{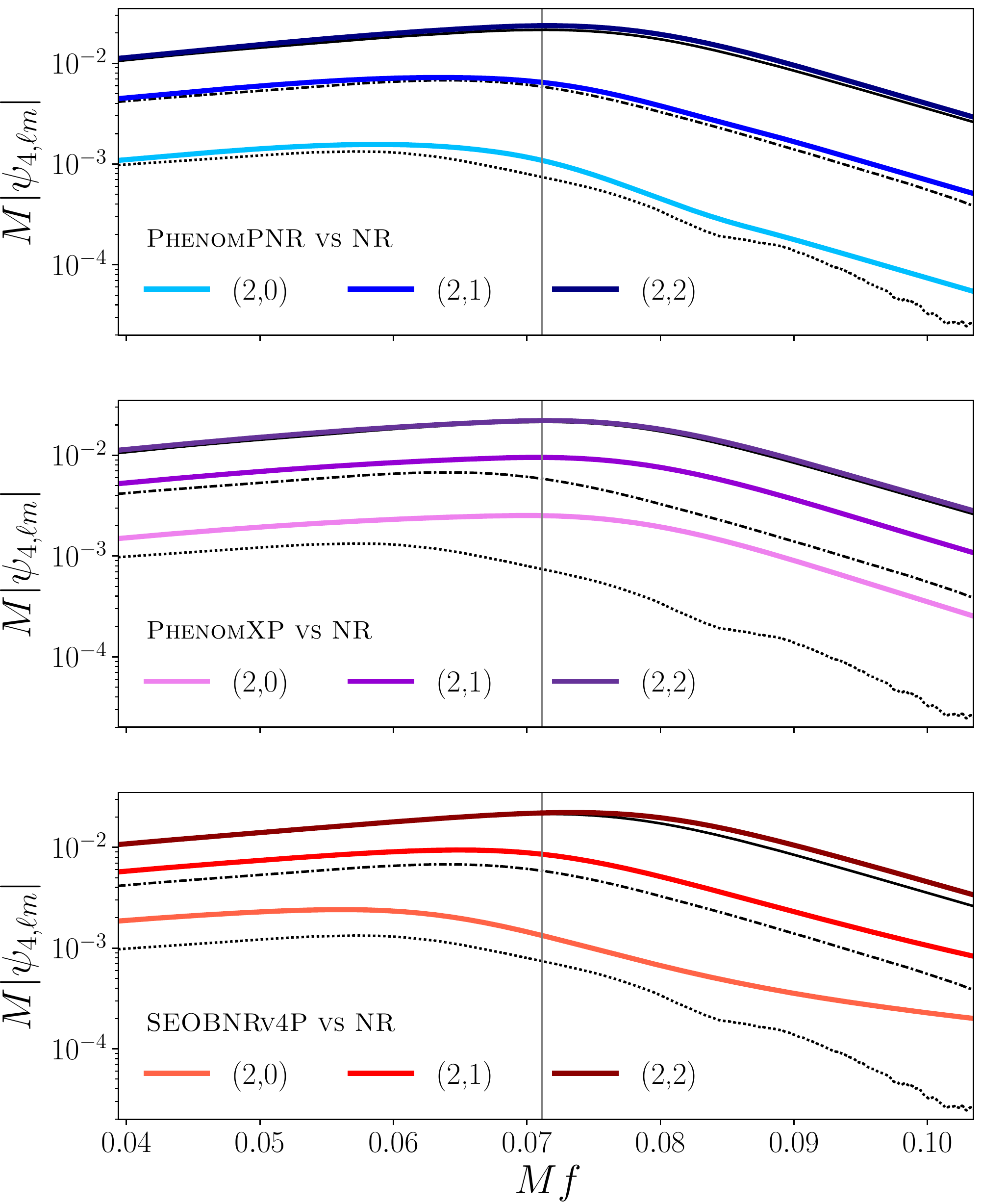}
   \caption{Amplitudes of the $\ell=2$ multipoles for the $(q,\chi,\theta_{\rm LS}) = (4,0.4,90^\circ)$ configuration at 100$\,M_\odot$. 
   The \nr{} data are shown in black on all panels, with \textsc{PhenomPNR} (top panel in blue), \textsc{PhenomPv3} and 
   \textsc{PhenomXP} (central panel in purple) and \textsc{SEOBNRv4P} (bottom panel in red). The vertical line indicates 
   the frequency of the peak (2,2)  amplitude. 
   }
   \label{fig: mode hierarchy}
\end{figure}

\begin{figure*}[t]
   \centering
   \includegraphics[width=\textwidth]{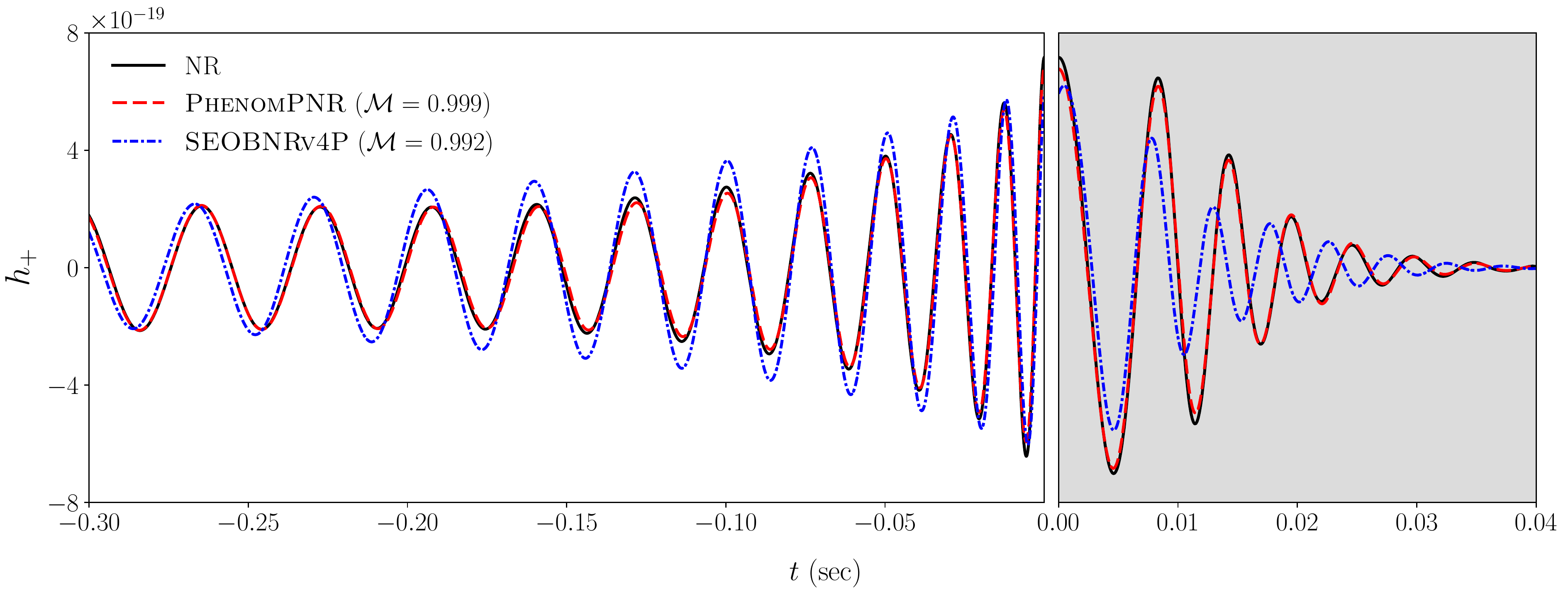} \\
   \includegraphics[width=\textwidth]{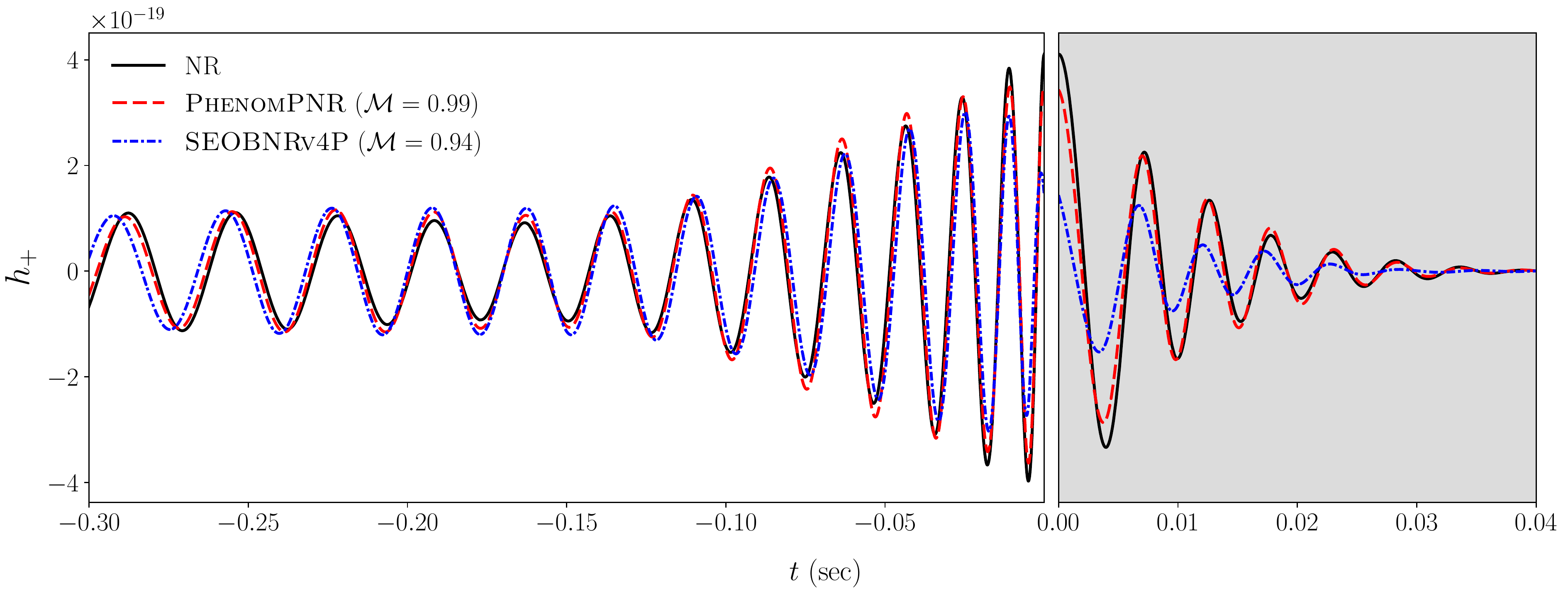}
   \caption{A comparison of the time domain obtained from \textsc{PhenomPNR} with the \nr{} data. The top panel shows the the 
   case $(q,\chi,\tls)=(4,0.8,60^\circ)$ while the bottom panel shows the case $(q,\chi,\tls)=(8,0.8,60^\circ)$. Both are for a face on 
   ($\theta_\text{LN}=0^\circ$) binary with a total mass of 100$M_\odot$. For comparison, we also show the waveform produced using 
   \textsc{SEOBNRv4P}. The match values for the specific configuration for each of the waveforms plotted are given in the legend.}
   \label{fig: td waveform}
\end{figure*}

The rapid drop in $\beta$ described in the previous section results in a key feature of precessing waveforms: a hierarchy in the turnover 
frequency of the $\ell=2$ multipoles. From Eq.~(\ref{eqn:easybeta}) we can see that $\beta$ is approximately given by the ratio of the 
amplitude of the (2,2) and (2,1) multipoles. The drop in $\beta$ therefore implies that the amplitude of the (2,1) multipole must have 
decreased relative to the (2,2) multipole and so the (2,1) multipole will begin to experience ringdown decay before the (2,2) multipole. 
Once both multipoles are decaying exponentially (at roughly the same rate) $\beta$ levels off. This trend continues for all of the $\ell=2$ 
multipoles. 

By capturing the drop in $\beta$ in our model, we succesfully model this hierarchy in the turnover frequency of the $\ell=2$ multipoles, 
as seen in the top panel of Fig.~\ref{fig: mode hierarchy}. This feature has not been modelled in previous precessing \textsc{Phenom} 
models, and the central panel of Fig.~\ref{fig: mode hierarchy} shows the multipole hierarchy for \textsc{PhenomXP}, which is also the
behaviour for \textsc{PhenomPv3}, since both use the same \msa{} angle model. 
We see that in these models each of the $\ell=2$ multipoles turn over at 
the same frequency. \textsc{SEOBNRv4P}, shown in the bottom panel, does capture this hierarchy but the amplitude of the higher order 
multipoles is not well modelled. This is due to modelling $\iota$ rather than $\beta$, which typically overestimates the amplitude as discussed in
\sect{sec: beta approx}.

\section{Time domain validation}\label{sec: td angles}
\subsection{Time domain waveform}

The improvements made in modelling precessing systems presented here --- both to the underlying co-precessing model and the precession
angles --- can also be clearly seen when inspecting the waveforms in the time domain. 
As can be seen in Fig.~\ref{fig: td waveform}, \textsc{PhenomPNR} correctly captures the precession envelope and the phasing of the 
waveform through inspiral, merger and ringdown. This figure also clearly shows that the frequency-domain modelling presented here does 
not introduce any strange artefacts in the time domain. For comparison we also show \textsc{SEOBNRv4P}, a naturally time domain precessing 
model. The configurations shown here are for a binary with the intrinsic properties $(q,\chi,\tls)=(4,0.8,60^\circ)$ (top panel) and 
$(q,\chi,\tls)=(8,0.8,60^\circ)$ (bottom panel). We have plotted the optimally aligned waveform for both waveform models for a face on 
($\theta_\text{LN}=0^\circ$) binary. We particularly note the good agreement between \textsc{PhenomPNR} and the \nr{} waveform after merger 
(the shaded region) due to accurately modelling the merger-ringdown precession angles and the effective ringdown frequency of the co-precessing 
waveform. 

The time and phase alignment of the waveforms plotted in Fig.~\ref{fig: td waveform} has been performed over the same range of frequencies 
as were used in calculating the matches detailed in Sec.~\ref{sec:matches} and quoted in the figure legend. This range is much greater than that 
shown in the plot so the deviations between the models and the \nr{} seen here do not contribute as much as might na\"ively be expected. We have 
plotted the waveform for the in-plane spin configuration and polarisation of the signal for which we get the maximum match. \textsc{PhenomPNR} 
agrees well with the \nr{} data from inspiral through merger and ringdown, capturing both the precession envelope and the phasing of the waveform 
correctly.

\subsection{Time domain angles}

\begin{figure}[t]
   \centering
   \includegraphics[width=0.47\textwidth]{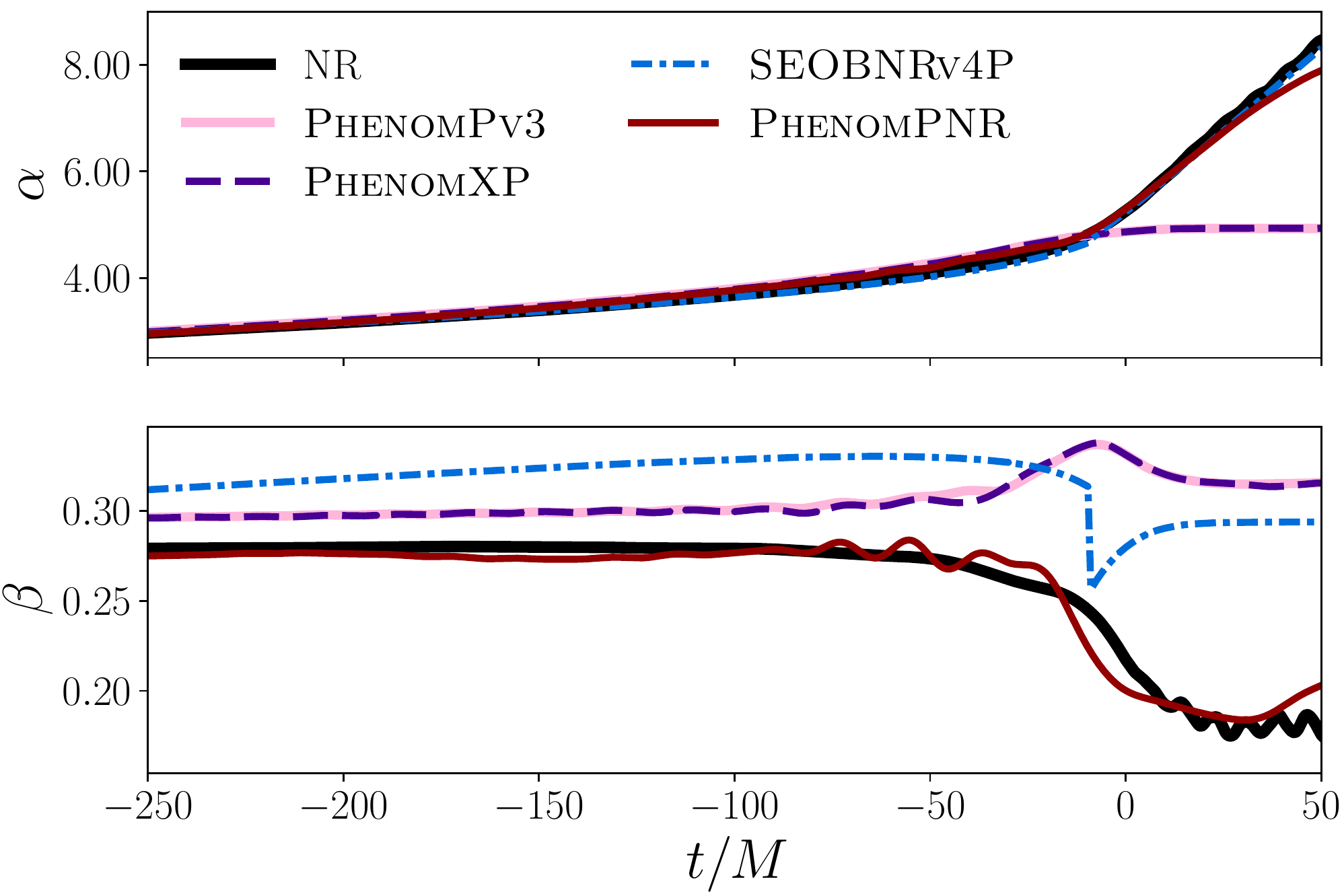}
   \caption{Comparison of the time domain precession angles for the \textsc{PhenomPv3}, \textsc{PhenomXP}, \textsc{SEOBNRv4P} and 
   \textsc{PhenomPNR} models with the \nr{} data. These angles are for the case with $\left(q,\chi,\tls\right) = \left(4, 0.4, 60^\circ\right)$. }
   \label{fig: q4a04t60 td angles}
\end{figure}

\begin{figure}[t]
   \centering
   \includegraphics[width=0.47\textwidth]{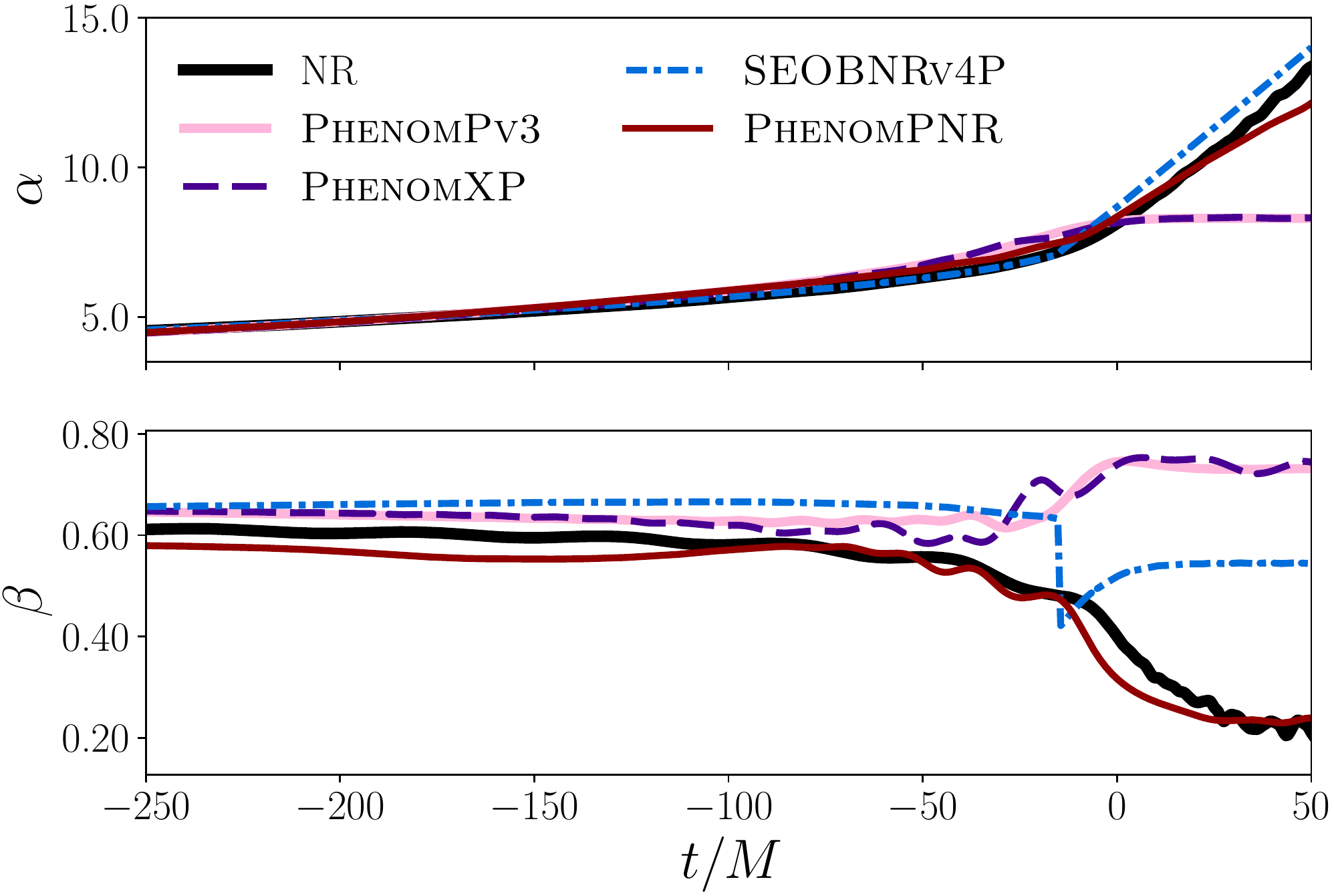}
   \caption{Comparison of the time domain precession angles for the \textsc{PhenomPv3}, \textsc{PhenomXP}, \textsc{SEOBNRv4P} and 
   \textsc{PhenomPNR} models with the \nr{} data. These angles are for the case with $\left(q,\chi,\tls\right) = \left(8, 0.8, 60^\circ\right)$. }
   \label{fig: q8a08t60 td angles} 
\end{figure}

Accurately modelling the merger-ringdown features of the angles in the frequency domain has also enabled us to reproduce key features of the 
angles in the time domain after merger. We compared the time domain angles for four models (\textsc{PhenomPv3}, \textsc{PhenomXP}, 
\textsc{SEOBNRv4P} and \textsc{PhenomPNR}) with the \nr{} angles. In order to avoid the introduction of artefacts due to unnecessary 
processing of the \nr{} data, we compare against the time domain angles calculated using the cleaned and symmetrised $\Psi_4$ data rather than $h$. 

The time domain angles for the frequency domain models (those belonging to the \textsc{Phenom} family) are calculated as follows. 
First we compute the $\psi_{4,\ell m}$ $\ell=2$ multipoles from the strain multipoles in the frequency domain using $\tilde{\psi}_{4,\ell m}(f) = (2\pi f)^2 \tilde{h}_{\ell m}(f)$. 
We then compute the time domain multipoles by performing the inverse Fourier transform each of the $\ell=2$ frequency domain multipoles. 
Finally we calculate the precession angles from the set of time domain $\ell=2$ multipoles. 

For \textsc{SEOBNRv4P}, a time domain model, we differentiated each of the $\ell=2$ time domain multipoles twice to get $\psi_{4,\ell m}$  from $h_{\ell m}$. 
We then calculated the precession angles using these multipoles. Since the connection between the inspiral and ringdown parts of the models 
for the multipole moments and the precession angles used in \textsc{SEOBNRv4P} is \(C^1\)-continuous, we see a discontinuity in the time 
domain angles presented here as a result of the double differentiation. 

The results of this comparison are shown in Figs.~\ref{fig: q4a04t60 td angles} and \ref{fig: q8a08t60 td angles}. Since \textsc{PhenomPv3} and 
\textsc{PhenomXP} use the same model for the precession angles with a different co-precessing model, the time domain angles presented here 
agree very closely. The two most notable features in the time domain angles are the continued rise in $\alpha$ after merger and the rapid drop in 
the value of $\beta$. If $\alpha$ takes a constant value it implies the precession of the optimum emission direction has stopped. As has been 
noted previously~\cite{OShaughnessy:2012iol}, this is clearly not seen in the \nr{} data. This feature of the precessional motion is captured by 
\textsc{SEOBNRv4P} and \textsc{PhenomPNR} but not by \textsc{PhenomPv3} and \textsc{PhenomXP}. The rapid drop in the value of $\beta$ 
is captured accurately only by \textsc{PhenomPNR}, although \textsc{SEOBNRv4P} does show some evidence of a drop in the value of $\beta$. 
This shows we have managed to capture the closing up of the opening angle as the angular momentum is radiated away through gravitational 
wave emission. The final feature to note is the amplitude of $\beta$ throughout inspiral is captured reasonably well by \textsc{PhenomPNR} 
whereas the other models all show a slight offset since (as previously discussed) they use the angles that describe the precessional dynamics 
rather than the precession of the direction of optimal emission. This can be seen more clearly at earlier times than are shown in 
Figs.~\ref{fig: q4a04t60 td angles} and  \ref{fig: q8a08t60 td angles} since here we chose to focus on the merger-ringdown region where data 
processing artefacts from the Fourier transform are stronger.

\section{Model validation: Matches}
\label{sec:matches}

We now wish to test the accuracy of our new precessing model in the context of \gw{} signal analysis. 
To do this we calculate the match (using the method detailed below in \sect{sec: match definitions}) between the \nr{} waveform and our 
model for a given configuration. We performed three sets of matches in order to inspect each of the components of our model individually 
as well as the complete final model. To assess the accuracy of the underlying co-precessing model we calculated the standard non-precessing 
match for a waveform containing only the (2,2)-multipole between the co-precessing model \textsc{PhenomDCP} and the co-precessing \nr{} 
waveform. In order to assess the accuracy of the angle model itself, we model the precessing waveform by twisting up the co-precessing \nr{} 
waveform with the \textsc{PhenomAngles} angles and match it against the corresponding $\mathbf{J}$-aligned \nr{} waveform. 
Finally, we assessed the accuracy of the complete tuned precessing model \textsc{PhenomPNR} by performing the SNR-weighted match 
between the model and the $\mathbf{J}$-aligned \nr{} waveforms, containing the $\ell=2$ multipoles. 
The match calculated between the \nr{} waveform and the complete model will contain errors introduced by inaccuracies in
both \dcp{} and \textsc{PhenomAngles}.
Since we do not aim to model asymmetries in the multipole moments in this work, our model does not capture them. 
We therefore perform matches testing the angle model using the symmetrised \nr{} waveform (in both the $\mathbf{J}$-aligned and co-precessing frames). 

\subsection{Match Definitions}\label{sec: match definitions}

The disagreement between two waveforms, a model template $h_\mt$ and an \nr{} signal $h_\ms$, is quantified using the standard inner 
product weighted by the power spectral density of the detector $S_n\left(f\right)$ \cite{Cutler:1994ys}, chosen for this work to be the noise 
spectrum of advanced LIGO at design sensitivity~\cite{aligopsd}:
\begin{align}
   \langle h_\ms | h_\mt \rangle = {}& 4 \text{Re} \int_{f_\text{min}}^{f_\text{max}} \frac{\tilde{h}_\ms\left(f\right)\tilde{h}^*_\mt\left(f\right)}{S_n\left(f\right)}\, \text{d}f.
\end{align} 
The \textit{match} is then given by the inner product between two normalised waveforms,
\begin{align} 
\label{eq:match}
   \mathcal{M}\left(h_\ms,h_\mt\right) = {}& \max_{\Xi_\text{t}} \frac{\langle h_\ms | h_\mt \rangle}{\sqrt{\langle h_\ms | h_\ms \rangle\langle h_\mt | h_\mt \rangle}},
\end{align}
maximised over a set of template parameters \(\Xi_\text{t}\) described below.

Time shifts and reference phase shifts have no physical effect on the signal; a time shift corresponds only to a change in the merger time of the 
binary, while a change in the phase corresponds to a change in the initial orientation of the binary's orbit.
For non-precessing waveforms containing only the (2,2)-multipole, the resulting match value is independent of the inclination and polarisation 
of the signal, as changes to the inclination simply re-scale the overall amplitude of both the signal and template, and the polarisation is degenerate 
with the reference phase and therefore optimised away. When computing the match for non-precessing signals, as is done in \sect{sec: cp matches}, 
the maximisation done in \eqn{eq:match} is done over time and phase shifts, \(\Xi_\text{t}=\{t_0,\phi_0\}\).

For precessing waveforms, both the inclination and polarisation must be taken into account. 
First, we compute the match outlined in \eqn{eq:match} whilst keeping the signal phase and polarisation fixed, and maximise over time shifts, 
reference phase and template polarisation following Ref.~\cite{Harry:2016ijz}. 
We further optimise over rotations to the in-plane spin components of the template at the reference frequency as in
Ref.~\cite{Pratten:2020ceb}, which effectively optimises the match over the initial precession phase 
\(\alpha_0\), \textit{i.e.,} \(\Xi_\text{t}=\{t_0,\phi_0,\psi_0, \alpha_0\}\). 
We then follow previous efforts to quantify precessing models~\cite{Schmidt:2014iyl,Khan:2018fmp,Pratten:2020ceb} and introduce an 
\textit{SNR-weighted match}. 
 
The SNR-weighted match is computed by averaging the match computed at each given signal phase and polarisation whilst volume-weighting 
with the SNR of the signal,
\begin{equation}
\label{eq:weightedMatch}
\mathcal{M}_\text{w}=
\left(
\frac{
	\sum_{\psi_\ms,\phi_\ms}\mathcal{M}^3 \langle h_\ms | h_\ms \rangle^{\frac32}
}
{
	\sum_{\psi_\ms,\phi_\ms}\langle h_\ms | h_\ms \rangle^{\frac32}
}
\right)^\frac13,
\end{equation}
where we have summed over the values of signal phase and polarisation, \(\phi_\ms\) and \(\psi_\ms\), respectively.
This is done to better account for the large variation in detectability and signal strength with sky location that occurs in precessing signals.

Finally, we compute the \textit{mismatch} between the signal and template for non-precessing signals as,
\begin{equation}
\label{eq:mismatch}
\mathfrak{M}=1-\mathcal{M},
\end{equation}
and similarly for precessing signals the SNR-weighted mismatch,
\begin{equation}
\label{eq:weightedMismatch}
\mathfrak{M}_\text{w}=1-\mathcal{M}_\text{w}.
\end{equation}

\subsection{Verification waveforms}\label{sec: verification wvfms}

We performed matches against 76 of the waveforms taken from the BAM catalogue described in~\sect{sec:NR}. 
We also considered an additional set of waveforms taken from the SXS~\cite{Mroue:2013xna, Boyle:2019kee} and Maya catalogues~\cite{Jani:2016wkt}. 
This enabled us to test the accuracy of the model for configurations for which it was not tuned, including two-spin configurations.
 A summary of the waveforms taken from the BAM catalogue are given in~\ctbl{tab: BAM cases}, while the details of those taken from the 
 SXS and Maya catalogues are in \ctbl{tab: SXS and GT cases}.  Only the subset of waveforms taken from the BAM catalogue were used to 
 study the accuracy of the individual components of the model; the underlying co-precessing model and the model for the precession angles. 
 The complete set of waveforms, taken from all three catalogues, was used to test the accuracy of the full model over a range of total masses for the system. 

\begin{figure*}[htbp]
   \centering
   \includegraphics[width=\textwidth]{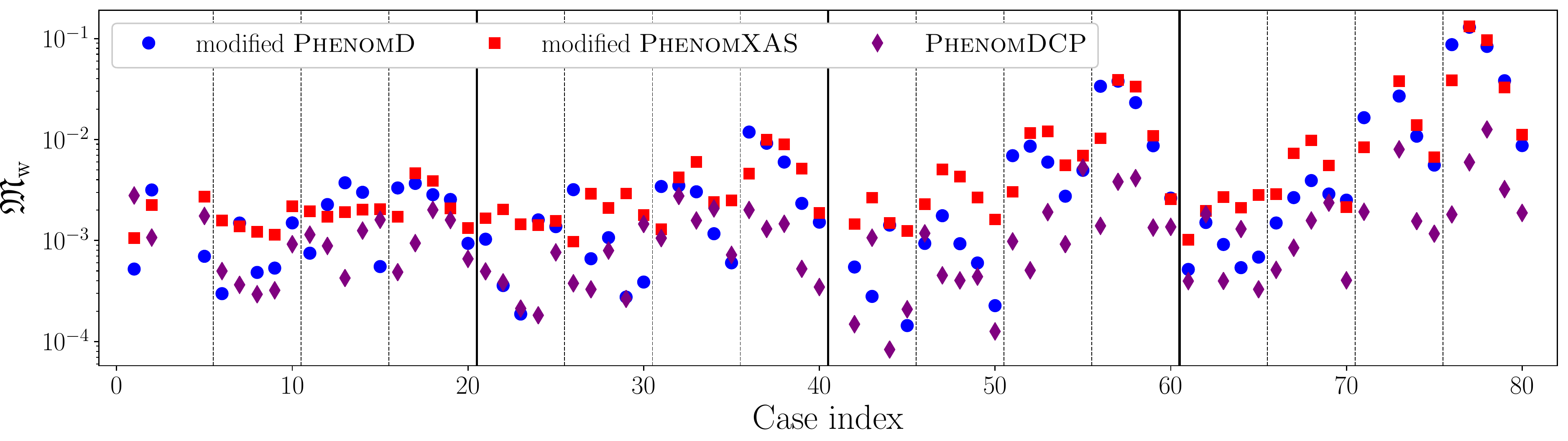} 
      \caption{Mismatches for each of the BAM calibration and verification waveforms, at a total mass of 100$M_\odot$. 
      Mismatches are between the symmetrised co-precessing \nr{} waveforms and \dcp{} (purple diamonds), modified \textsc{PhenomD} 
      (blue circles) and modified \textsc{PhenomXAS} (red squares). The configuration mass ratio increases from left to right (with $q \in \{1,2,4,8\}$). 
      Solid black lines separate cases mass ratios and dotted lines separate spin magnitudes.}
   \label{fig: cp matches}
\end{figure*}

\begin{figure*}[htbp]
   \centering
   \includegraphics[width=\textwidth]{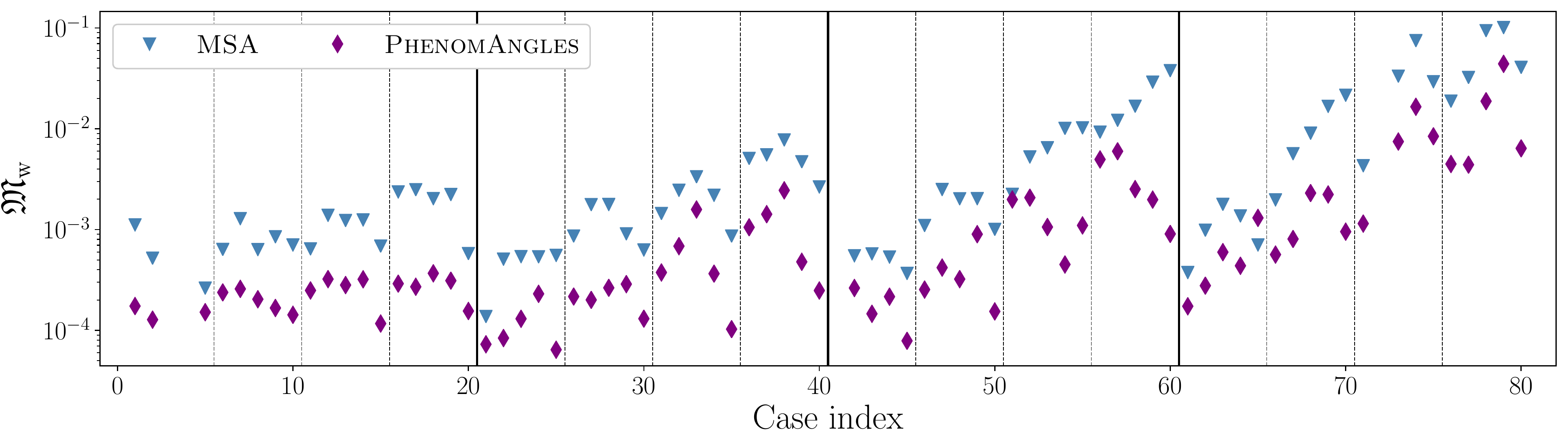} 
   \caption{SNR-weighted mismatches for the same configurations as in Fig.~\ref{fig: cp matches}, averaged over inclination. These mismatches are 
   between the symmetrised \nr{} waveforms in the $\mathbf{J}$-aligned frame and the co-precessing \nr{} waveform twisted up with the angle model 
   presented here (purple diamonds) and twisted up with the angle model used by \textsc{PhenomPv3} (steel blue triangles). 
   }
   \label{fig: angle matches}
\end{figure*}

\subsection{Matches: Accuracy of the co-precessing model}\label{sec: cp matches}

We computed the match between various models for the co-precessing waveform and the co-precessing \nr{} 
waveform. We considered a system of total mass 100M$_\odot$ and performed the match over the frequency range for which the \nr{} data 
was available; from $\left( f_\text{ref} + 5 \right)$Hz to 244Hz.  The value of the reference frequency $f_\text{ref}$ for each simulation is 
given in~\ctbl{tab: BAM cases}. The co-precessing-frame models we consider are \textsc{PhenomPv3}, \textsc{PhenomXP} and \textsc{PhenomDCP}.

As can be seen from Fig.~\ref{fig: cp matches}, the assumptions that go into producing the aligned-spin mapping used in the production 
of modified \textsc{PhenomD} and modified \textsc{PhenomXAS} become less accurate as both mass ratio and spin are increased. 
\textsc{PhenomDCP} performs better than both modified \textsc{PhenomD} and modified \textsc{PhenomXAS} for almost all cases, with 
the most noticeable improvement for the higher mass ratio, high-spin cases where we are in greatest need of a tuned co-precessing model. 
In the cases where \textsc{PhenomDCP} has a similar or slightly worse performance than either of the other two models the match is 
generally already comparable to the accuracy level of our input \nr{} waveforms.

\subsection{Matches: Accuracy of the angle model}
\label{sec: matches angles}

In order to test the accuracy of the angle model we constructed a set of precessing waveforms by calculating the symmetrised frequency-domain 
co-precessing \nr{} waveform containing only the $\ell=2$ multipoles and ``twisting'' this waveform up with the modelled precession angles. We 
constructed two sets of precessing waveforms in this fashion; one using the model for the angles presented in this paper,
and the other using the \msa{} angles, in order to quantify the effect of modelling the merger-ringdown behaviour of 
the angles. We then calculated the SNR-weighted match between these waveforms and the symmetrised \nr{} waveforms in the 
$\mathbf{J}$-aligned frame comprising only the $\ell=2$ multipoles. As with the co-precessing matches described above, these matches were 
calculated at a fixed total mass $M=100$M$_\odot$ and performed over a frequency range from $\left( f_\text{ref} + 5 \right)$Hz to 244Hz 
(the frequency range for which the \nr{} data was available). 

\begin{figure}[htbp]
   \centering
   \includegraphics[width=0.47\textwidth]{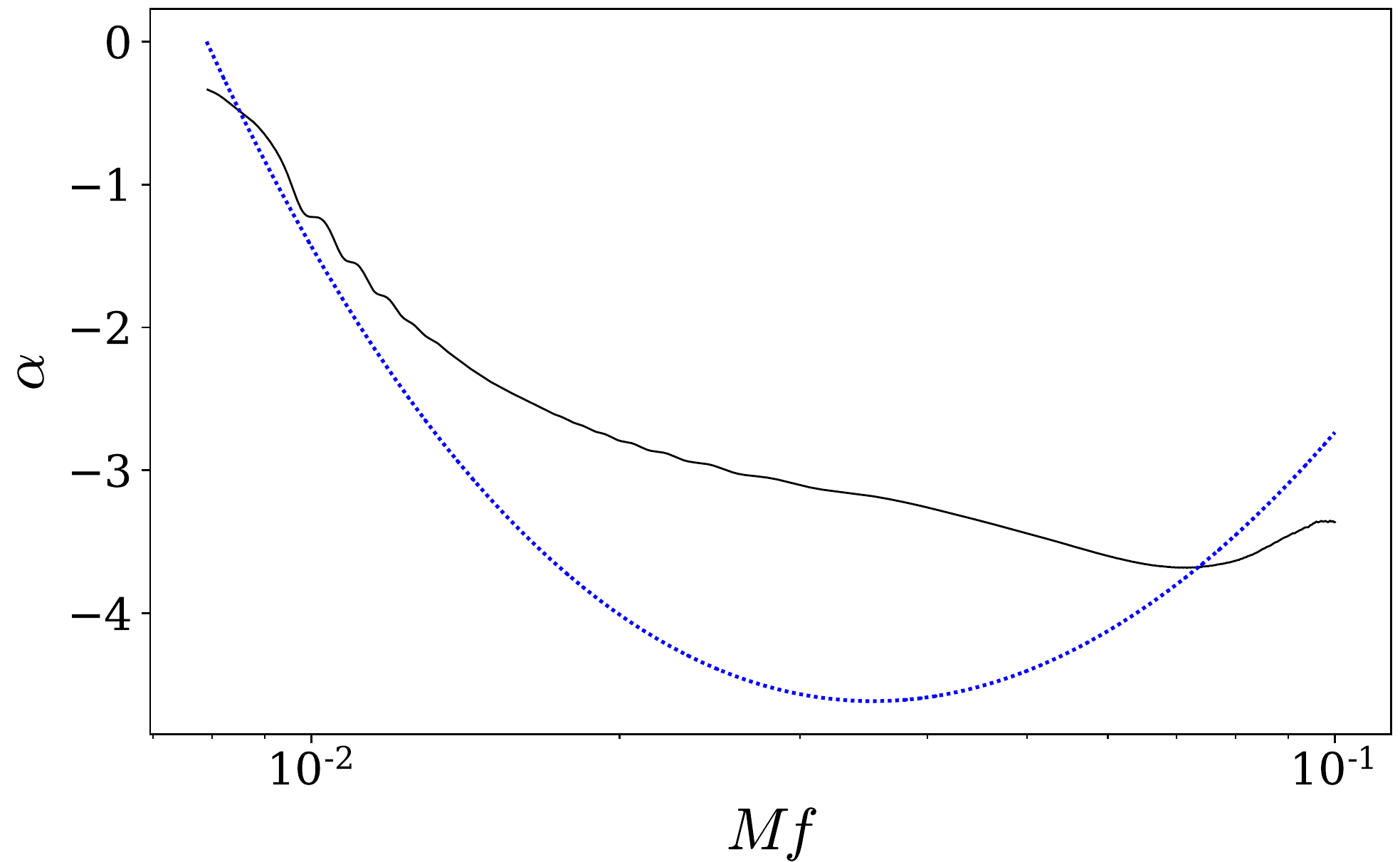} 
   \caption{Comparison \msa{} $\alpha$ (blue dashed line) with the value calculated from the \nr{} waveform (black solid line) for the 
   $(q, \chi, \theta_{\rm LS}) = (8,0.8,120^\circ)$ configuration. In order to see a 
   region over which the two values agree well we would need a longer \nr{} waveform; see text for more details. 
   }
   \label{fig: alpha q8a08t120}
\end{figure}

In Fig.~\ref{fig: angle matches}, we have shown the inclination average of the full precessing match for ease of presentation. 
We can see that the matches using the improved angle model are above 0.99 across the majority of the parameter space. 
The only cases for which this is not true are in the most extreme corner of the parameter space we modelled; cases with $q=8$, $\chi=0.6$ and $\tls \ge 90^\circ$.  
In these cases we find the \pn{} expressions used for $\alpha$ during inspiral deviate from those calculated from the \nr{} waveform at reasonably low frequencies. 
In the case of 
$(q,\chi,\tls)=(8,0.8,120^\circ)$ this is before the start of the \nr{} waveform, as shown in Fig.~\ref{fig: alpha q8a08t120}. 
Improving the model for these cases would require a model for the intermediate region between where the \pn{} expression ceases to be accurate 
and where the current model begins, which may require longer \nr{} waveforms to be produced. 
Additionally, we expect that modelling this intermediate region will improve matches for several other cases as well, where the \pn{} expressions 
for the angles deviate from what we see in the \nr{} data at lower frequencies than are covered by our current merger-ringdown model for the angles. 
Nonetheless, in all cases we see significant improvement over the previous model.

\begin{figure*}[htbp]
   \centering
   \includegraphics[width=\textwidth]{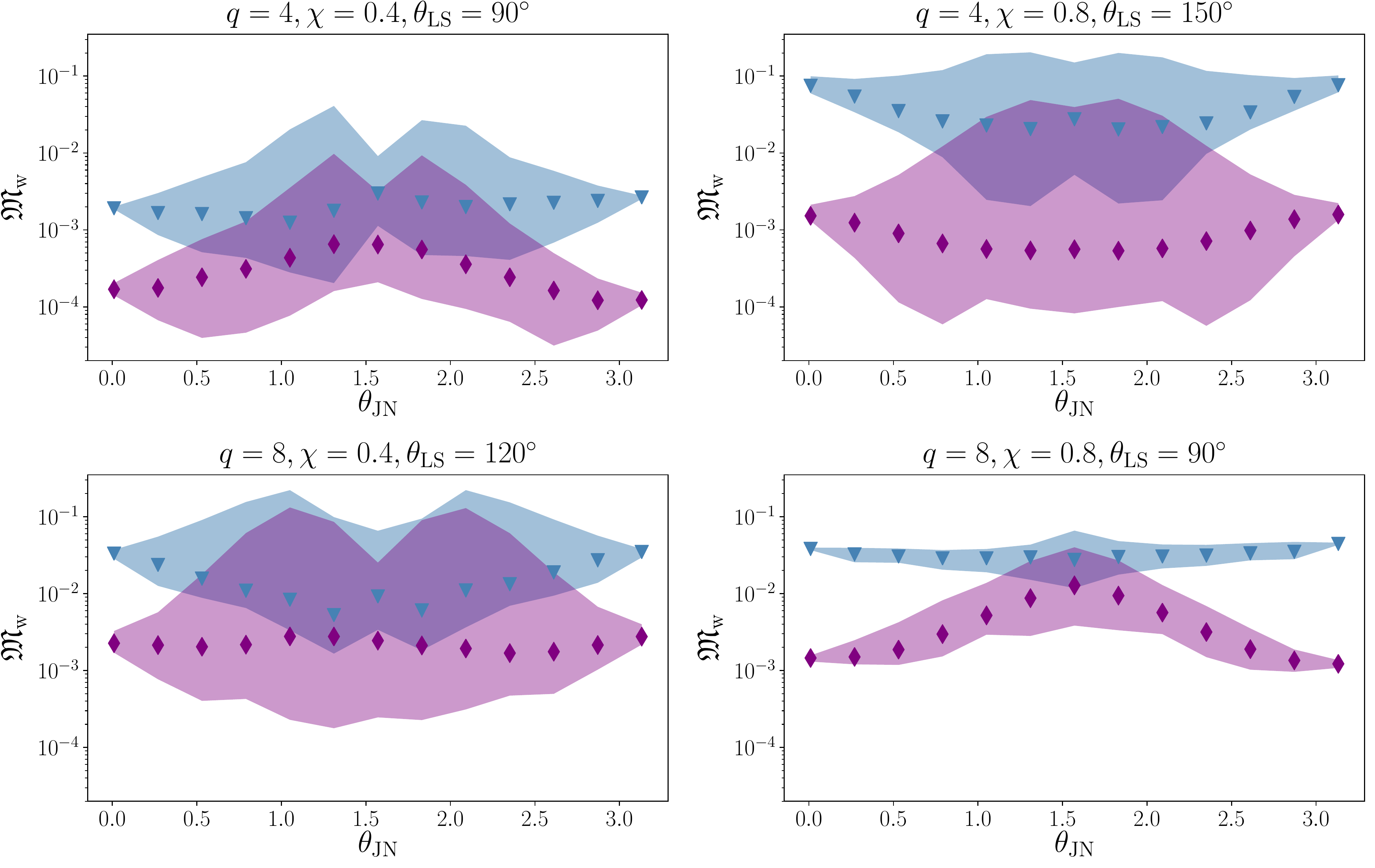} 
   \caption{Mismatch as a function of the inclination of the binary, quantified by the angle between the line of sight and the total angular 
   momentum $\theta_\text{JN}$, for four cases, at 100$M_\odot$. These mismatches consider the co-precessing NR waveform twisted 
   up with the angle model used by \textsc{PhenomPv3} (steel blue) and \textsc{PhenomPNR} (purple). The solid markers show the SNR-weighted 
   average mismatch while the shaded regions show the variation with respect to signal polaristation and phase. 
   }
   \label{fig: subset of matches}
\end{figure*}

The best matches are seen in the least extreme part of parameter space; namely for low mass ratio systems. This is the region of parameter
 space where existing models for the angles already perform reasonably well. The biggest improvement in the matches as a result of the 
improved model for the angles is seen at higher mass ratios, particularly for larger $\tls$. 

For a selection of these cases we show the mismatch as a function of $\theta_{\rm JN}$ in Fig.~\ref{fig: subset of matches}. 
The figure shows both the SNR-weighted average, and the range of mismatches with respect to signal polarisation and phase. 
We see that the mismatches against symmetrised \nr{} waveforms are approximately 
symmetric about $\theta_\text{JN} = \pi/2$. 
The \msa{} angles generally give the worst SNR-weighted average mismatch for systems with $\theta_\text{JN} = 0, \pi$, although this is not always the
case, and the variations with respect to different choices of polarisation and phase are often larger than those with respect to 
inclination. This mismatch then typically improves as it approaches $\theta_\text{JN}=\pi /2$ systems, with a slight increase 
for systems at exactly $\pi/2$ in most cases. In contrast, the SNR-weighted average mismatches involving the new angle model show one of two main 
behaviours with respect to inclination: the first gives the lowest mismatches for systems with $\theta_\text{JN} = 0, \pi$ with a marked degradation 
towards $\theta_\text{JN} = \pi/2$, while the second shows approximately constant values for the mismatch with respect to inclination, with a
possible slight improvement for systems with $\theta_\text{JN} = \pi/2$. However, we do not observe any clear pattern in how these two trends manifest 
themselves across the parameter space. The most important result to note is that 
in comparing the new angle model with the \msa{} angles in this figure, for the new angle model the lowest mismatch is always better 
(the lower edge of the envelopes), the highest mistmatch is always better (the upper edge of the envelopes), and the SNR-weighted 
mismatch is always better.

In general we might expect errors in the angle models to lead to worse mismatches for edge-on configurations, 
since at these orientations the contributions of the subdominant $\ell=2$ multipoles are largest, and the strength of those multipoles 
in our model is directly related to the precession angles, in particular $\beta$. However, $\theta_{\rm JN} = \pi/2$ doesn't necessarily
correspond to the binary being edge-on to the detector, unless $\beta$ is close to zero; in general, a system viewed from 
$\theta_{\rm JN} = \pi/2$ is \emph{never} edge on. Because of this, and because of the large variation in matches across the
cases shown in Fig.~\ref{fig: subset of matches}, we revisit this question in the full-model mismatches in the next section, where
we specify the binary orientation at the beginning of the waveform (so $\theta_\text{LN} = \pi/2$ corresponds to edge-on at least at
one point in the inspiral), and consider an exhaustive set of masses, orientations and polarisations for every \nr{} waveform.

\subsection{Matches: Accuracy of \pnr}\label{sec: full model matches}

In this section we compare the accuracy of the complete \pnr{} model to existing precessing waveform models by computing 
SNR-weighted mismatches between these approximants and the various \nr{} waveforms detailed in \sect{sec: verification wvfms}.
Each SNR-weighted mismatch is computed over a range of total masses \(M_\text{total}\in[100,120,140,160,180,200,220,240]\)M\(_\odot\) 
and at four inclination values, \(\theta_\text{LN}\in[0,\pi/6,\pi/3,\pi/2]\), specified at the reference frequencies given in the waveform tables. 
The choice to sample in \(\theta_\text{LN}\), rather than \(\theta_\text{JN}\) as done above, was motivated partially by the frame convention 
of \texttt{LALsuite}~\cite{lalsuite}, which specifies that the LAL inertial frame~\cite{Schmidt:2017btt} in which the waveforms are generated be 
instantaneously \(\hat{\mathbf{L}}\)-aligned at the given reference frequency. 
This choice allows for comparisons with match results already present in the literature.
As was also noted in the previous section, the conventional wisdom gleaned from non-precessing signals regarding model performance 
and the importance of higher multipoles for configurations with $\theta_\text{LN}\sim\pi/2$ also holds for precessing cases where \(\beta\) 
remains small throughout most of the inspiral, as is the case with most of the \nr{} waveforms we consider.
The matches were performed starting at a frequency of 20Hz or \(f_\text{ref}+5\)Hz, whichever was higher, with \(f_\text{ref}\) listed for 
each \nr{} waveform in Tables~\ref{tab: BAM cases} and \ref{tab: SXS and GT cases}. The nominal starting frequency of 20\,Hz was chosen
to match the approximate low-frequency cut-off of typical signal analysis, and dictated our choice of 100\,$M_\odot$ as the lowest total
mass we consider.

\subsubsection{Match variation with inclination}

\begin{figure}[htbp]
   \centering
   \includegraphics[width=0.49\textwidth]{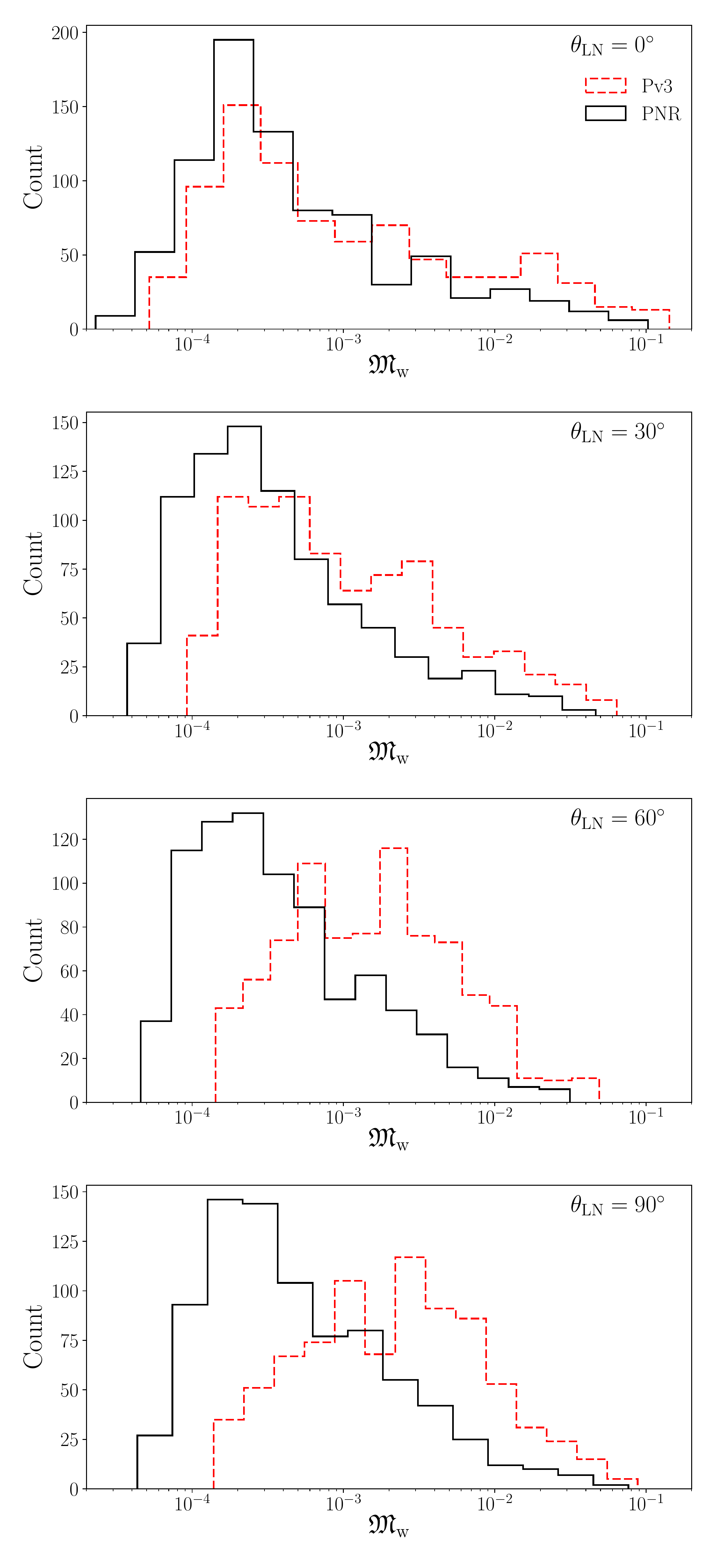} 
   \caption{Histograms of the SNR-weighted mismatches between the \nr{} waveforms listed in Tables~\ref{tab: BAM 
   cases}-\ref{tab: SXS and GT cases} and the waveform models \pnr{} and \pv3. Each subplot contains the SNR-weighted mismatches 
   for all total masses separated by inclination descending as \(\theta_\text{LN}\in[0^\circ,30^\circ,60^\circ,90^\circ]\). The mismatches for 
   all total mass values listed in \sect{sec: full model matches} are included at each inclination. The results for \pnr{} are present in solid 
   black while the results for \pv3{} are given in dashed red.}
   \label{fig: inc varied histograms}
\end{figure}

As discussed in the previous section, we would expect improvement in \pnr{} to be most apparent when trying to replicate
highly-precessing signals at high inclination, \(\theta_\text{LN}\sim\pi/2\), where the modulations in the signal due to precession grow stronger 
as more power is distributed across the \(\ell=2\) multipoles. 
We therefore compare the performance of \pnr{} with the earlier precessing model \pv3{}, plotting the SNR-weighted mismatches between
these two models and the \nr{} waveforms for each inclination value used. We choose \pv3{}, since it was the base model that we modified to 
produce \pnr{}, and this comparison provides the most direct measure of the level of improvement achieved by including \nr{}-tuned precession 
effects in both the co-precessing-frame and angle models.  
The overall distribution of SNR-weighted mismatches is shown in \cfig{fig: inc varied histograms}. 
For \pv3{}, which uses the uncalibrated \msa{} angles, the performance noticably degrades as the signal inclination increases, whereas the 
mismatches for \pnr{} remain relatively unchanged with respect to changes in inclination. This is largely consistent with 
\cfig{fig: subset of matches}, where the average match shows little variation with respect to inclination for three out of the four 
configurations shown, and suggests that the behaviour of the model for the $(q, \chi, \theta_{\rm {LS}}) = (8, 0.8, 90^\circ)$ case is atypical.

\subsubsection{General match results}

\begin{figure*}%[htbp]
   \centering
   \includegraphics[width=\textwidth]{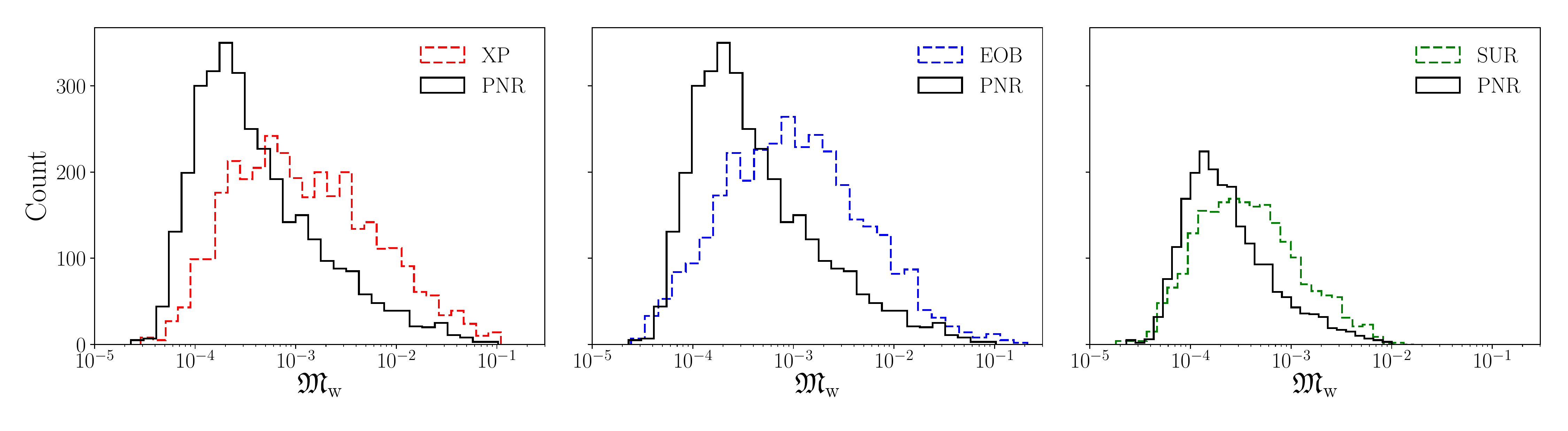} 
   \caption{Histograms of the SNR-weighted mismatches between various models in comparison and the \nr{} waveforms 
   listed in Tables~\ref{tab: BAM cases}-\ref{tab: SXS and GT cases}. The mismatches for all inclination and total mass values listed in 
   \sect{sec: full model matches} are included. In all three subplots, the results for \pnr{} (``PNR'') are presented with a solid black outline, 
   with the other model results given with dashed outlines from left to right as \xp{} (``XP'') in red, \eobnr{} (``EOB'') in blue, and \nrsur{} 
   (``SUR'') in green. For the comparison plot between \pnr{} and \nrsur{} we only include results of \nr{} waveforms for which both models are run. 
   }
   \label{fig: mismatch histograms}
\end{figure*}

\begin{figure*}%[htbp]
   \centering
   \includegraphics[width=\textwidth]{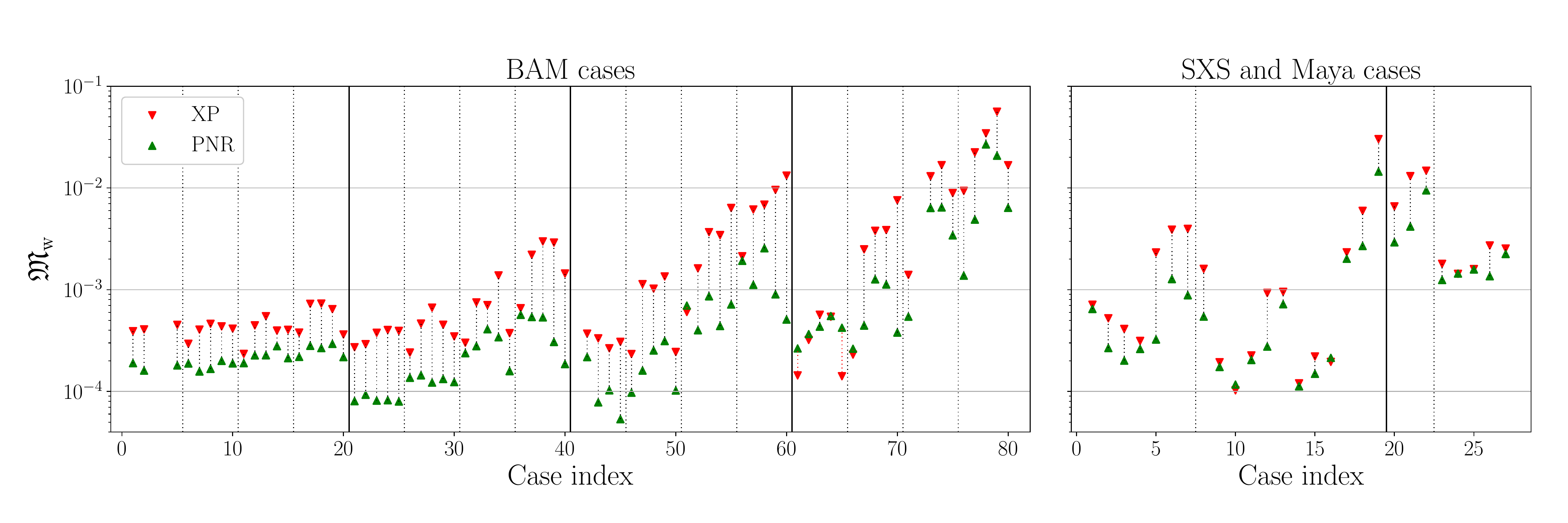} 
   \includegraphics[width=\textwidth]{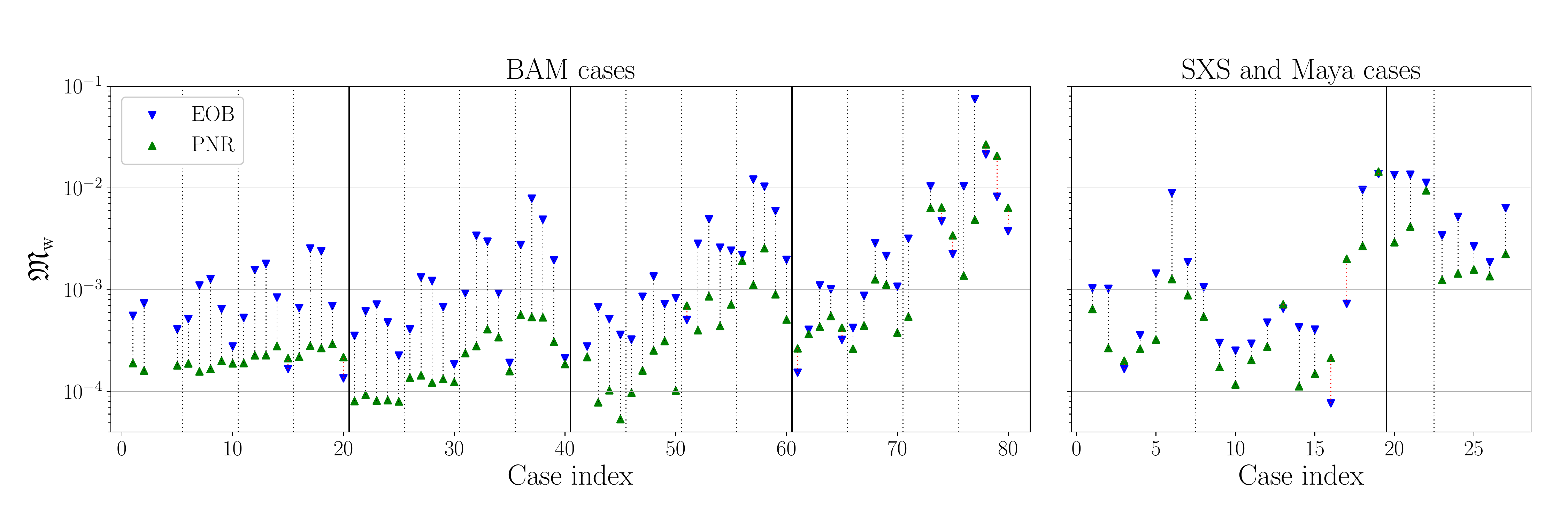} 
   \caption{SNR-weighted mismatches averaged over total mass and inclination between the precessing waveform models \pnr{} (``PNR''), 
   \xp{} (``XP''), and \eobnr{} (``EOB''), and the \nr{} waveforms listed in Tables~\ref{tab: BAM cases}-\ref{tab: SXS and GT cases}, shown 
   in order of the table listings. For the BAM cases, the solid vertical lines separate cases by mass ratio, and the dashed vertical lines separate 
   spin magnitude. For the SXS and Maya cases, the solid vertical line splits by \nr{} catalogue, and the dashed vertical line indicates a transition 
   from single-spin to two-spin cases.
   }
   \label{fig: mismatch scatters}
\end{figure*}

We compare the performance or \pnr{} against the precessing waveform approximants \xp, \eobnr, and \nrsur.
The full results are shown for all inclinations and total masses in \cfig{fig: mismatch histograms}, and the mass- and inclination-averaged 
SNR-weighted mismatches are shown per waveform in \cfig{fig: mismatch scatters}.
The model \nrsur{} was calibrated only up to \(q=4\), and while its implementation in \texttt{LALsuite} allows for extrapolation beyond this, 
we choose to limit the comparison with this model to the subset of the available \nr{} waveforms with~\(q\le4\) to ensure accuracy is maintained.

Overall we see an improvement in the mismatches between \pnr{} and the \nr{} waveforms compared to \xp{} and \eobnr. 
The mismatch results show comparable performance between \pnr{} and \nrsur{}, but we caution a reminder that \nrsur{} is a 
model that does not make the simplifying assumptions outlined in \sect{sec:prelims}, and while the effects of these additional 
physical features are generally small, we expect that their presence in \nrsur{} compared to the \nr{} data used for this comparison 
would bias the results toward slightly higher mismatches. Nonetheless, it is encouraging to observe that the \pnr{} model, while tuned 
to a comparatively small number of waveforms over a large configuration parameter space, and using a simple set of model 
ans\"atze, and several simplifying assumptions, in general has comparable mismatches to the \nrsur{} model.

From \cfig{fig: mismatch scatters} it is apparent that the overall mismatch increases with mass-ratio, and for each mass ratio the mismatch 
generally worsens with increasing spin magnitude. 
Such a trend is also visible in \figs{fig: cp matches}{fig: angle matches}.
A simple explanation for this observation arises from the \pn{} scaling of the opening angle \(\iota\) with symmetric mass ratio and spin 
magnitude in quasi-circular binaries with simple precession~\cite{Apostolatos:1994mx},
\begin{equation}
\sin\iota=\frac{S_\perp}{\sqrt{(\eta\sqrt{MR}+S_\parallel)^2+S_\perp^2}},
\end{equation}
where \(M\) is the system's total mass and \(R\) its orbital separation.
A larger opening angle increases the impact of precession modulations on the signal, and these are where model inaccuracies will be most
apparent.
One would similarly expect to see worsening mismatches as \(S_\perp\) is maximised, \textit{i.e.}, \(\tls=90^\circ\) for single-spin cases; 
however this trend is not as apparent in the results.
The results in \cfig{fig: mismatch scatters} show that \pnr{} is an improvement over \xp{} in the most extreme region of parameter space 
for \(q=8, \chi\in[0.6,0.8]\), while \eobnr{} yields better results in this region when \(\tls>90^\circ\).

Regarding the performance of \pnr{} for the two-spin \nr{} cases listed in \ctbl{tab: SXS and GT cases}, specifically cases 8-19 and cases 23-27, 
we observe that \pnr{} and \xp{} perform surprisingly similarly for these cases, both for the SXS and MAYA cases, whereas \pnr{} provides a 
general improvement over \eobnr{} for the two-spin cases. 
These results provide a reassuring validation of the single-spin mapping detailed in \sect{sec: spin parameterisation}.

\begin{figure}
   \centering
   \includegraphics[width=0.49\textwidth]{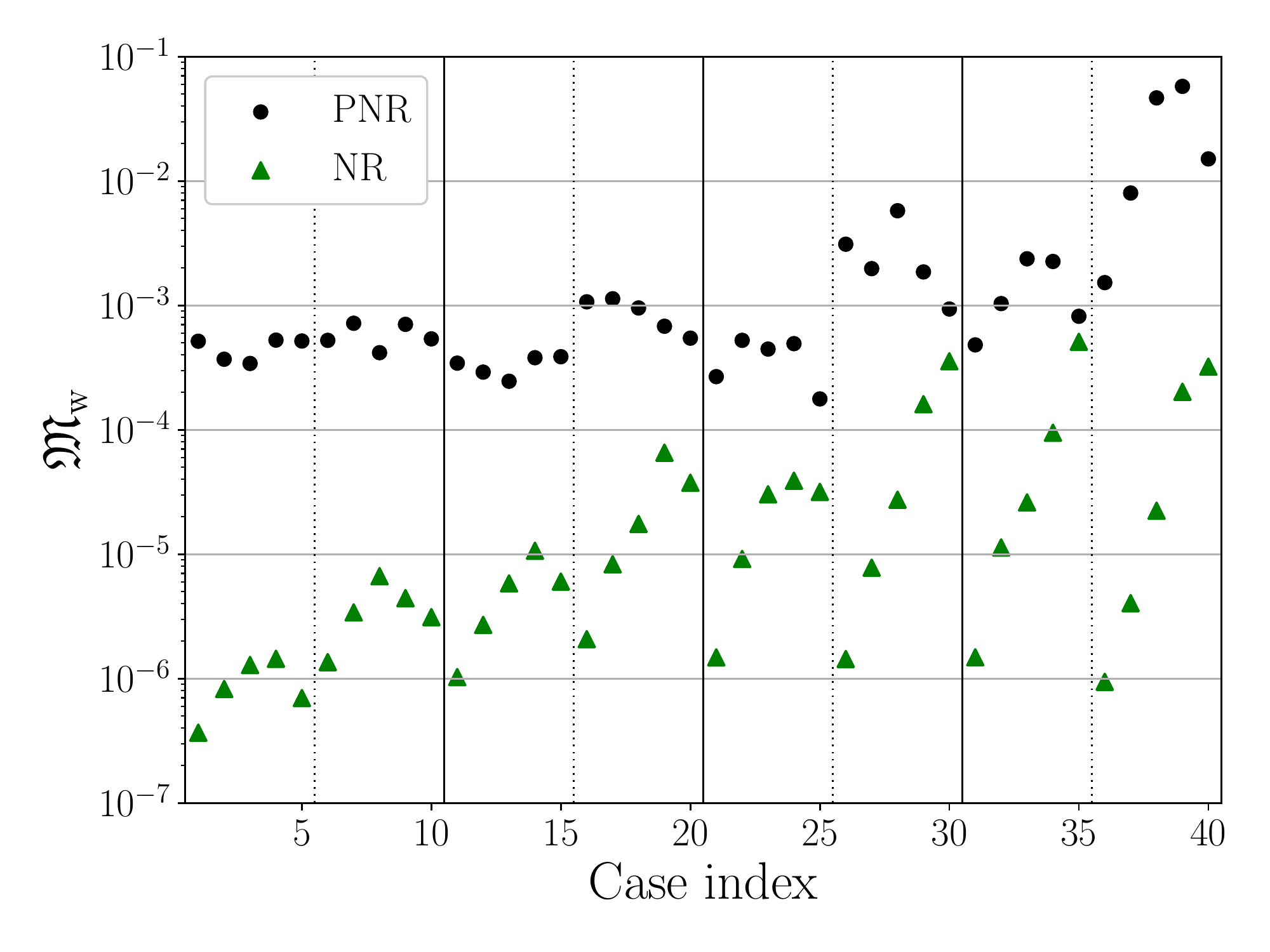} 
   \caption{SNR-weighted mismatches computed at 100M\(_\odot\) and averaged over inclination between the precessing waveform model 
   \pnr{} (``PNR'') and the \nr{} waveforms listed in Table~\ref{tab: BAM cases}, shown in order of the table listings. Alongside these results 
   are plotted the SNR-weighted mismatches computed between the \nr{} waveforms in the initially \(\hat{\mathbf{J}}\)-aligned frame and the 
   fixed-\(\hat{\mathbf{J}}\) frame (``NR''). The solid vertical lines separate cases by mass ratio, and the dashed vertical lines separate spin 
   magnitude.
   }
   \label{fig: jt vs jt scatters}
\end{figure}

Finally we remark on the impact of the fixed-\(\hat{\mathbf{J}}\) assumption used in the modeling of \pnr{} and outlined in \csec{sec:motivation}.
We computed the SNR-weighted mismatches between the raw \nr{} signals in an initially \(\hat{\mathbf{J}}\)-aligned frame and those in the 
fixed-\(\hat{\mathbf{J}}\) frame and find that the resulting mismatches are more than an order of magnitude lower than the mismatches between 
\pnr{} and the fixed-\(\hat{\mathbf{J}}\) frame \nr{} signals presented in this section, and in all cases lower than \(5.1\times10^{-4}\) at \(100 M_\odot\).
The full comparison is displayed in \cfig{fig: jt vs jt scatters}, and shows that the fixed-\(\hat{\mathbf{J}}\) approximation remains valid over a 
broad range of parameter space where \(\tls < 90^\circ\) but begins to break down for systems with higher mass ratio and opening angle, implying 
that future modelling efforts should take care to re-evaluate the validity of this approximation in more extreme regions of parameter space.

\section{Conclusion}

We have presented a new model of the GW signal from the inspiral, merger and ringdown of precessing non-eccentric 
black-hole binaries, \textsc{PhenomPNR}. This is the first model to explicitly calibrate precession effects through merger and 
ringdown to \nr{} simulations, and to use higher-order \pn{} amplitude terms to consistently define a signal-based co-precessing 
frame (the ``quadrupole aligned'' (QA), or ``optimal emission direction'') throughout the model. 

The model is calibrated to 40 \nr{} simulations of binaries where only the larger black hole is spinning; the simulations cover four mass
ratios ($q = 1, 2, 4, 8$), two spin magnitudes ($\chi = 0.4, 0.8$), and five values of the spin misalignment angle 
($\theta_{\rm LS} = 30^\circ, 60^\circ, 90^\circ, 120^\circ, 150^\circ$). In the frequency domain we separately model the 
co-precesing-frame signal, $h_{2,2}^{\rm CP}(f)$, and the precession angles, $(\alpha, \beta, \gamma)$. We model only the dominant
$(\ell=2,|m|=2)$ multipoles in the co-precessing frame, and neglect $\pm m$ asymmetries in the multipoles. 

The co-precessing-frame model, \textsc{PhenomDCP}, is an extension of the earlier aligned-spin model \textsc{PhenomD}, which was
calibrated to 19 \nr{} simulations of either single-spin or equal-spin binaries, up to $q=18$ and spins of $|\chi| \leq 0.85$. Our extension 
captures the effect of in-plane spin on the amplitude and phase of the co-precessing-frame signal in the late inspiral and merger-ringdown. 
We note for the first time that the
final black hole's ringdown frequency is shifted to an \emph{effective ringdown frequency} in the co-precessing frame. For this
reason we explicitly model the effective ringdown frequency across the single-spin parameter space, and do not make use of estimates 
of the final black hole's mass and spin. 

The angle model, \textsc{PhenomAngles}, uses during inspiral the \msa{} \pn{} angles used in previous \textsc{Phenom} models.
These angles describe the dynamics of the orbital plane of the binary, which is only approximately equal to the QA direction that we require
to correctly model the signal. We find that
this approximation holds well throughout the inspiral for the angles $\alpha$ and $\gamma$, but is not sufficiently accurate for $\beta$. 
However, it is possible to use higher-order \pn{} amplitude expressions, and reduction to a single-spin subspace, to transform the 
\msa{} binary inclination into a good approximation of the QA $\beta$. This is discussed in \sect{sec: beta approx}. Note that 
current \textsc{EOBNR} models also use the orbital precession dynamics as an approximation to the signal precession dynamics, and so
an approach like the one used here is likely to also improve the accuracy of those models. 

The \textsc{PhenomAngles} precession angles through merger and ringdown are where our model differs most significantly from previous
\textsc{Phenom} and \textsc{EOBNR} models. We observe and model a ``dip'' in $\alpha(f)$ (and therefore $\gamma(f)$) around the effective 
ringdown frequency, similar to that found in the phase derivative when constructing \textsc{PhenomD}~\cite{Husa:2015iqa,Khan:2015jqa}. 
Most importantly, we also model the 
steep collapse of $\beta$ through merger. As we discuss in \sect{sec:physfeat}, this feature is quite distinct from the asymptotic 
ringdown behaviour of $\beta$, and results in a shift in the frequency location of the peak amplitude for each of the 
$\ell=2$ multipoles; see Fig.~\ref{fig: mode hierarchy}. 

Our precession model is tuned to single-spin \nr{} simulations, but we make use of a non-bijective mapping between the six spin
components required to describe a two-spin system, and the two components required in our single-spin fits to \nr{} data; see 
\sect{sec: spin parameterisation}. In some parts
of our model, this is equivalent to using the $\chi_{\rm p}$ parameter from earlier \textsc{Phenom} models, but we also introduce modifications to
produce a mapping with greater physical fidelity near $q=1$, and in the precession dynamics we taper away two-spin oscillations as 
the system approaches merger. The result is an approximate \imr{} model for two-spin systems. 

In \sect{sec:matches} we demonstrate the accuracy of \textsc{PhenomDCP}, \textsc{PhenomAngles}, and the complete \imr{} 
model \textsc{PhenomPNR}, by calculating matches against \nr{} waveforms. The matches are calculated against not
only the 40 calibration waveforms, but an additional 36 \textsc{BAM} verification waveforms from across the same single-spin
parameter space, plus 27 SXS and Maya waveforms, which include two-spin systems. We find that our model in general improves significantly 
over previous \textsc{Phenom} and \textsc{EOBNR} models, as illustrated in Fig.~\ref{fig: mismatch scatters}. 

There are several immediate directions for future work. \textsc{PhenomPNR} does not model subdominant multipoles in the 
co-precessing frame, but these will be essential for measuring the properties of observations at larger mass ratios, which is the
very region of parameter space where \textsc{PhenomPNR} shows the greatest improvement over previous models. This could be 
achieved through directly modelling each of the multipoles, and including mode-mixing effects as in Ref.~\cite{Garcia-Quiros:2020qpx}.
Alternatively, one could estimate the subdominant multipoles through the approximation used in Ref.~\cite{London:2017bcn}.

Beyond this, the model needs to be extended to include explicit \nr{} calibration to two-spin systems, and to model $\pm m$ 
multipole asymmetries.
Our results also suggest that the angle modelling needs to be improved at lower frequencies for cases with large mass ratios, large
spins, and large values of $\theta_{\rm LS}$; it is possible that this will require longer \nr{} simulations. 

\textsc{PhenomPNR} models the signal in a frame where the direction of the total angular momentum is constant, by first transforming the 
calibration \nr{} waveforms to a frame that tracks the evolution of $\hat{\mathbf{J}}(t)$. The error incurred by this approximation is 
evaluated in Fig.~\ref{fig: jt vs jt scatters}, and we note that this error is in general well below the other sources of modelling error. 
However, in future, if we wish to construct models with mismatch errors below $10^{-4}$, this approximation will need to be removed.

Finally, although most GW
observations to date have been of systems with comparable masses, there has been one observation (GW190814~\cite{Abbott:2020khf}) 
where the mass ratio is likely outside the calibration region of this model, and so it is necessary that the calibration region be extended to higher 
mass ratios. All of these areas are the subject of ongoing work.

\section{Acknowledgements}

We would like to thank other members of the Cardiff Gravity Exploration Institute who performed simulations that were used in this 
project: Shrobana Ghosh, Charlie Hoy, Panagiota Kolitsidou and David Yeeles.

The authors were supported in part by Science and Technology Facilities Council (STFC) grant ST/V00154X/1 and European Research Council (ERC)
Consolidator Grant 647839.
E. Hamilton was supported in part through the COST Action CA18108, supported by COST (European Cooperation in Science and Technology), 
by Swiss National Science Foundation (SNSF) grant IZCOZ0-189876. 
L. London was supported at Massachusetts Institute of Technology~(MIT) by National Science Foundation Grant No. PHY-1707549 
as well as support from MIT’s School of Science and Department of Physics.
A. Vano-Vinuales also thanks the PhD researcher Decree-Law no. 57/2016 of August 29 (Portugal) and Project No. UIDB/00099/2020 for support.

M. Hannam thanks Universit\`a di Roma ``Sapienza'' for hospitality while part of this work was completed. 

Simulations used in this work were performed on the DiRAC@Durham facility, managed by the Institute for Computational Cosmology 
on behalf of the STFC DiRAC HPC Facility (www.dirac.ac.uk). The equipment was funded by BEIS capital funding via STFC capital grants 
ST/P002293/1 and ST/R002371/1, Durham University and STFC operations grant ST/R000832/1. In addition, several of the simulations used in this 
work were performed as part of an allocation graciously provided by Oracle to explore the use of our code on the Oracle Cloud Infrastructure. 

This research also used the supercomputing facilities at Cardiff University operated by Advanced Research Computing at Cardiff (ARCCA) on 
behalf of the Cardiff Supercomputing Facility and the HPC Wales and Supercomputing Wales (SCW) projects. We acknowledge the support 
of the latter, which is part-funded by the European Regional Development Fund (ERDF) via the Welsh Government. In part the computational 
resources at Cardiff University were also supported by STFC grant ST/I006285/1.

Various plots and analyses in this paper were made using the Python software packages \texttt{LALSuite}~\cite{lalsuite}, 
\texttt{Matplotlib}~\cite{Hunter:2007}, \texttt{Numpy}~\cite{harris2020array}, \texttt{PyCBC}~\cite{alex_nitz_2021_4849433}, 
and \texttt{Scipy}~\cite{2020SciPy-NMeth}.

\appendix

%%%%%%%%%%%%%%%%%%%%%%%%%%%%%%%%%%%%%%%%%%%%%%%%%%%%%%%%%%%%%%
\section{Calculation of Precession Angles}
\label{app:calc_angles}
Here we outline the calculation of the \oed{} quantified by $\alpha(f)$, $\beta(f)$ and $\gamma(f)$; see \cfig{fig: LaboutJ}.
To calculate each angle we use the rotationally invariant eigenvalue method~\cite{OShaughnessy:2011pmr,Boyle:2011gg}.
As shown in Ref.~\cite{Boyle:2011gg}, when multipole moments are limited to cases where $\ell=2$ this is equivalent to the 
original \qa{} method~\cite{Schmidt:2010it}. The result is independent of the initial inertial frame 
when the minimum rotation condition is imposed~\cite{Boyle:2011gg}. 

The eigenvalue method and minimal rotation convention are described in Refs.~\cite{OShaughnessy:2011pmr,Pekowsky:2013ska} and 
\cite{Boyle:2011gg}. 
Ref.~\cite{OShaughnessy:2011pmr} introduces the eigenvalue method. 
Ref.~\cite{Pekowsky:2013ska} details the practical structure of this method in its Appendix A, and Ref.~\cite{Boyle:2011gg} 
adds the minimal rotation convention which defines the optimal emission direction in a frame invariant way. 
Here we provide a self-contained description of the algorithm to calculate 
$\alpha(f)$, $\beta(f)$ and $\gamma(f)$.

Starting with the discrete Fourier transform of $\Psi_4$ decomposed into spin weight $-2$ spherical harmonics, $\tilde{\psi}_{\ell m}$, we compute the effect of all pair-wise angular momentum generators averaged about the binary's centre of mass. This is $\langle \mathcal{L}_{(a}\mathcal{L}_{b)} \rangle$, where
\begin{align}
  \label{Lab}
  \langle \mathcal{L}_{(a}\mathcal{L}_{b)} \rangle \; =& \;  \frac{1}{2} \langle  \mathcal{L}_{a}\mathcal{L}_{b} + \mathcal{L}_{b}\mathcal{L}_{a} \rangle
  \\ \nonumber
                                                   \; =& \; \frac{ \int_{\Omega} \tilde{\Psi}_4^*(f) \mathcal{L}_{(a}\mathcal{L}_{b)} \tilde{\Psi}_4(f) {d}\Omega } { \int_{\Omega} |\tilde{\Psi}_4(f)|^2 {d}\Omega }\;,
\end{align}
with $a$ and $b$ over $\{x,y,z\}$, and where
\def\mcL{\mathcal{L}}
\def\Lab{\langle \mathcal{L}_{(a}\mathcal{L}_{b)} \rangle}
\begin{align}
  \label{Lpm}
  \mcL_x &=   \frac{1}{2}(\mcL_+ + \mcL_-)\text{,  }\mcL_y = -i\frac{1}{2}(\mcL_+ - \mcL_-)\;,
  \\ \nonumber
  \mcL_\pm &= e^{\pm i \varphi} \left[ \pm i \partial_\theta - \cot \theta \partial_\varphi - i s \csc \theta \right]\;,
\end{align}
and
\begin{align}
  \label{Lz}
  \mcL_z = \partial_\varphi \; .
\end{align}
In \eqn{Lpm}, $s$ is the spin weight of the object being acted upon \cite{NP66,Shah:2015sva}.
As we are only interested in outgoing \gw{} radiation, $s=-2$.

In practice, evaluation of \eqn{Lab} need not require direct integration when $\tilde{\Psi}_4$ is written in terms of its multipole 
moments, $\tilde{\psi}_{\ell m}$; see \eqn{ylm}.
That is, as the operation of $\mcL_\pm$ and $\mcL_z$ on ${_{-2}}Y_{\ell m}$ are known \cite{Ruiz.0707.4654,NP66}, one finds that
\begin{subequations}
\label{I}
\begin{align}
  \Lab &=& \small{\frac{1}{\sum_{\ell,m}|\tilde{\psi}_{\ell m}|^2}
              \begin{bmatrix}
               I_0 + \text{Re}(I_2) & \text{Im} I_2  &   \text{Re} I_1 \\
                &   I_0 - \text{Re}(I_2) & \text{Im} I_1 \\
               & & I_{zz}
             \end{bmatrix}}
\end{align}
where 
\begin{eqnarray}
I_2&\equiv&  \frac{1}{2}\,(\tilde{\Psi},\mcL_+\mcL_+\tilde{\Psi}) \nonumber \\
 &=& \frac{1}{2}\,\sum_{\ell,m} c_{\ell m}c_{\ell m+1} \tilde{\psi}_{\ell m+2}^*\tilde{\psi}_{\ell m} \\
I_1 & \equiv &(\tilde{\Psi},\mcL_+(\mcL_z+1/2)\tilde{\Psi})  \nonumber \\
 &=& \sum_{\ell m} c_{\ell m}(m+1/2) \tilde{\psi}_{\ell m+1}^*\tilde{\psi}_{\ell m} \\
I_0 &\equiv& \frac{1}{2}\left(\tilde{\Psi}| \ell(\ell+1) - \mcL_z^2 |\tilde{\Psi}\right) \nonumber\\
 &=& \frac{1}{2}\sum_{\ell m} [\ell(\ell+1)-m^2]|\tilde{\psi}_{\ell m}|^2  \\
I_{zz} &\equiv& (\tilde{\Psi},\mcL_z \mcL_z \tilde{\Psi}) = \sum_{\ell m} m^2 |\tilde{\psi}_{\ell m}|^2
\end{eqnarray}
and where $c_{\ell m} = \sqrt{\ell(\ell+1)-m(m+1)}$.  \\

\end{subequations}

The resulting tensor, $\Lab$, is analogous to the Cauchy stress tensor in continuum mechanics, and describes infinitesimal changes 
in momenta (linear and angular) associated with $\tilde{\Psi}_4(f)$ averaged about the source.

From the discussion in \sect{sec:prelims}, we see that 
 $\Lab$ is unchanged when considering $\tilde{h}(f)$ rather than $\tilde{\Psi}_4(f)$, as the factor of $1/2\pi f$ amounts to a simple overall rescaling that 
 does not affect normalized eigenvectors.
From these points, it follows that the eigenvector of $\Lab$ with the largest eigenvalue describes the direction about the source that 
experiences the largest strain (or curvature) and strain-rate (or curvature-rate). This is the \oed{}.
\def\V{\hat{V}}

If we label $\Lab$'s dominant normalised eigenvector as $\V=(v_x,v_y,v_z)$, then the angles associated with the \oed{} are given by
\begin{align}
  \label{fda}
  \alpha(f) &= \arctan{\left(\frac{v_y(f)}{v_x(f)}\right)},
  \\
  \label{fdb}
  \beta(f)  &= \arccos{\left(v_z(f)\right)},
  \\ 
  \label{fdc}
  \gamma(f) &= - \int^{f} (\partial_{f'} \alpha(f') )\cos \beta(f') \; \text{d}f' \; .
\end{align}
\eqns{fda}{fdb} follow from the use of a source centered spherical polar coordinate system in the asymptotically flat decomposition frame.
Equivalently, this is related to the frame of a distant observer.
\eqn{fdc} is the minimum rotation condition presented in Ref.~\cite{Boyle:2011gg}, which removes secular changes in phase due to the 
evolution of $\alpha$ and $\beta$.

\section{Waveforms used in analysis}

The \nr{} waveforms used in the analysis of the model are listed in Tables~\ref{tab: BAM cases} and \ref{tab: SXS and GT cases}. \ctbl{tab: BAM cases} 
contains the 80 waveforms which comprise the BAM catalogue~\cite{BAM-catalog} of single spin precessing systems up to mass ratio \(q=8\) and single-spin magnitude \(\chi=0.8\).  
A subset of 40 of these waveforms were also used in tuning the model. \ctbl{tab: SXS and GT cases} lists the additional waveforms taken from the 
SXS~\cite{Mroue:2013xna, Boyle:2019kee} and Maya catalogues~\cite{Jani:2016wkt} used in assessing the accuracy of the model and ensuring it 
was not over-fitted. This selection of waveforms includes two-spin cases.

\begin{table*}[htbp]
\centering
\begin{tabular}{@{} cccccc @{}}
\toprule
Simulation ID & $\frac{100 M_\odot}{M} f_\text{ref}$ (Hz) & $q$ & $\chi$ & $\theta_\text{LS} (^\circ)$  \\
\hline
CF21-1 & 14.8 & 1 & 0.2 & 30 \\
CF21-2 & 14.8 & 1 & 0.2 & 60 \\
CF21-3 & -- & -- & -- & -- \\
CF21-4 & -- & -- & -- & -- \\
CF21-5 & 14.7 & 1 & 0.2 & 150 \\
CF21-6 & 14.8 & 1 & 0.4 & 30 \\
CF21-7 & 14.8 & 1 & 0.4 & 60 \\
CF21-8 & 14.9 & 1 & 0.4 & 90 \\
CF21-9 & 14.8 & 1 & 0.4 & 120 \\
CF21-10 & 14.8 & 1 & 0.4 & 150 \\
CF21-11 & 18.2 & 1 & 0.6 & 30 \\
CF21-12 & 14.8 & 1 & 0.6 & 60 \\
CF21-13 & 14.9 & 1 & 0.6 & 90 \\
CF21-14 & 14.8 & 1 & 0.6 & 120 \\
CF21-15 & 14.8 & 1 & 0.6 & 150 \\
CF21-16 & 14.8 & 1 & 0.8 & 30 \\
CF21-17 & 14.7 & 1 & 0.8 & 60 \\
CF21-18 & 14.9 & 1 & 0.8 & 90 \\
CF21-19 & 14.9 & 1 & 0.8 & 120 \\
CF21-20 & 15.2 & 1 & 0.8 & 150 \\
CF21-21 & 14.7 & 2 & 0.2 & 30 \\
CF21-22 & 14.7 & 2 & 0.2 & 60 \\
CF21-23 & 14.8 & 2 & 0.2 & 90 \\
CF21-24 & 15.2 & 2 & 0.2 & 120 \\
CF21-25 & 14.8 & 2 & 0.2 & 150 \\
CF21-26 & 14.8 & 2 & 0.4 & 30 \\
CF21-27 & 14.6 & 2 & 0.4 & 60 \\
CF21-28 & 14.7 & 2 & 0.4 & 90 \\
CF21-29 & 14.8 & 2 & 0.4 & 120 \\
CF21-30 & 14.8 & 2 & 0.4 & 150 \\
CF21-31 & 14.7 & 2 & 0.6 & 30 \\
CF21-32 & 14.9 & 2 & 0.6 & 60 \\
CF21-33 & 14.5 & 2 & 0.6 & 90 \\
CF21-34 & 14.9 & 2 & 0.6 & 120 \\
CF21-35 & 14.4 & 2 & 0.6 & 150 \\
CF21-36 & 14.9 & 2 & 0.8 & 30 \\
CF21-37 & 14.9 & 2 & 0.8 & 60 \\
CF21-38 & 14.7 & 2 & 0.8 & 90 \\
CF21-39 & 15.0 & 2 & 0.8 & 120 \\
CF21-40 & 15.0 & 2 & 0.8 & 150 \\
\botrule
\end{tabular}
\qquad
\begin{tabular}{@{} cccccc @{}}
\toprule
Simulation ID & $\frac{100 M_\odot}{M} f_\text{ref}$ (Hz) & $q$ & $\chi$ & $\theta_\text{LS} (^\circ)$  \\
\hline
CF21-41 & -- & -- & -- & -- \\
CF21-42 & 16.0 & 4 & 0.2 & 60 \\
CF21-43 & 16.8 & 4 & 0.2 & 90 \\
CF21-44 & 15.3 & 4 & 0.2 & 120 \\
CF21-45 & 15.2 & 4 & 0.2 & 150 \\
CF21-46 & 16.6 & 4 & 0.4 & 30 \\
CF21-47 & 16.3 & 4 & 0.4 & 60 \\
CF21-48 & 14.7 & 4 & 0.4 & 90 \\
CF21-49 & 14.8 & 4 & 0.4 & 120 \\
CF21-50 & 15.0 & 4 & 0.4 & 150 \\
CF21-51 & 17.0 & 4 & 0.6 & 30 \\
CF21-52 & 16.2 & 4 & 0.6 & 60 \\
CF21-53 & 15.8 & 4 & 0.6 & 90 \\
CF21-54 & 15.1 & 4 & 0.6 & 120 \\
CF21-55 & 14.0 & 4 & 0.6 & 150 \\
CF21-56 & 17.5 & 4 & 0.8 & 30 \\
CF21-57 & 16.8 & 4 & 0.8 & 60 \\
CF21-58 & 14.9 & 4 & 0.8 & 90 \\
CF21-59 & 14.8 & 4 & 0.8 & 120 \\
CF21-60 & 14.9 & 4 & 0.8 & 150 \\
CF21-61 & 18.4 & 8 & 0.2 & 30 \\
CF21-62 & 18.1 & 8 & 0.2 & 60 \\
CF21-63 & 17.8 & 8 & 0.2 & 90 \\
CF21-64 & 17.3 & 8 & 0.2 & 120 \\
CF21-65 & 17.2 & 8 & 0.2 & 150 \\
CF21-66 & 19.0 & 8 & 0.4 & 30 \\
CF21-67 & 18.6 & 8 & 0.4 & 60 \\
CF21-68 & 17.8 & 8 & 0.4 & 90 \\
CF21-69 & 17.0 & 8 & 0.4 & 120 \\
CF21-70 & 16.5 & 8 & 0.4 & 150 \\
CF21-71 & 19.7 & 8 & 0.6 & 30 \\
CF21-72 & -- & -- & -- & -- \\
CF21-73 & 17.9 & 8 & 0.6 & 90 \\
CF21-74 & 16.7 & 8 & 0.6 & 120 \\
CF21-75 & 17.0 & 8 & 0.6 & 150 \\
CF21-76 & 20.5 & 8 & 0.8 & 30 \\
CF21-77 & 19.5 & 8 & 0.8 & 60 \\
CF21-78 & 18.0 & 8 & 0.8 & 90 \\
CF21-79 & 16.0 & 8 & 0.8 & 120 \\
CF21-80 & 15.2 & 8 & 0.8 & 150 \\
\botrule
\end{tabular}

\caption{BAM single-spin configurations used in tuning the co-precessing and angle models as well as in the assessment of the accuracy of the model.}
\label{tab: BAM cases}
\end{table*}

\begin{table*}[htbp]
\centering
\begin{tabular}{@{} cccccccc @{}}
\toprule
Simulation ID & $\frac{100 M_\odot}{M} f_\text{ref}$ (Hz) & $q$ & $\chi$ & $\theta_\text{LS} (^\circ)$ & $\chi_1$ & $\chi_2$ \\
\hline
%SXS0077 & 9.3 & 1.5 & 0.5 & 90 & $\left(-0.493,0,0.082\right)$ & $\left(0,0,0\right)$ \\
SXS0097 & 9.2 & 1.5 & 0.5 & 90 & $\left(-0.493,0,0.083\right)$ & $\left(0,0,0\right)$ \\
%SXS0080 & 9.3 & 1.5 & 0.5 & 90 & $\left(-0.493,0,0.082\right)$ & $\left(0,0,0\right)$ \\
SXS0018 & 7.9 & 1.5 & 0.5 & 90 & $\left(-0.494,0,0.078\right)$ & $\left(0,0,0\right)$ \\
SXS0092 & 9.4 & 1.5 & 0.5 & 150 & $\left(-0.29,0,-0.407\right)$ & $\left(0,0,0\right)$ \\
SXS0033 & 11.2 & 3.0 & 0.5 & 30 & $\left(-0.19,0,0.463\right)$ & $\left(0,0,0\right)$ \\
SXS0035 & 8.5 & 3.0 & 0.5 & 90 & $\left(-0.476,0,0.154\right)$ & $\left(0,0,0\right)$ \\
SXS1109 & 10.2 & 5.0 & 0.5 & 90 & $\left(-0.435,0,0.246\right)$ & $\left(0,0,0\right)$ \\
%SXS0058 & 10.2 & 5.0 & 0.5 & 90 & $\left(-0.435,0,0.246\right)$ & $\left(0,0,0\right)$ \\
SXS0062 & 14.1 & 5.0 & 0.5 & 116 & $\left(-0.492,0,0.088\right)$ & $\left(0,0,0\right)$ \\
SXS0161 & 9.2 & 1.0 & 1.199 & 120 & $\left(-0.579,0,-0.158\right)$ & $\left(-0.579,0,-0.158\right)$ \\
SXS0115 & 10.2 & 1.07 & 0.246 & 74 & $\left(-0.027,-0.016,-0.203\right)$ & $\left(-0.236,0.018,0.304\right)$ \\
SXS0116 & 10.2 & 1.08 & 0.167 & 40 & $\left(0.022,-0.099,0.032\right)$ & $\left(-0.143,0.115,0.106\right)$ \\
SXS0124 & 10.2 & 1.26 & 0.412 & 44 & $\left(-0.247,-0.041,0.091\right)$ & $\left(-0.079,0.065,0.294\right)$ \\
SXS0102 & 9.3 & 1.5 & 0.5 & 90 & $\left(-0.486,0,0.116\right)$ & $\left(-0.486,0,0.116\right)$ \\
SXS1397 & 5.1 & 1.56 & 0.299 & 111 & $\left(-0.242,0.037,-0.172\right)$ & $\left(0.458,-0.089,0.102\right)$ \\
SXS0135 & 10.3 & 1.64 & 0.186 & 128 & $\left(-0.059,0.095,0.025\right)$ & $\left(0.003,-0.257,-0.228\right)$ \\
SXS0143 & 10.2 & 1.92 & 0.441 & 28 & $\left(-0.072,0.041,0.443\right)$ & $\left(-0.413,-0.15,-0.056\right)$ \\
SXS0144 & 10.2 & 1.94 & 0.214 & 146 & $\left(-0.135,0.011,-0.281\right)$ & $\left(0.05,-0.04,0.213\right)$ \\
SXS0049 & 11.2 & 3.0 & 0.527 & 72 & $\left(-0.474,0,0.159\right)$ & $\left(0.159,0,0.474\right)$ \\
SXS1160 & 11.1 & 3.0 & 0.658 & 63 & $\left(-0.455,-0.024,0.531\right)$ & $\left(0.41,0.212,-0.384\right)$ \\
SXS0165 & 18.1 & 6.0 & 0.93 & 125 & $\left(-0.648,0.003,0.639\right)$ & $\left(-0.186,-0.094,-0.216\right)$ \\
GT0745 & 22.3 & 6.0 & 0.6 & 91 & $\left(-0.437,0,0.411\right)$ & $\left(0,0,0\right)$ \\
GT0742 & 21.9 & 7.0 & 0.6 & 91 & $\left(-0.404,0,0.444\right)$ & $\left(0,0,0\right)$ \\
GT0834 & 20.9 & 7.0 & 0.8 & 168 & $\left(-0.375,0,0.706\right)$ & $\left(0,0,0\right)$ \\
GT0880 & 22.0 & 4.5 & 0.537 & 52 & $\left(-0.269,0,0.536\right)$ & $\left(0.269,0,-0.536\right)$ \\
GT0887 & 21.8 & 5.0 & 0.543 & 51 & $\left(-0.256,0,0.542\right)$ & $\left(0.256,0,-0.542\right)$ \\
GT0889 & 21.6 & 6.0 & 0.552 & 50 & $\left(-0.234,0,0.552\right)$ & $\left(0.234,0,-0.552\right)$ \\
GT0888 & 21.4 & 7.0 & 0.559 & 49 & $\left(-0.216,0,0.56\right)$ & $\left(0.216,0,-0.56\right)$ \\
GT0886 & 21.4 & 8.0 & 0.564 & 49 & $\left(-0.2,0,0.566\right)$ & $\left(0.2,0,-0.566\right)$ \\
\botrule
\end{tabular}
\caption{Additional configurations from the SXS and MAYA catalogues used in the assessment of the accuracy of the model.}
\label{tab: SXS and GT cases}
\end{table*}

\section{Parameter-space fits} \label{app:fitsurfaces}

\begin{figure*}[htb]
   \begin{tabular}{cc}
      
      \includegraphics[width=0.5\textwidth]{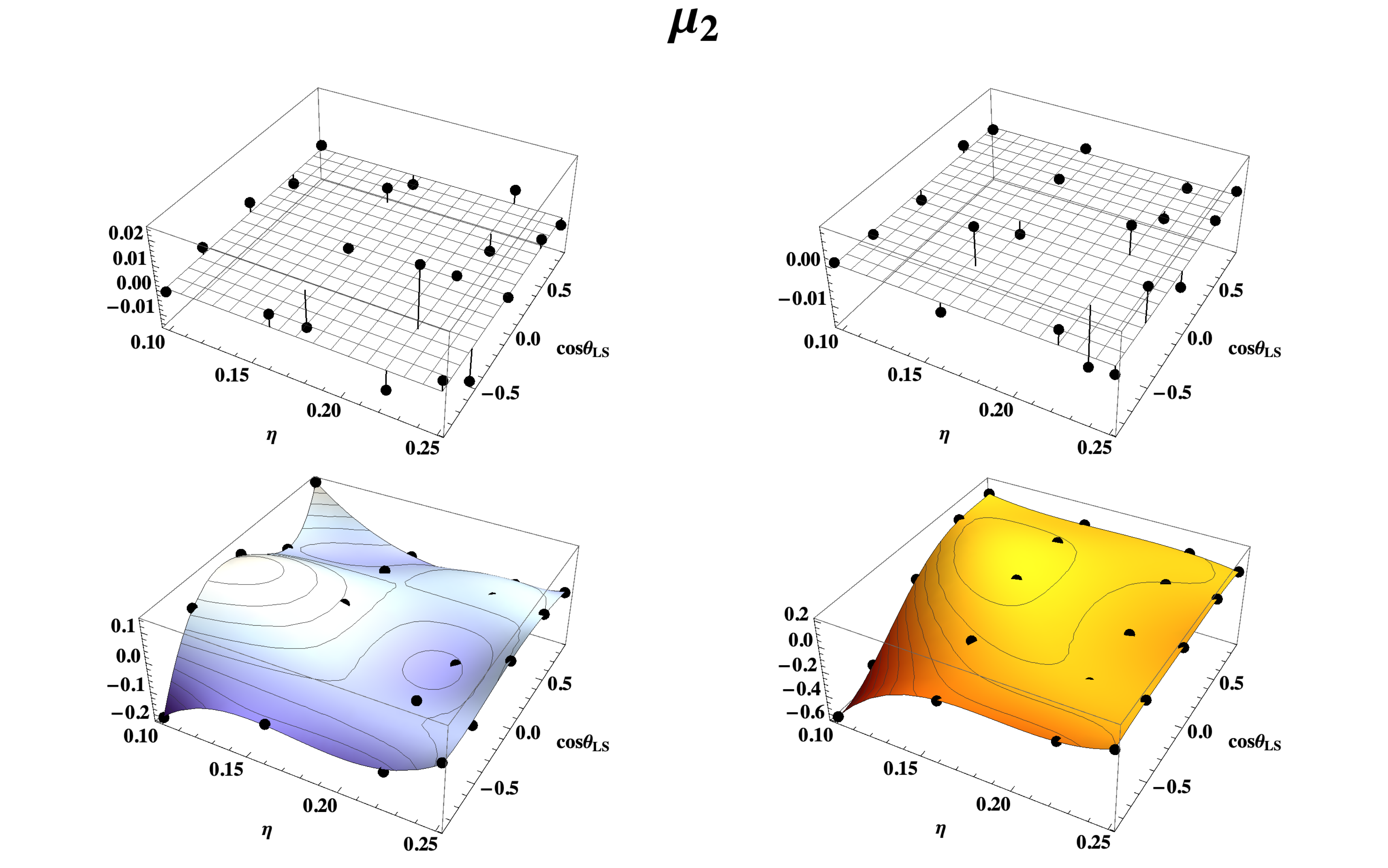}
      &
      \includegraphics[width=0.5\textwidth]{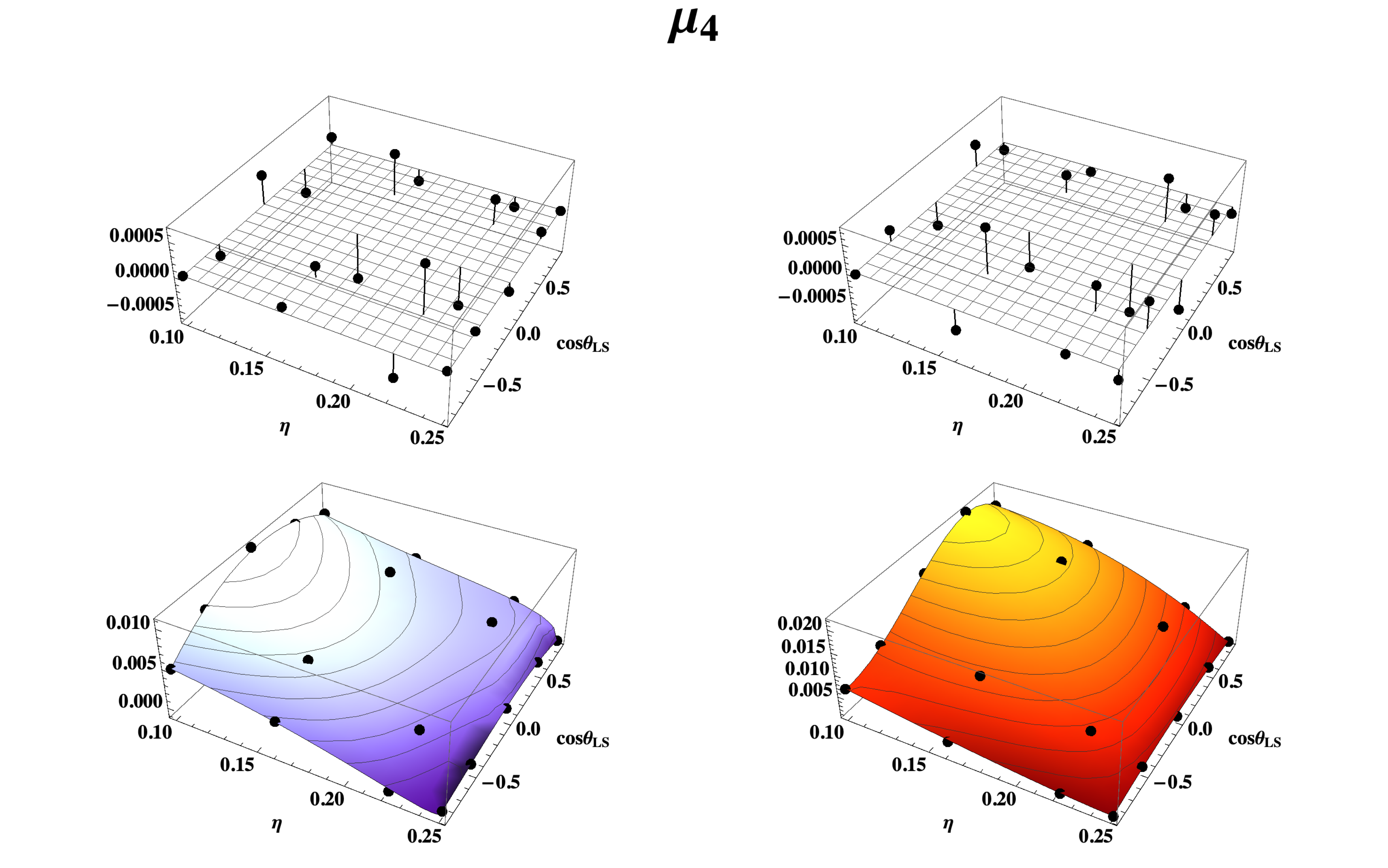}
      
   \end{tabular}
   \caption{Amplitude parameters for tuned co-precessing waveform model, \textsc{PhenomDCP}. The fits are shown as two-dimensional surfaces 
   covering the parameter space described by $\eta$ and $\cos\tls$. On the left in blue are the fits for the simulations with $\chi=0.4$ and on the right
    in red are the fits for $\chi=0.8$. Above each of these surfaces are shown the residuals.}
   \label{fig:dcpmu} 
\end{figure*}

\begin{figure*}[htb]
   \begin{tabular}{cc}
      
      \includegraphics[width=0.5\textwidth]{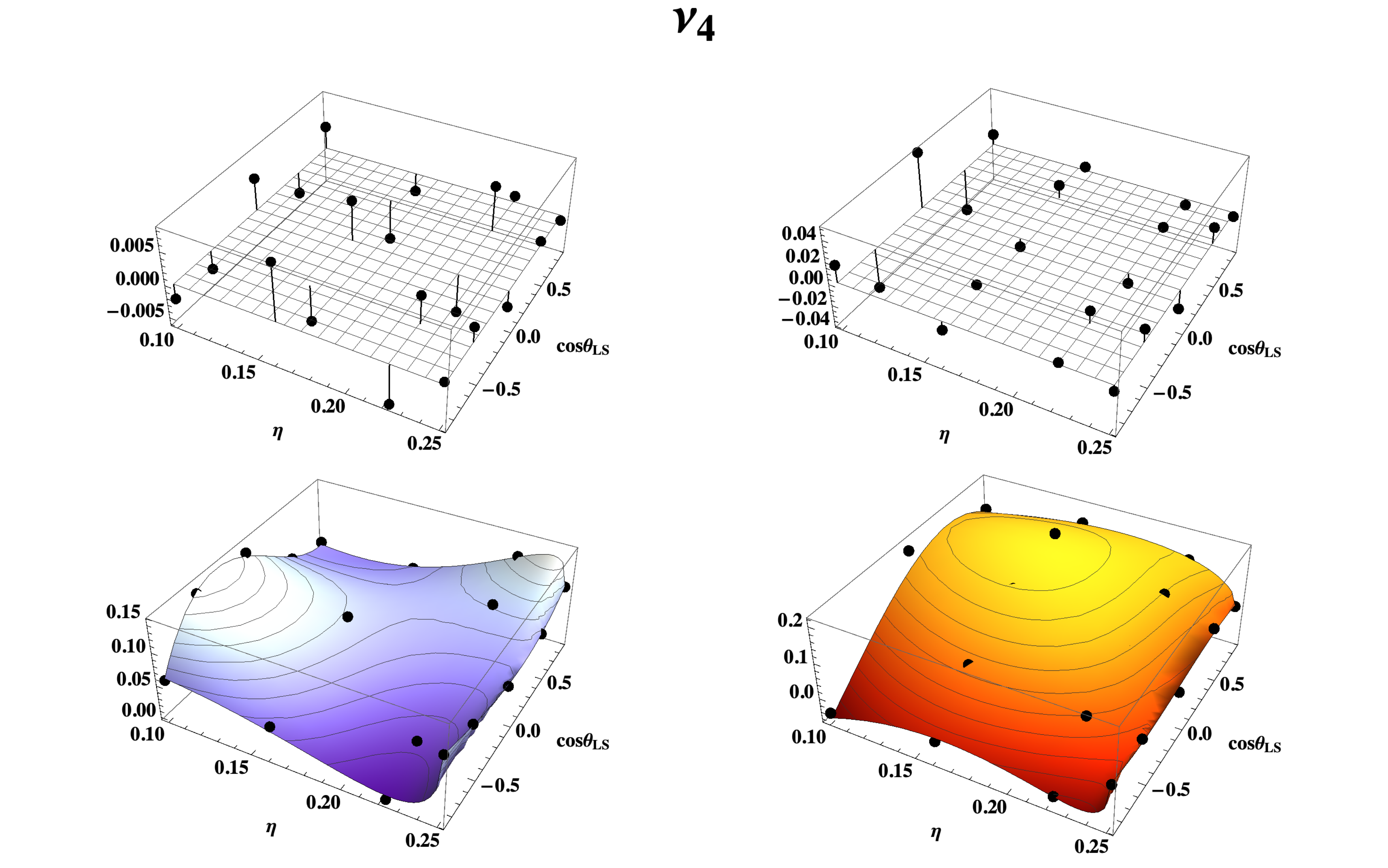}
      &
      \includegraphics[width=0.5\textwidth]{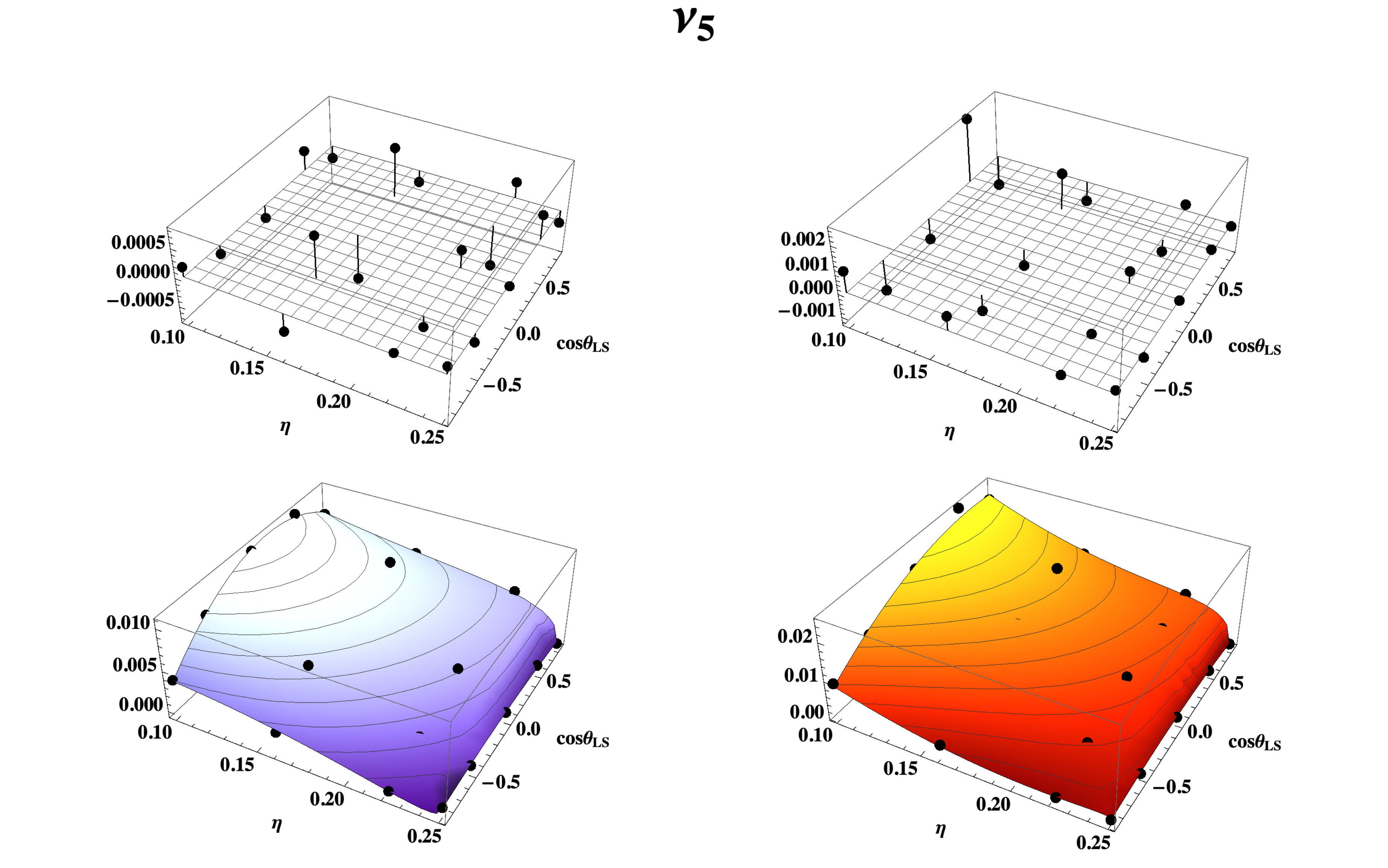}
      
      \\
      
      \includegraphics[width=0.5\textwidth]{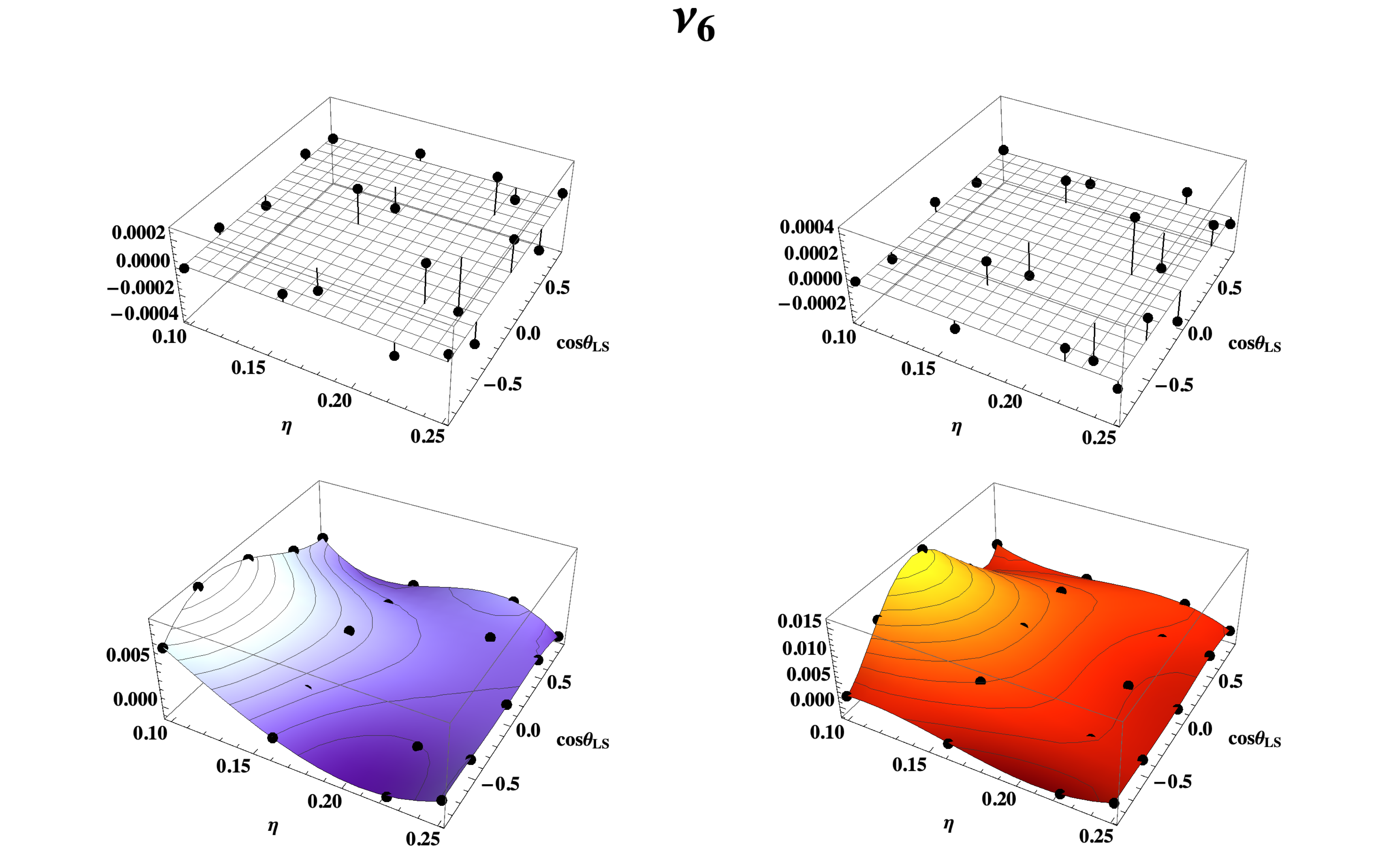}
      &
      \includegraphics[width=0.5\textwidth]{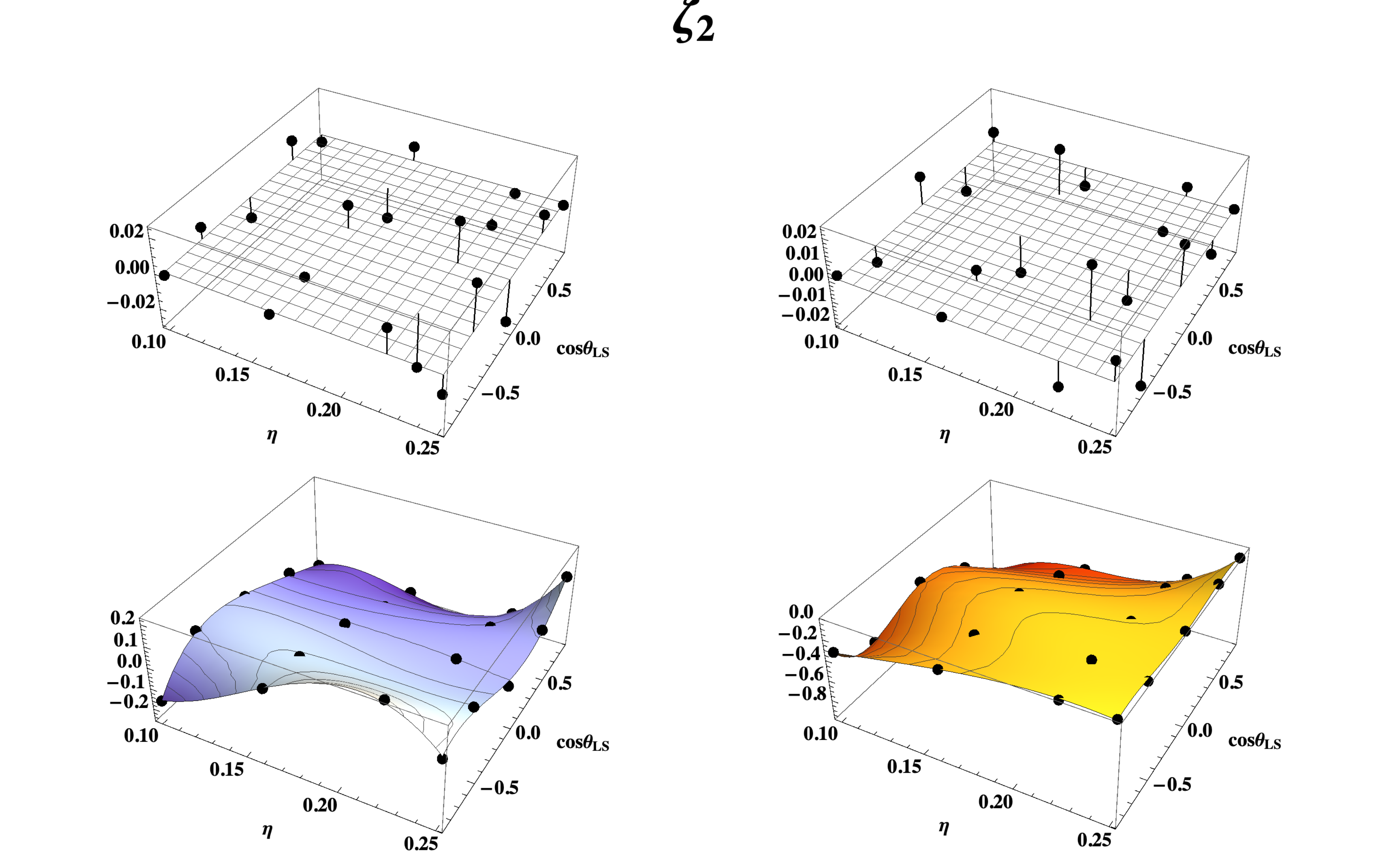}
      
   \end{tabular}
   \caption{Phase parameters for tuned co-precessing waveform model, \textsc{PhenomDCP}. The fits are shown as two-dimensional surfaces
    covering the parameter space described by $\eta$ and $\cos\tls$. On the left in blue are the fits for the simulations with $\chi=0.4$ and on the 
    right in red are the fits for $\chi=0.8$. Above each of these surfaces are shown the residuals.}
   \label{fig:dcpnu} 
\end{figure*}

\begin{figure*}[htbp]
   \centering
   \includegraphics[width=\textwidth]{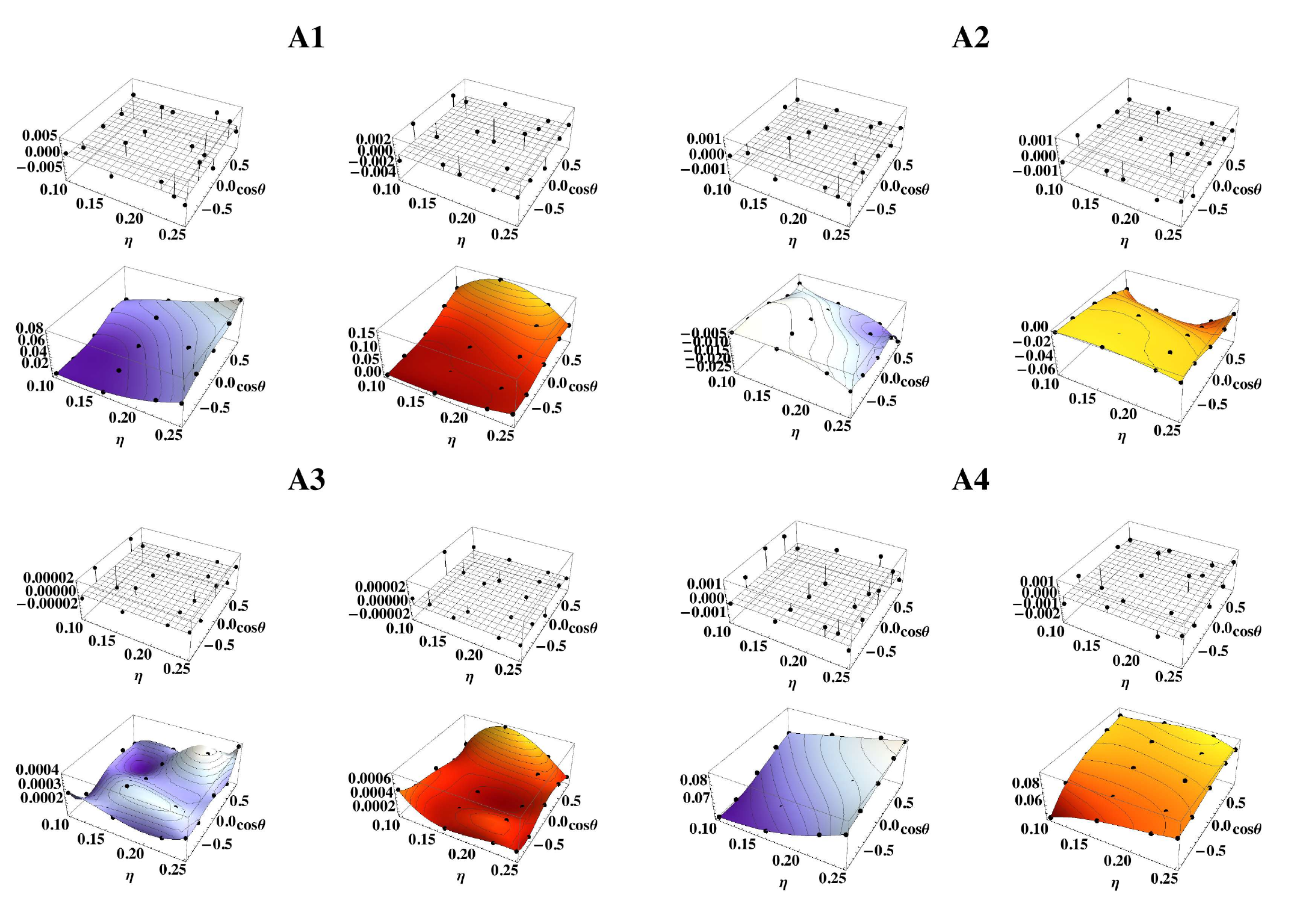}
   \caption{Comparison of the fits for each of the co-efficients for the ansatz for $\alpha$ given in equation \ref{eqn: alpha} with the co-efficients found 
   from the data as described in \sect{sec:angle model MR}. The fits are shown as two-dimensional surfaces covering the parameter space described 
   by $\eta$ and $\cos\tls$. On the left in blue are the fits for the simulations with $\chi=0.4$ and on the right in red are the fits for $\chi=0.8$. Above each 
   of these surfaces are shown the residuals.
   }
   \label{fig: alpha coefficients}
\end{figure*}

\begin{figure*}[htbp]
   \centering
   \includegraphics[width=\textwidth]{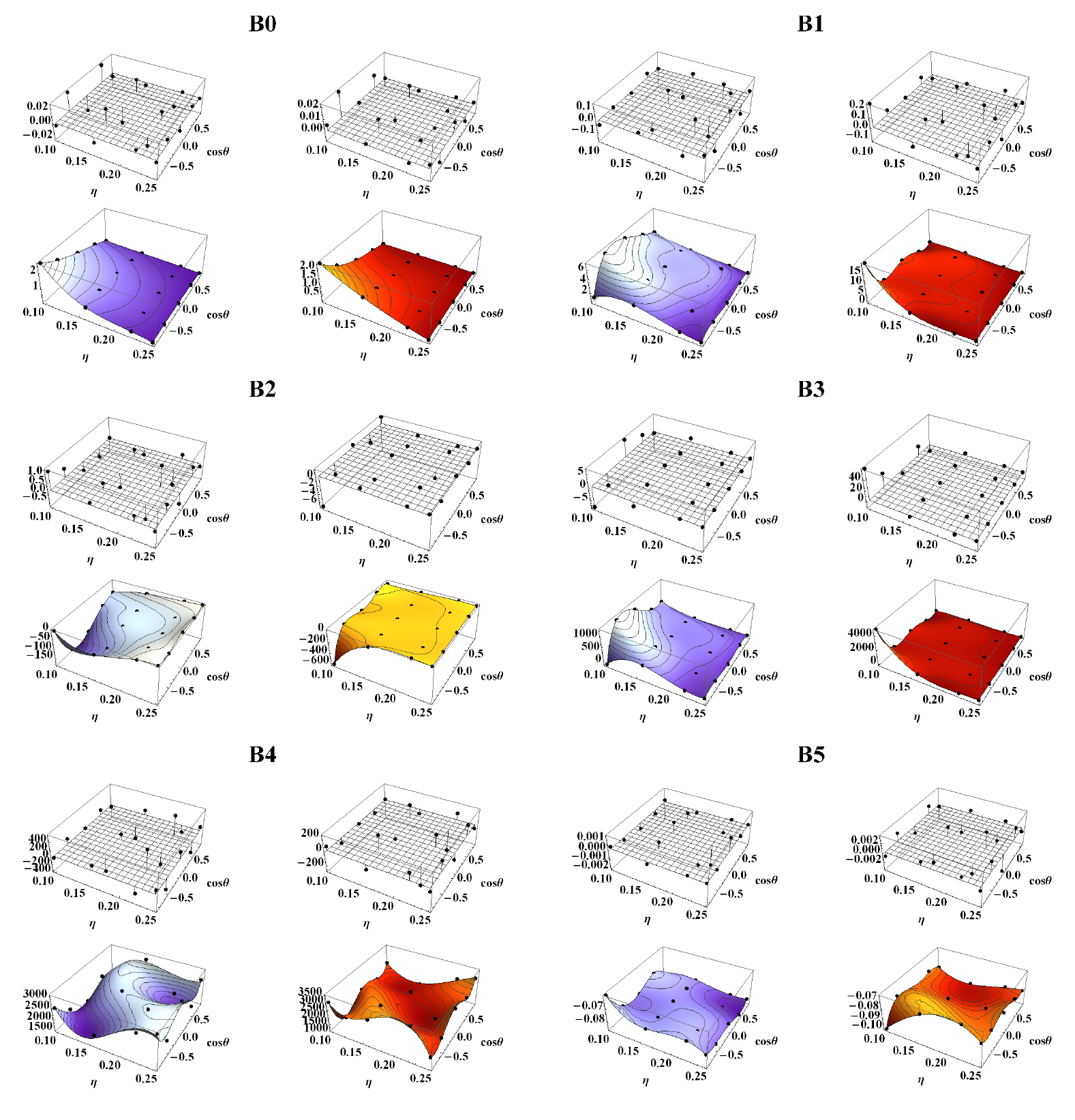}
   \caption{Comparison of the fits for each of the co-efficients for the ansatz for $\beta$ given in equation \ref{eqn: beta} with the co-efficients found 
   from the data as described in \sect{sec:angle model MR}. The fits are shown as two-dimensional surfaces covering the parameter space described 
   by $\eta$ and $\cos\tls$. On the left in blue are the fits for the simulations with $\chi=0.4$ and on the right in red are the fits for $\chi=0.8$. Above 
   each of these surfaces are shown the residuals.
   }
   \label{fig: beta coefficients}
\end{figure*}

Here we show how each of the co-efficients that appear in \pnr{} vary across the parameter space. Figs.~\ref{fig:dcpmu} and \ref{fig:dcpnu} show the 
variation of the co-efficients which appear in \dcp{}, as described in~\sect{sec:coprecessing model}. Fig.~\ref{fig: alpha coefficients} shows the 
co-efficients in the ansatz for $\alpha$ and Fig.~\ref{fig: beta coefficients} shows those in the ansatz for $\beta$, which are presented 
in~\sect{sec:angle model MR}. As can be seen from these figures, the co-efficients vary smoothly across the parameter space.

\bibliography{references.bib}

\end{document}